\documentclass[letter,11pt]{article}
\pdfoutput=1
\usepackage{heppub} 
\allowdisplaybreaks
\usepackage{hyperref}
\usepackage{graphicx}
\usepackage{amsmath}
\usepackage{amssymb} 
\usepackage[normalem]{ulem}
\usepackage{xspace}
\usepackage{slashed}
\usepackage[utf8]{inputenc}
\usepackage[T1]{fontenc}
\usepackage{subcaption}
\usepackage{scalerel,stackengine}

\newcommand*\diff{\mathop{}\!\mathrm{d}}

\newcommand{\nn}{\nonumber}

\newcommand{\Pythia}{\textsc{Pythia}\xspace}
\newcommand{\Vincia}{\textsc{Vincia}\xspace}

\DeclareRobustCommand{\Tab}[1]{Table~\ref{#1}}

\DeclareRobustCommand{\Fig}[1]{Fig.~\ref{#1}}

\DeclareRobustCommand{\Eq}[1]{Eq.~(\ref{#1})}
\DeclareRobustCommand{\Eqs}[2]{Eqs.~(\ref{#1}) and (\ref{#2})}

\DeclareRobustCommand{\Ref}[1]{Ref.~\cite{#1}}

\newcommand\pig[1]{\scalerel*[5pt]{\big#1}{%
    \ensurestackMath{\addstackgap[1.5pt]{\big#1}}}}
\def\eqn#1{Eq.~(\ref{eq:#1})}
\def\eqns#1#2{Eqs.~(\ref{eq:#1}) and~(\ref{eq:#2})}
\def\fig#1{Fig.~{\ref{fig:#1}}}
\renewcommand{\sec}[1]{Section~\ref{sec:#1}}

\newcommand{\ma}{\mathrm}
\newcommand{\ml}{\mathcal}
\newcommand{\be}{\begin{equation}}
\newcommand{\ee}{\end{equation}}
\newcommand{\bea}{\begin{eqnarray}}
\newcommand{\eea}{\end{eqnarray}}
\newcommand{\Dms}{\Delta m^2}

\newcommand{\zcuta}{z_{{\rm cut}1}}
\newcommand{\zcutb}{z_{{\rm cut}2}}

\newcommand{\Qcuti}{Q_{{\rm cut}\,i}}
\newcommand{\Qcuta}{Q_{{\rm cut}1}}
\newcommand{\Qcutb}{Q_{{\rm cut}2}}

\preprint{\vbox{\hbox{MIT--CTP 5365}}}

\title{Pure Quark and Gluon Observables in Collinear Drop}

\author{Iain W. Stewart and Xiaojun Yao}

\affiliation{Center for Theoretical Physics, Massachusetts Institute of Technology, Cambridge, MA 02139, USA}

\emailAdd{iains@mit.edu, xjyao@mit.edu}

\abstract{
We construct a class of pure quark and gluon observables by using the collinear drop grooming technique. The construction is based on linear combinations of multiple cumulative distributions of the jet mass in collinear drop, whose specific weights are fully predicted perturbatively.
This yields observables which obtain their values purely from quarks (or purely from gluons) in a wide region of phase space. 
We demonstrate this by showing that these observables are effective in two phase space regions, one dominated by perturbative resummation and one dominated by nonperturbative effects. 
The nonperturbative effects are included using shape functions which only appear as a common factor in the linear combinations constructed.
We test this construction using a numerical analysis with next-to-leading logarithmic resummation and various shape function models, as well as analyzing these observables with \Pythia and \Vincia. 
Choices for the collinear drop parameters are optimized for experimental use.
}

\begin{document}
\maketitle

\section{Introduction}
\label{sec:intro}

Collimated sprays of energetic particles called jets are produced in high energy lepton and hadron collisions. Because of the large total energy the jets contain, they are produced by initial hard scattering processes that involve quarks and gluons, followed by further partonic radiation and finally hadronization. Thus by studying jets in collider experiments, we can learn about both the perturbative and nonperturbative aspects of Quantum Chromodynamics (QCD), the theory of strong interaction. Jet observables are important for testing our understanding of the perturbative aspect of QCD and factorization as well as for measurements of the strong coupling constant and parton distribution functions.

In recent years, jet substructure observables have drawn a great amount of interest and been investigated widely~\cite{Butterworth:2008iy,Ellis:2009su,Abdesselam:2010pt,Altheimer:2012mn,Larkoski:2013eya,Altheimer:2013yza,Adams:2015hiv,Larkoski:2015kga,Moult:2016cvt,Moult:2017okx,Larkoski:2017jix,Andrews:2018jcm,Dasgupta:2018nvj,Asquith:2018igt,Marzani:2019hun}. These observables look into the internal structure of the jet and study how the energy and particles are distributed. The construction of jet substructure observables usually takes two steps: applying jet grooming to remove soft radiation and then defining interesting observables to measure on the groomed jet, such as prong-finders or other distributions. By reducing the sensitivity of the observables to the soft physics we also reduce the impact of nonperturbative effects, and thus increase the reliability of perturbative calculations. Our understanding of the groomed jet dynamics can then be tested by comparing calculations with experimental measurements. Furthermore, jet substructure observables can serve as useful probes of the quark-gluon plasma in heavy-ion collisions~\cite{Mehtar-Tani:2016aco,Casalderrey-Solana:2019ubu,Brewer:2021hmh}, see e.g. Ref.~\cite{Qin:2015srf} for a recent review. Finally, jet substructure observables provide tools to distinguish quark- and gluon-initiated jets, which we will explore here.

In general, jet and jet substructure observables $x$ contain contributions from both quark- and gluon-initiated jets:
\be
p(x) = f_q\,  p_q(x) + f_g\, p_g(x)\,,
\ee
where $f_{q,g}$ denotes the quark and gluon fraction in the jet sample and $p_{q,g}(x)$ is the distribution of the observable for a quark- or gluon-initiated jet (we will call them quark and gluon jets for simplicity from now on). A given experimental measurement only gives access to the sum of these two contributions, $p(x)$.  The physics goal of disentangling quark and gluon jets aims at separating the two contributions from each other and extract the individual fractions and underlying quark and gluon distributions. By disentangling quark and gluon jets, we can increase the sensitivity in searches of the physics beyond Standard Model which may couple more strongly to either quarks or gluons~\cite{FerreiradeLima:2016gcz,Bhattacherjee:2016bpy}. Furthermore, we can use separated quark and gluon jet samples to better constrain parton shower generators, and better probe the quark-gluon plasma~\cite{Brewer:2020och}. Many studies have been devoted to realize the discrimination between quark and gluon jets~\cite{Gallicchio:2011xq,Gallicchio:2012ez,Larkoski:2013eya,Larkoski:2014pca,ATLAS:2014vax,Komiske:2016rsd,Frye:2017yrw,Gras:2017jty,Davighi:2017hok,Larkoski:2019nwj,CMS:2020plq}. Among these studies, two main categories of tools have been explored: jet shapes such as angularities and energy correlation functions~\cite{Gallicchio:2011xq,Gallicchio:2012ez,Larkoski:2013eya} and multiplicity based observables such as the ``soft drop multiplicity''~\cite{Frye:2017yrw}. In the former case, observables exhibit Casimir scaling at leading logarithmic accuracy. Deviations from the Casimir scaling start at next-leading logarithmic accuracy and have been systematically studied in~\cite{Larkoski:2014pca,Gras:2017jty}. In the latter case, the leading logarithmic behavior is already beyond Casimir scaling and becomes Poisson-like. Both tools are well motivated from theoretical analysis but neither can provide a 100\% efficiency in the discrimination. More recently, a data-driven method called jet topics~\cite{Metodiev:2018ftz,Komiske:2018vkc} has been proposed, which can extract quark- and gluon-initiated jets from experimental jet samples under certain general conditions.

\subsection{Review of Jet Topics}

Jet topics~\cite{Metodiev:2018ftz,Komiske:2018vkc} provide a method to assign an operational definition to the meaning of quark- or gluon-initiated jet samples, and study when these operational definitions agree with the fundamental distributions one would infer from a quantum field theory calculation that factorizes the jet dynamics from that of the hard collision. We start with two samples of jets $A$ and $B$. For example, the sample $A$ can be taken from a $Z$-jet event (tagged by a $Z$-boson) while the sample $B$ is a dijet event. Instead of looking at the observable $x$ at a particular value, we study the experimentally measured distributions of the observable for both samples
\bea \label{pApB}
p_A(x) &=& f_q^A p_q(x) + f_g^A p_g(x) \,,  \\
p_B(x) &=& f_q^B p_q(x) + f_g^B p_g(x) \,.  \nonumber
\eea
Here $f_{q,g}^{A,B}$ is the quark or gluon fraction in each sample, and $p_{q,g}(x)$ denotes the individual quark or gluon jet substructure distribution which is assumed to be independent of the hard processes $A$ and $B$ producing the quark or gluon jets. The fractions are normalized such that
\be
\label{eq:norm}
f_q^A + f_g^A = 1\,, \quad \quad f_q^B + f_g^B = 1 \,.
\ee
Without loss of generality, we can assume $f_q^A > f_q^B$. 

By just using the experimental data, jet topics give operational definitions for quark and gluon jets:
\begin{align}
p_{T1}(x) &= \frac{ p_A(x) - \kappa(A|B) p_B(x) }{ 1-\kappa(A|B) } \,, 
 && p_{T2}(x)  = \frac{ p_B(x) - \kappa(B|A) p_A(x) }{ 1-\kappa(B|A) } \,, 
\end{align}
where the reducibility factors are defined by
\begin{align}
\kappa(A|B) &= \min_x \frac{p_A(x)}{p_B(x)} \,,
 && \kappa(B|A) = \min_x \frac{p_B(x)}{p_A(x)} \,. 
\end{align}
The same formulas also define reducibility factors for quarks and gluons, $\kappa(q|g)$ and $\kappa(g|q)$. 

If the condition of mutual irreducibility is satisfied:
\be
\kappa(q|g) = \kappa(g|q) = 0 \,,
\ee
then using \Eq{pApB} one finds $\kappa(A|B) = f_g^A/f_g^B$ and $\kappa(B|A)= f_q^B/f_q^A$~\cite{Metodiev:2018ftz,Komiske:2018vkc}. This then implies that the two operationally defined distributions $p_{T1}(x)$ and $p_{T2}(x)$ exactly correspond to the fundamental quark and gluon distributions respectively, i.e., 
\be
p_{T1}(x) = p_q(x)\,, \quad\quad  p_{T2}(x) = p_g(x) \,,
\ee
The mutual irreducibility implies the existence of ``anchor'' bins of the observable $x$ at which either the quark or the gluon contribution vanishes.

In practice, however, most jet and jet substructure observables do not satisfy the mutual irreducibility. For example, the jet mass after soft drop  (SD)~\cite{Dasgupta:2013ihk,Larkoski:2014wba} grooming at leading logarithmic (LL) accuracy gives
\be
\kappa(g|q) = 0 \,, \quad\quad \kappa(q|g) = \frac{C_F}{C_A} \,,
\ee
where $C_{F,A}$ is the Casimir in the fundamental or the adjoint representation of SU$(3)$. Without the mutual irreducibility, we can still connect the operational definitions with the fundamental objects. In the example of the soft drop jet mass, $p_{T2}(x)$ still corresponds to the pure gluon distribution $p_g(x)$. But $p_{T1}(x)$ gives the gluon-subtracted quark distribution:
\be
p_{q|g}(x) = \frac{p_q(x) - \kappa(q|g) p_g(x)}{1 - \kappa(q|g)}\,,
\ee
which can be solved to obtain $p_q(x)$ if we know $\kappa(q|g)$. Therefore, in practice, without the mutual irreducibility one must take a theoretical input calculation in order to carry out the disentangling procedure.

Another practical challenge when applying jet topics is that the ``anchor'' bins are usually defined in a very limited phase space region involving the tail regions of the distributions, where experimental uncertainties are often large. To help address these practical difficulties (tails of distributions and experimental uncertainties) in applying the jet topics, we want to explore improving the method by finding pure quark and gluon observables in a region of phase space. By this we mean that if the jet sample were to only contain quark (gluon) jets, it will lead to a vanishing result for the pure gluon (quark) observable.

\subsection{Disentangling Jets with Pure Quark and Gluon Observables}

To see how distinguishing quark and gluon jets can be simplified in this case, assume we can construct pure quark and gluon observables $\ml{Q}$ and $\ml{G}$ which are active over a significant region of phase space for an observable $y$. By measuring these observables in both jet samples $A$ and $B$ mentioned above, we obtain
\bea
&& \ml{Q}_A(y) = f_q^A \ml{Q}(y)\,, \quad\quad \ml{G}_A(y) = f_g^A\ml{G}(y) \,, \nn \\
&& \ml{Q}_B(y) = f_q^B \ml{Q}(y)\,, \quad\quad \ml{G}_B(y) = f_g^B\ml{G}(y) \,.
\eea
By taking ratios of the experimentally measured observables $\ml{Q}_{A,B}$ and $\ml{G}_{A,B}$, we can then obtain the ratios of the quark and gluon fractions in the samples $A$ and $B$
\be
\label{eq:fraction_ratio}
\frac{f_q^A}{f_q^B} = \frac{\ml{Q}_A(y)}{\ml{Q}_B(y)} \,, \quad \quad \frac{f_g^A}{f_g^B} = \frac{\ml{G}_A(y)}{\ml{G}_B(y)} \,.
\ee
Note that this strategy comes with a built in consistency test by confirming that these ratios are $y$ independent in the expected phase space region.
Together with the normalization conditions (\ref{eq:norm}), we can then solve for the quark and gluon fractions in both samples. Once we have obtained the quark and gluon fractions, we can then also solve for individual distributions $p_q(x)$ and $p_g(x)$ for a given observable $x$ from \Eq{pApB}.

This motivates us to think about constructing pure quark and gluon observables. In this paper, we present a construction  using the collinear drop (CD) grooming procedure~\cite{Chien:2019osu}. The construction is based on the jet mass observable and relies on both the perturbative and nonperturbative features of the CD jet mass.
A novel feature of the spectrum in the CD jet mass $\Delta m^2$ that we will exploit is that it perturbatively goes to a constant as the CD cumulative jet mass $\Delta m_c^2\to 0$. In contrast most jet distributions vanish in this limit due to the presence of Sudakov exponential that gives zero probability for the emission of no radiation. For the CD jet mass the measurement is made on an intermediate soft region of phase space and a finite number of different events have $\Delta m^2=0$, and furthermore this number depends sensitively on both the CD parameters and whether the jet was initiated by a quark or gluon. Our construction of pure quark and gluon observables will exploit factorization based predictions for the full $\Delta m_c^2$ spectrum, as we will explain in the following. 

The paper is organized as follows: In Sec.~\ref{sec:jetmass}, we will review the observable of jet mass in collinear drop. In Sec.~\ref{sec:nonper} nonperturbative effects are discussed for the cumulative collinear drop jet mass.  The construction of the pure quark and gluon observables is given in Sec.~\ref{sec:obs}, together with results and analysis. Finally, we will summarize and draw conclusions in Sec.~\ref{sec:conclusions}.

\section{Review of Jet Mass in Collinear Drop}
\label{sec:jetmass}
\subsection{Observable}
In this paper, we consider collinear drop observables in proton-proton ($pp$) collisions that are defined by two soft drop procedures with the parameters $(\zcuta,\beta_1)$ and $(\zcutb,\beta_2)$. 

In the soft drop grooming procedure with the parameters $(z_{{\rm cut}},\beta)$~\cite{Larkoski:2014wba}, we start with a jet of radius $R$ constructed from a jet algorithm such as the anti-$k_T$ algorithm, and recluster all the particles in the jet using the Cambridge-Aachen (C/A) algorithm. The C/A algorithm first recombines particles $i$ and $j$ with the smallest relative angular distance
\be
\Delta R_{ij} = \frac{2p_i\cdot p_j}{p_{Ti}\, p_{Tj}} = 2\cosh(y_i-y_j) - 2\cos(\phi_i - \phi_j) \approx (\phi_i - \phi_j)^2 + (y_i-y_j)^2 \,,
\ee
where $y_i$ and $y_j$ denote the rapidity of the two particles and $\phi_i$ and $\phi_j$ are their azimuthal angles. The approximation sign is valid in the limit of small $R$ and often is just used as the definition for all $R$ values. In the same limit,
\be
\Delta R_{ij}   \approx \theta_{ij} \cosh \eta_J\,,
\ee
where $\theta_{ij}$ is the angle between particles $i$ and $j$, and $\eta_J$ is the pseudorapidity of the jet. The reclustering leads to a tree of particles that are ordered by the relative angular distance, so that the branching with the largest relative angular distance occurs earliest in the tree. This is also consistent with the branching tree of the jet at leading logarithmic (LL) accuracy. Then we sweep through each branching point in the reclustered tree and keep removing the softer sub-branch until the following condition is satisfied:
\be
\frac{{\rm min}(p_{Ti},p_{Tj})}{p_{Ti} + p_{Tj}} > z_{{\rm cut}} \Big( \frac{\Delta R_{ij}}{R_0} \Big)^\beta \approx \tilde{z}_{\rm{cut}}\, \theta_{ij}^\beta\,,
\ee
where 
\be
\tilde{z}_{\rm{cut}} = z_{\rm{cut}} \Big( \frac{\cosh\eta_J}{R_0} \Big)^\beta
\ee
and $R_0$ is another parameter that sets the typical angular distance in the soft drop grooming procedure. We choose $R_0=1$ throughout the paper.

The collinear drop~\cite{Chien:2019osu} sample is then constructed from a jet defined by a jet finding algorithm by first applying a soft drop grooming with the parameters $(\zcuta,\beta_1)$, and then applying an anti-soft drop step, by removing the particles that pass a second soft drop procedure with the parameters $(\zcutb,\beta_2)$. To guarantee a finite number of particles are left after the two steps, we require the second soft drop grooming is more aggressive than the first one, which can be implemented by taking $\zcuta \leq \zcutb$ and $\beta_1 \geq \beta_2$. The collinear drop groomed jet is then the complement of the second soft drop groomed jet in the first soft drop groomed jet: ${\rm jet}_{\rm CD} = {\rm jet}_{{\rm SD}1} \setminus {\rm jet}_{{\rm SD}2} $. The observable we consider in this paper is the jet mass in collinear drop, which is defined by
\begin{align}
\label{eq:jet_mass}
\Dms &= m_{{\rm SD}1}^2 - m_{{\rm SD}2}^2 \,,
  \qquad \textrm{where }\ 
  m_{{\rm SD}\,i}^2 = p_{{\rm SD}\,i}^2 
     = \Big( \sum_{k \in {\rm jet}_{{\rm SD}\,i}} p_k^\mu \Big)^2\,.
\end{align}
The definition in Eq.~(\ref{eq:jet_mass}) is consistent with the following definition
\be
\Dms = \Big( \sum_{k \in {\rm jet}_{\rm CD}} p_k^\mu \Big)^2\,,
\ee
since the jet mass is a linear observable up to power corrections
\be
\Dms = (p_{{\rm SD}1} + p_{{\rm SD}2}) \cdot (p_{{\rm SD}1} - p_{{\rm SD}2}) = Qn \cdot (p_{{\rm SD}1} - p_{{\rm SD}2}) + \cdots\,,
\ee
where $Q$ is the energy scale in the initial hard vertex that generates the jet and $n=(1,\hat n_J)$ is a light-like four-vector encoding the jet direction $\hat n_J$. The value of $Q$ is given by twice the jet energy $Q=2E_J$ or for $pp$ collisions one has $Q=2 p_T \cosh\eta_J$ where $p_T$ is the jet transverse momentum. 

Within collinear drop the goal of having soft jet grooming in addition to the removal of collinear particles is to limit the impact of underlying event and other soft contamination in the jet. Such radiation is also impacted by the choice of the jet radius $R$, and we will find it useful to consider jets with $R=0.2$ for our construction.

\subsection{Relevant Modes in SCET}
\label{eq:modes}

We primarily consider the collinear jet mass in the hierarchical limit
\begin{align}
  \label{eq:CDhierarchy}
\frac{\Dms}{(p_TR)^2} \ll z_{{\rm cut}\,i} \ll 1 \,.
\end{align}
Due to this hierarchy of scales and parameters, large logarithms need to be resummed. To carry out the resummation, a factorization formula for the collinear drop jet mass cross section has been constructed~\cite{Chien:2019osu} by using the Soft-Collinear Effective Theory (SCET)~\cite{Bauer:2000ew,Bauer:2000yr,Bauer:2001ct,Bauer:2001yt,Bauer:2002nz}, which is based on the factorization formula of energy correlators in soft drop~\cite{Frye:2016okc,Frye:2016aiz}.  SCET has been widely applied in studies of jet physics, see e.g.~\cite{Fleming:2007xt,Becher:2008cf,Stewart:2010tn,Stewart:2010pd,Ellis:2010rwa,Abbate:2010xh,Bauer:2011uc,Stewart:2013faa,Stewart:2014nna,Chien:2015cka,Chien:2015ctp,Becher:2015hka,Kang:2016mcy,Kang:2016ehg,Kolodrubetz:2016dzb,Moult:2016fqy,Chien:2016led,Moult:2017jsg,Kang:2018jwa,Ebert:2018lzn,Moult:2018jjd,Chien:2018lmv,Kang:2018vgn,Hoang:2019ceu,Kang:2019prh,Chien:2019gyf,Vaidya:2020cyi,Pathak:2020iue,Vaidya:2020lih,Vaidya:2021mjl}.

Our analysis at next-to-leading logarithmic (NLL) order will be set up such that it remains valid when the hierarchies in \eqn{CDhierarchy} are relaxed.
This is achieved following Ref.~\cite{Chien:2019osu}.

\begin{figure}[t]
a) \hspace{7.6cm} b) \\[-5pt]
        \includegraphics[height=1.8in]{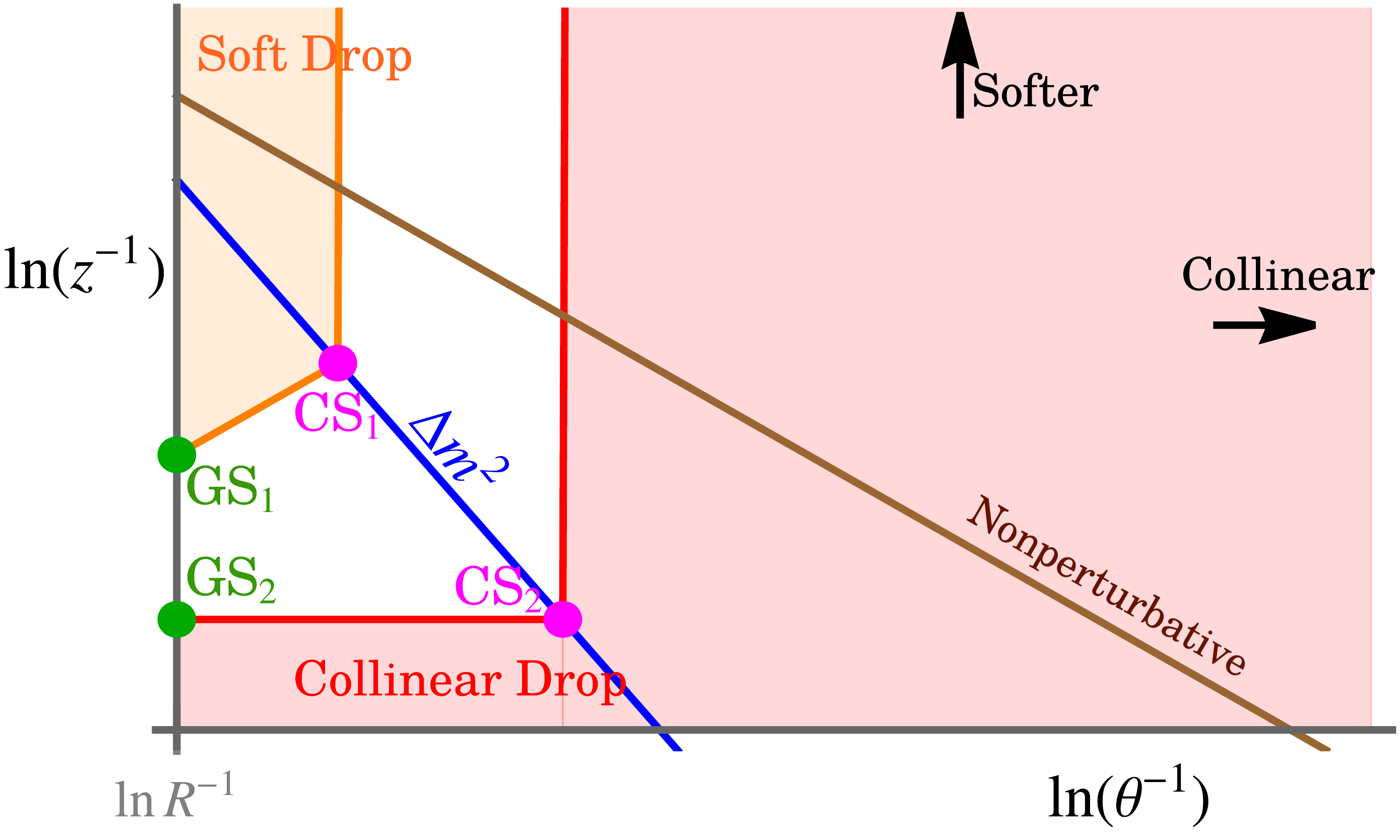}
   \hspace{0.5cm}
        \includegraphics[height=1.8in]{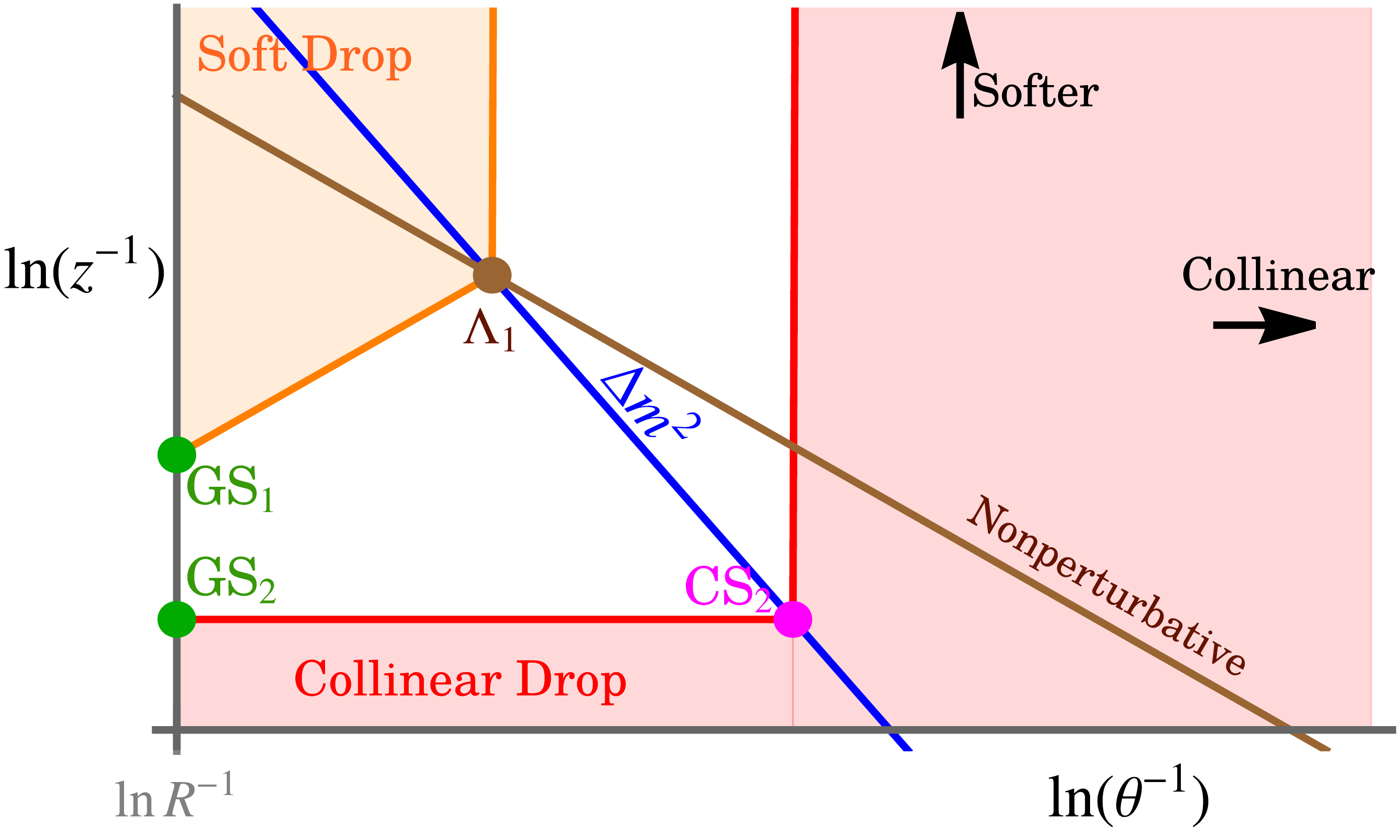} \\
c) \hspace{7.6cm} d) \\[-5pt]    
        \includegraphics[height=1.8in]{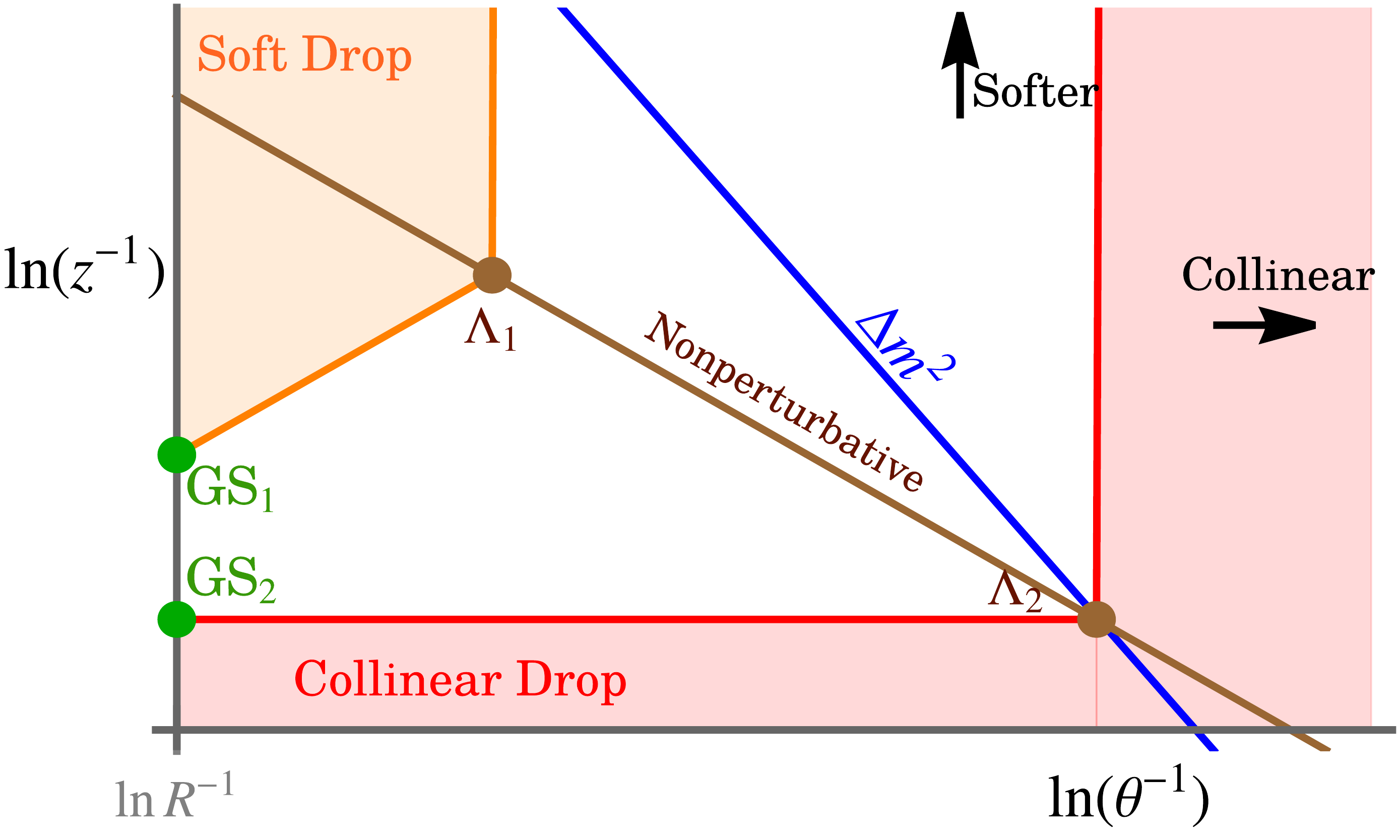}
  \hspace{0.5cm}
        \includegraphics[height=1.8in]{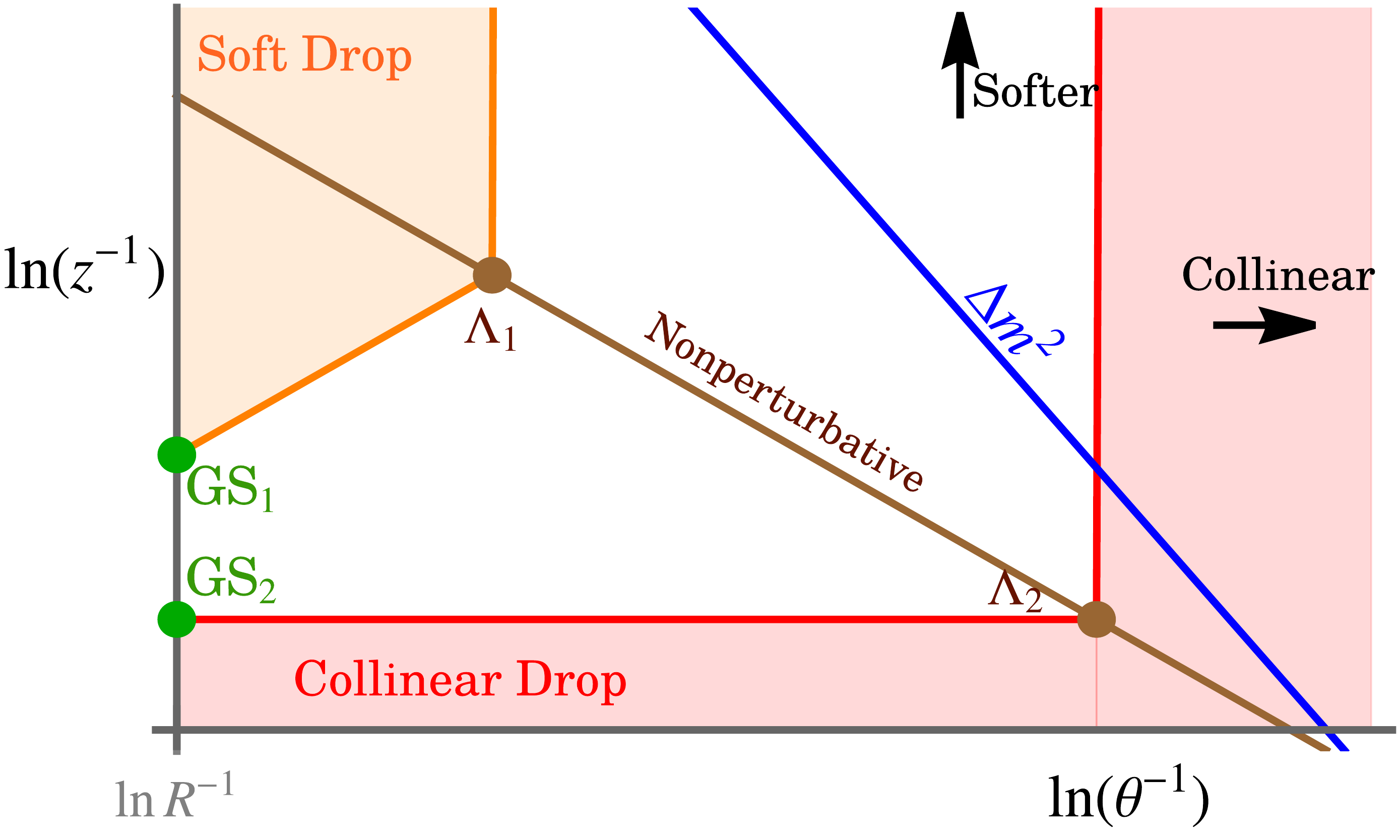}
\caption{Relevant modes of the factorization formula of jet mass in collinear drop in the $(\ln(\theta^{-1}),\ln(z^{-1}))$ plane.  The orange and red shaded areas indicate phase space regions removed by the collinear drop procedure. The solid brown line indicates the onset of nonperturbative contributions. The solid blue line represents the measurement of the collinear drop squared jet mass $\Delta m^2$. The value of $\Delta m^2$ decreases as we go from panel a) to d). In panel a) it is in a perturbative regime, in panel b) the CS$_1$ mode transitions to being nonpertubative, in panel c) the CS$_2$ mode transitions to being nonperturbative, and in panel d) we are deep in the nonperturbative regime.}
\label{fig:modes}
\end{figure}

The factorization formula consists of several functions that represent specific modes contributing to the collinear drop jet mass. Each mode is specified by the scaling of its lightcone momentum $p^\mu = (p^+,p^-,p_\perp) = (n\cdot p,\, \bar{n}\cdot p,\, p_\perp)$ where $n=(1,\hat{n}_J)$ is the light-like vector pointing in the direction of the jet and $\hat{n}_J$ is along the jet axis in space. The four vector $\bar{n}$ is auxiliary and satisfies $n\cdot \bar{n}=2$. It is usually chosen as $\bar{n} = (1, -\hat{n}_J)$. By construction, the large component of the jet momentum is along the $p^-$ direction. We now discuss all modes relevant for the collinear drop jet mass. Each of them is depicted in the $(\ln \theta^{-1} ,\ln z^{-1} )$ plane in Fig.~\ref{fig:modes}, where $z$ represents the fraction of the lightcone energy carried by a parton $k$ in the jet $z=\frac{p^-_k}{Q}$, and $\sin\theta=\frac{|{p}_k^\perp|}{p_k^0}\approx \frac{2|{p}_k^\perp|}{p_k^-}\approx \theta$ measures the angle of the parton with respect to the jet axis. Since the axes are logarithmic, the relative locations in the plot indicate parametric scaling.  The modes for perturbative soft drop were determined in Ref.~\cite{Frye:2016aiz}, for perturbative collinear drop in Ref.~\cite{Chien:2019osu}, and for nonperturbative corrections to soft drop in Ref.~\cite{Hoang:2019ceu}. Here we review these results and introduce the relevant nonperturbative modes for collinear drop. 
Note that the simplicity of the soft drop grooming algorithm is important for ensuring that the enumeration of modes is not affected by the order of perturbation theory. 

Different situations arise depending on the value of the collinear drop jet mass observable $\Delta m^2$, as depicted by the four panels in Fig.~\ref{fig:modes}. 
The typical momentum of a parton inside the jet scales as
\be
p^\mu \sim zQ\Big( \frac{\theta^2}{4}, 1, \frac{\theta}{2} \Big)\,.
\ee
The parton is collinear if $\theta\ll1$ and soft if $z\ll1$. Since the jet is constructed initially with a radius $R$, the maximum value of $\theta$ is $\theta_{\rm max} \approx \frac{R}{\cosh \eta_J}$. For jets with zero pseudorapidity $\eta_J=0$, we have $\theta_{\rm max} \approx R$. 

The blue line in \fig{modes} represents the measurement of the collinear drop jet mass, and is determined by
\be
\label{eq:measure}
p^+Q \simeq zQ^2 \frac{\theta^2}{4} \simeq \Dms \,.
\ee
Here $z$ appears with a linear power since the definition of the jet mass involves a summation over all partons inside the jet. Rewriting \Eq{eq:measure} leads to the equation for the blue line
\be
\label{eq:measure2}
\ln\frac{1}{z} \simeq \ln\frac{Q^2}{4\Dms} - 2\ln\frac{1}{\theta} \,.
\ee
If there were no grooming, the intercept of the measurement line with the vertical axis gives the soft mode while the intercept with the horizontal axis gives the collinear mode. The collinear and soft modes appear in the factorization formula of the jet mass cross section, in the absence of any collinear or soft jet grooming.

The brown line in \fig{modes} represents where nonperturbative effects become important and is determined by when $p_\perp \sim \Lambda_{\rm QCD}$, so
\be
\label{eq:nonperturb}
zQ\frac{\theta}{2} \simeq \Lambda_{\rm QCD}\,.
\ee
Rewriting \Eq{eq:nonperturb} gives the equation for the brown line
\be
\ln\frac{1}{z} \simeq \ln\frac{Q}{2\Lambda_{\rm QCD}} - \ln\frac{1}{\theta} \,.
\ee

The orange and red lines in \fig{modes} represent the constraints introduced by the two soft drop procedures, and can be estimated from the criterion to just pass the soft drop procedure:
\be
\label{eq:sd_stop}
z \simeq \tilde{z}_{{\rm cut}\,i}\, \theta^{\beta_i} \,.
\ee
Rewriting \Eq{eq:sd_stop} leads to an equation for the soft drop or collinear drop grooming lines
\be
\label{eq:sd_stop2}
\ln\frac{1}{z} \simeq \ln\frac{1}{\tilde{z}_{{\rm cut}\,i}}
+ \beta_i \ln\frac{1}{\theta} \,.
\ee
For $i=1$ ($i=2$) only phase space that is below (above) this line can contribute to a collinear drop observable. However, as $\theta$ decreases and $\ln (\theta^{-1})$ increases, the grooming lines will cross either the line for the jet mass measurement or the nonperturbative line, and the picture will be modified by whichever of these happens first. After this point the space above the grooming lines can now contribute as indicated by the vertical orange and red lines. This is because in the C/A reclustering, the branching history is ordered by angles, from large to small. When the soft drop criterion (\ref{eq:sd_stop}) is reached, all remaining branches with smaller pairwise angles that are part of the subjets being compared are kept.

The collinear-soft (CS$_{1,2}$) modes indicated by magenta dots in \fig{modes} are located at the intercepts between the lines of soft drop constraints and the line of jet mass measurement.  The scaling of the collinear-soft momentum can be obtained by solving \Eqs{eq:measure}{eq:sd_stop}, which leads to
\begin{align}
\frac{\theta_{{\rm cs}\,i}}{2} &\sim \Big( \frac{1}{\tilde{z}_{{\rm cut}\,i}}\frac{\Dms}{2^{\beta_i}Q^2} \Big)^{\frac{1}{2+\beta_i}} = \Big( \frac{\Dms}{QQ_{{\rm cut}\,i}} \Big)^{\frac{1}{2+\beta_i}}
\,, 
& z_{{\rm cs}\,i} &\sim\frac{\Dms}{Q^2} \frac{4}{\theta_{{\rm cs}\,i}^2}
\,,
\end{align}
where we have defined $Q_{{\rm cut}\,i} = 2^{\beta_i} Q \tilde{z}_{{\rm cut}\,i}$. The collinear-soft scale with $Q_{{\rm cs}\,i}^2\sim p_{cs}^2\sim p_{cs}^{\perp\,2}$ is therefore given by
\be
\label{eq:mu_cs}
Q_{{\rm cs}\,i} = z_{{\rm cs}\,i} \,Q\, \frac{\theta_{{\rm cs}\,i}}{2} = \Big( \frac{\Dms}{Q} \Big)^{\frac{1+\beta_i}{2+\beta_i}} \Qcuti^{\frac{1}{2+\beta_i}} \,.
\ee
The intercepts of the orange and purple lines with the vertical axis are at $\theta = \theta_{\rm max} = \frac{R}{\cosh\eta_J}$, and give the global-soft (GS$_{1,2}$) modes, indicated by green dots. Plugging $\theta=\theta_{\rm max}$ into \Eq{eq:sd_stop} gives
\begin{align}
\theta_{{\rm gs}\,i} &= \frac{R}{\cosh\eta_J} 
 \,,
& z_{{\rm gs}\,i} &= \tilde{z}_{{\rm cut}\,i} \Big( \frac{R}{\cosh\eta_J} \Big)^{\beta_i} = z_{{\rm cut}\,i}\Big( \frac{R}{R_0} \Big)^{\beta_i} \,.
\end{align}
The global-soft scale is given by
\be
\label{eq:mu_gs}
Q_{{\rm gs}\,i} = z_{{\rm gs}\,i}\, Q\, \frac{\theta_{{\rm gs}\,i}}{2}
 = p_{T} R \, z_{{\rm cut}\,i}\Big( \frac{R}{R_0} \Big)^{\beta_i} \,.
\ee
Although it is possible to work with various scenarios for hierarchies or non-hierarchies between the two soft drop parameters (see Ref.~\cite{Chien:2019osu}), here we will only consider the hierarchical case where $z_{{\rm cut}1} \ll z_{{\rm cut}2}$ and $\theta_{{\rm cs}2} \ll \theta_{{\rm cs}1}$. This implies that there are individual collinear-soft and global-soft modes for $i=1$ and $i=2$ as shown in \fig{modes}.

Nonperturbative effects will be power suppressed if the line of the jet mass measurement is far away from the line of nonperturbative regime, as shown in \fig{modes}a. This occurs when the perturbative CS modes are far away from the nonperturbative regime.
\be
Q_{{\rm cs}\,i} \gg \Lambda_{\rm QCD} \,.
\ee
Nonperturbative effects on the $i$-th CS mode will become important when $Q_{{\rm cs}\,i} \sim \Lambda_{\rm QCD}$, which happens when the collinear drop jet mass becomes smaller than the critical value,  $\Delta m \lesssim \Delta m_{\Lambda\,i}$, where
\be
\label{eq:mass_in_nonpert}
\Delta m_{\Lambda\,i}^2 = \Lambda_{\rm QCD}\, Q\, \Big( \frac{\Lambda_{\rm QCD}}{Q_{{\rm cut}\,i}} \Big)^{\frac{1}{1+\beta_i}} \,.
\ee
We note that the scale of the jet mass in Eq.~(\ref{eq:mass_in_nonpert}) can still be much bigger than $\Lambda_{\rm QCD}$
\be
\frac{\Delta m_{\Lambda\,i}^2}{\Lambda_{\rm QCD}^2} = \Big( \frac{Q}{\Lambda_{\rm QCD}} \Big)^{\frac{\beta_i}{1+\beta_i}} \Big( \frac{1}{2^{\beta_i} \tilde{z}_{{\rm cut}\,i}}  \Big)^{\frac{1}{1+\beta_i}} \gg 1\,.
\ee
Since the collinear drop with $i=2$ has stronger grooming than the soft drop with $i=1$, as we decrease $\Delta m^2$ it is the CS$_1$ mode which will become nonperturbative first, as shown in \fig{modes}b. At this and smaller values of $\Delta m^2$ the CS$_1$ mode is replaced by the $\Lambda_1$ mode given by the intersection of the brown and orange lines, since there are always particles with such nonperturbative momenta available to stop soft drop. At a smaller value of $\Delta m^2$ the CS$_2$ mode becomes nonperturbative, and is replaced by the $\Lambda_2$ mode as shown in \fig{modes}c. 
The scaling of these nonperturbative collinear-soft modes can be obtained by setting $Q_{{\rm cs}\,i} \sim \Lambda_{\rm QCD}$ in \Eq{eq:mu_cs}, which gives
\be
p_{\Lambda\,i}^\mu \sim \Lambda_{\rm QCD} \Big(\frac{\theta_{{\rm cs}\,i}}{2}, \frac{2}{\theta_{{\rm cs}\,i}}, 1 \Big) \,.
\ee
The scale of these nonperturbative modes is of course $p_{\Lambda\,i}^2 \sim \Lambda_{\rm QCD}^2$.

Our explicit construction of pure quark and gluon observables will be carried out for both the perturbative regions in \fig{modes}a and the nonperturbative regions in \fig{modes}c,d. Our construction of these observables will be valid for the final case in \fig{modes}b, by interpolation from the surrounding cases.

\subsection{Factorization and Summation of Large Logs} 
\label{sec:resum}

In this subsection, we will review the factorization of the collinear drop jet mass cross section in the purely perturbative regime $Q_{{\rm cs}\,i} \gg \Lambda_{\rm QCD}$, which corresponds to the case in \fig{modes}a. We will discuss the other cases where the jet mass acquires ${\cal O}(1)$ nonperturbative corrections in more detail in Section~\ref{sec:nonper}. We also only consider the case with fully hierarchical global-soft and collinear-soft modes, and assume that $R/2\ll 1$, which is a reasonable approximation for most realistic values of the jet radius $R$ (even working well in practice for the case $R=1$). 

Since the global-soft modes are far away from the line of the collinear drop jet mass measurement, the shape of the jet mass cross section is independent of the GS modes. The GS modes only modify the overall normalization of the cross section. This is also manifest in \Eq{eq:mu_gs} where the GS scale is independent of the jet mass. The factorization formula of the collinear drop jet mass differential cross section is~\cite{Frye:2016aiz,Chien:2019osu}:
\be
\label{eq:differential_sigma}
\frac{\diff \sigma}{\diff \Dms} = \sum_{j=q,g} N_j^{\rm CD}(p_T,\eta_J, R, \tilde{z}_{{\rm cut}\,i},\beta_i,\mu)\, P_j^{\rm CD} (\Dms, Q, \tilde{z}_{{\rm cut}\,i},\beta_i,\mu) \,,
\ee
where the sum over $j$ adds contributions from the quark- and gluon-initiated jets. In different processes (such as dijets and $Z$ boson-jet events), the fraction of quark- and gluon-initiated contributions are different in general, and this process dependence is carried by the normalization factors $N_j^{\rm CD}$, which are independent of the collinear drop jet mass $\Dms$. 

For an explicit process the normalization factor $N_j^{\rm CD}$ can also be further factorized into hard, global-soft, and other contributions.  For collisions producing an identified groomed jet of radius $R$  factorized into a hard function and two global-soft functions we have
\be
\label{eq:Ncd}
N_j^{\rm CD}(p_T,\eta_J, R, \tilde{z}_{{\rm cut}\,i},\beta_i,\mu) 
  = H_j(p_T,\eta_J, R) \otimes_{\Omega} S_{G_j}(Q_{{\rm gs}1},R,\beta_1,\mu) 
   \otimes_{\Omega} S_{\overline{G}_j}(Q_{{\rm gs}2},R,\beta_2,\mu) \,.
\ee
The hard function $H_j$ determines the fraction of quark and gluon contributions (with the presence of gluon jets starting at ${\cal O}(\alpha_s)$). The global-soft functions $S_{G_j}$ and $S_{\overline{G}_j}$ encode contributions from the first and second global-soft modes respectively, as well as unmeasured soft function contributions from outside the jet. The overline in the second GS function emphasizes that in collinear drop, it is particles removed by the second soft drop that are kept, which results in an expression of $S_{\overline{G}_j}$ that is different from $S_{G_j}$. Finally the $\otimes_{\Omega}$ indicates that the angular integrals in the functions cannot be done independently since the modes have the same angular scaling, i.e., they cannot be distinguished by their angular separation. This generically leads to the presence of so-called non-global logarithms (NGL) that start at ${\cal O}(\alpha_s^2 \ln^2)$~\cite{Dasgupta:2001sh}.  
For both soft drop and collinear drop jet mass observables with hierarchical scales, the non-global logarithmic effects only appear in the normalization factors $N_j$. This occurs due to the fact that the spectrum dependent $P_j$ functions are factorized in terms of single scale functions~\cite{Frye:2016aiz,Chien:2019osu}.  In our observable these non-global effects only appear in the quark and gluon fractions, which are  treated as fixed numbers to be determined experimentally from our analysis.  Therefore we refrain from going into further detail about these 
effects.~\footnote{We also remark that for $R/2\ll 1$, there can be large $\ln R$ terms in the $N_j^{\rm CD}$ factors, which for the same reason we do not elaborate on here. All such $\ln R$ terms in $P_j^{\rm CD}$ are resummed together with other logarithms at the order of our analysis, again due to the single scale nature of the factorization.}
We refer the interested reader to the literature for further details on the calculation of these non-global effects in normalization factors, see~\cite{Kang:2019prh,Pathak:2020iue} and references therein.
For $e^+e^-$ collisions, $H_j(p_T,\eta_J,R)$ is given by a purely perturbative series.
For proton-proton collisions it contains convolutions with parton distribution functions, such as $H_j(p_T,\eta_J,R)= \sum_{a,b} f_a\otimes f_b\otimes H_{abj}(p_T,\eta_J,R)$, where $H_{abj}$ describes the hard dynamics of the partonic process $a+b\to j+X$.

At one-loop, the bare in-jet global-soft functions can be evaluated in dimensional regularization ($d=4-2\epsilon$) as~\cite{Chien:2019osu}
\bea \label{eq:SG-oneloop}
S_{G_j}(Q_{{\rm gs}1},\beta_1,\epsilon) &=& 
1+\frac{4g^2C_j\mu^{2\epsilon}e^{\epsilon\gamma_E}}{(4\pi)^\epsilon} \int \frac{\diff^dk}{(2\pi)^d} \frac{1}{k^+k^-} 2\pi\delta^+(k^2)\, \overline{\Theta}_{{\rm SD}1}^{({\rm gs})}\, \Theta_{\rm alg}
  \,, \\
S_{\overline{G}_j}(Q_{{\rm gs}2},\beta_2,\epsilon) &=& 1+\frac{4g^2C_j\mu^{2\epsilon}e^{\epsilon\gamma_E}}{(4\pi)^\epsilon} \int \frac{\diff^dk}{(2\pi)^d} \frac{1}{k^+k^-} 2\pi\delta^+(k^2) \big(-\overline{\Theta}_{{\rm SD}2}^{({\rm gs})}\big) \Theta_{\rm alg} \,,  \nn
\eea
where $\delta^+(k^2) = \delta (k^2) \theta(k^0)$, the color factor $C_q=C_F$ and $C_g=C_A$,  $\overline{\Theta}_{{\rm SD}\,i}^{({\rm gs})}$ means the global-soft mode fails the $i$-th SD criterion, and $\Theta_{\rm alg}$ represents the jet finding algorithm. In collinear drop, the first global-soft mode fails the SD criterion while the second passes. This is why the signs of the SD kinematic constraints in the two GS functions are opposite. The kinematic constraints of SD and the jet finding algorithm are given by
\begin{align}
\overline{\Theta}_{{\rm SD}\,i}^{({\rm gs})} &= \theta\Big( Q\tilde{z}_{{\rm cut}\,i} \Big(\frac{2k^+}{k^-}\Big)^{\frac{\beta_i}{2}} - k^+ - k^- \Big) \,,
 &\Theta_{\rm alg} &= \theta\Big(R^2 - 4\cosh^2\!\eta_J \, \frac{k^+}{k^-} \Big) \,.
\end{align}
Explicit calculations give
the $\overline{\rm MS}$ renormalized GS functions as~\cite{Frye:2016aiz}  
\bea
S_{G_j}(Q_{{\rm gs}1},\beta_1,\mu) &=& 1+\frac{\alpha_s(\mu)C_j}{\pi(1+\beta_1)} \Big( 
\ln^2\frac{\mu}{Q_{{\rm gs}1}} - \frac{\pi^2}{24}\Big) \,, \\
S_{\overline{G}_j}(Q_{{\rm gs}2},\beta_2,\mu) &=& 1 - \frac{\alpha_s(\mu)C_j}{\pi(1+\beta_2)} \Big( 
\ln^2\frac{\mu}{Q_{{\rm gs}2}} - \frac{\pi^2}{24}\Big) \,. \nonumber
\eea
They satisfy the following general renormalization group (RG) equations that are valid even at higher loops
\bea
\frac{\diff}{\diff\ln\mu} \ln S_{G_j}(Q_{{\rm gs}1},\beta_1,\mu)
  &=& \frac{2C_j}{1+\beta_1}\Gamma_{\rm cusp}[\alpha_s]
  \: \ln\frac{\mu}{Q_{{\rm gs}1}} + \gamma_{S_{G_j}}[\alpha_s] \\
\frac{\diff}{\diff\ln\mu} \ln S_{\overline{G}_j}(Q_{{\rm gs}2},\beta_2,\mu) 
  &=& - \frac{2C_j}{1+\beta_2}\Gamma_{\rm cusp}[\alpha_s]
  \: \ln\frac{\mu}{Q_{{\rm gs}2}} + \gamma_{{S}_{\overline{G}_j}}[\alpha_s] \,,\nn
\eea
where $\Gamma_{\rm cusp}$ denotes the cusp anomalous dimension, and $\gamma_{S_{G_j}}$ and $\gamma_{{S}_{\overline{G}_j}}$ are non-cusp anomalous dimensions. (At 4-loops the cusp anomalous dimension starts to depend on the index $j$ in a manner different from the overall $C_j$ factor pulled out here, but this is beyond the order needed for our analysis.) The solutions to the RG equations are given by
\begin{align}  \label{eq:rg_gs12}
S_{G_j}(Q_{{\rm gs}1},\beta_1,\mu) 
  &= S_{G_j}(Q_{{\rm gs}1},\beta_1,\mu_1) \exp\bigg( \frac{2C_j}{1+\beta_1} K(\mu_1,\mu) + \omega_{S_{G_j}}(\mu_1,\mu) \bigg)
\Big( \frac{\mu_1}{Q_{{\rm gs}1}} \Big)^{\frac{2C_j}{1+\beta_1}\omega(\mu_1,\mu)} \,, \nn\\ 
S_{\overline{G}_j}(Q_{{\rm gs}2},\beta_2,\mu) 
  &= S_{\overline{G}_j}(Q_{{\rm gs}2},\beta_2,\mu_2) \exp\bigg( \frac{-2C_j}{1+\beta_2} K(\mu_2,\mu) + \omega_{S_{\overline{G}_j}}(\mu_2,\mu) \bigg) \Big( \frac{\mu_2}{Q_{{\rm gs}2}} \Big)^{\frac{-2C_j}{1+\beta_2}\omega(\mu_2,\mu)}\,, 
\end{align}
where
\begin{align}
\label{eq:K}
K(\mu_1,\mu_2) &= \int_{\alpha_s(\mu_1)}^{\alpha_s(\mu_2)} \diff\alpha \frac{\Gamma_{\rm cusp}(\alpha)}{\beta(\alpha)} \int_{\alpha_s(\mu_1)}^{\alpha} \frac{\diff\alpha'}{\beta(\alpha')} 
\,, \\
\omega(\mu_1,\mu_2) &= \int_{\alpha_s(\mu_1)}^{\alpha_s(\mu_2)} \diff\alpha \frac{\Gamma_{\rm cusp}(\alpha)}{\beta(\alpha)}
\,,
& \omega_X(\mu_1,\mu_2) &=  \int_{\alpha_s(\mu_1)}^{\alpha_s(\mu_2)} \diff\alpha \frac{\gamma_X(\alpha)}{\beta(\alpha)}
 \,, \nn
\end{align}
for $X=S_{G_j},S_{\overline{G}_j}$. For LL accuracy, we only need the one-loop result of the cusp anomalous dimension:
\be
\Gamma_{\rm cusp}(\alpha_s) = 4\Big(\frac{\alpha_s}{4\pi}\Big) \,.
\ee
For NLL accuracy, we need the two-loop result of the cusp anomalous dimension:
\be
\Gamma_{\rm cusp}(\alpha_s) = 4\Big(\frac{\alpha_s}{4\pi}\Big) + 4\bigg[ \Big( \frac{67}{9} - \frac{\pi^2}{3} \Big)C_A - \frac{20}{9}T_F n_f \bigg] \Big(\frac{\alpha_s}{4\pi}\Big)^2\,,
\ee
and the one-loop results of the non-cusp anomalous dimensions, which happen to vanish
\be
\gamma_{S_{G_j}} = \gamma_{\overline{S}_{G_j}}=0 \,.
\ee
To minimize the logarithmic term in the boundary term of the RG equation, we choose the scales of evaluation as $\mu_1\simeq Q_{{\rm gs}1}$ and $\mu_2\simeq Q_{{\rm gs}2}$ for ${S}_{G_j}$ and ${S}_{\overline{G}_j}$ respectively.

At the perturbative level the function that determines the shape of the collinear drop jet mass cross section, $\hat P_j^{\rm CD}$ can be written as a convolution of two collinear-soft functions $\hat S_{C_j}$ and $\hat D_{C_j}$~\cite{Chien:2019osu}:
\begin{align}
\label{eq:convol}
&\hat P_j^{\rm CD} (\Dms, Q, \tilde{z}_{{\rm cut}\,i},\beta_i,\mu) \\
 &\quad = \Qcuta^{\frac{1}{1+\beta_1}} 
\Qcutb^{\frac{1}{1+\beta_2}} \int \diff \ell_1^+ \diff \ell_2^+ \delta \big(\Dms - Q\ell_1^+ - Q\ell_2^+ \big)
\hat S_{C_j}\big( \ell_1^+\Qcuta^{\frac{1}{1+\beta_1}} ,\beta_1,\mu \big)
\hat D_{C_j}\big( \ell_2^+\Qcutb^{\frac{1}{1+\beta_2}} ,\beta_2,\mu\big) \,.\nn
\end{align}
At one loop, the dimensionally regularized ($d=4-2\epsilon$) perturbative CS functions are obtained from the integrals
\begin{align}
\hat S_{C_j}( \ell_1^+ ,\beta_1,\mu ):\qquad
 &
 \frac{4g^2 C_j \mu^{2\epsilon}e^{\epsilon\gamma_E}}{(4\pi)^\epsilon} \int \frac{\diff^d k}{(2\pi)^d} \frac{2\pi\delta^+(k^2)}{k^+k^-} \Big( \delta(k^+-\ell^+_1) - \delta(\ell^+_1) \Big) \Theta_{{\rm SD}1}^{({\rm cs})}
\,, \\
\hat D_{C_j}( \ell_2^+ ,\beta_2,\mu ):\qquad
&
 \frac{4g^2 C_j \mu^{2\epsilon}e^{\epsilon\gamma_E}}{(4\pi)^\epsilon}
\int \frac{\diff^d k}{(2\pi)^d} \frac{2\pi\delta^+(k^2)}{k^+k^-} \Big( \delta(k^+-\ell^+_2) - \delta(\ell^+_2) \Big)\big( - \Theta_{{\rm SD}2}^{({\rm cs})}  \big) \,, \nonumber
\end{align}
where the kinematic constraint $\Theta_{{\rm SD}\,i}^{({\rm cs})}$ gives the phase space passing the SD criterion
\be
\Theta_{{\rm SD}\,i}^{({\rm cs})} = \theta\Big( k^+ + k^- - Q_{{\rm cut}\,i} \Big( \frac{k^+}{k^-}\Big)^{\frac{\beta_i}{2}} \Big) \,.
\ee
Only collinear-soft radiation that passes the first SD grooming and fails the second SD grooming contributes to the jet mass in collinear drop, which is reflected in the signs of the theta functions in the one-loop expressions of the CS functions (note that $\Theta_{{\rm SD}1}^{({\rm cs})}(1-\Theta_{{\rm SD}2}^{({\rm cs})}) = \Theta_{{\rm SD}1}^{({\rm cs})} - \Theta_{{\rm SD}2}^{({\rm cs})}$). 
Explicit calculations give~\cite{Frye:2016aiz,Chien:2019osu}
\bea
\hat S_{C_j}\big( \ell_1^+\Qcuta^{\frac{1}{1+\beta_1}} ,\beta_1,\mu \big) &=& \delta\big( \ell_1^+\Qcuta^{\frac{1}{1+\beta_1}}\big) + \frac{\alpha_s C_j}{\pi}\Bigg( \delta\big( \ell_1^+\Qcuta^{\frac{1}{1+\beta_1}} \big) \frac{2+\beta_1}{1+\beta_1}\Big( -\frac{1}{2\epsilon^2} + \frac{\pi^2}{24}\Big)  \\
&+& \frac{1}{\epsilon}\mu^{-\frac{2+\beta_1}{1+\beta_1}} \ml{L}_0\bigg( \frac{\ell_1^+\Qcuta^{\frac{1}{1+\beta_1}}}{\mu^{\frac{2+\beta_1}{1+\beta_1}}} \bigg) - \frac{2(1+\beta_1)}{2+\beta_1} \mu^{-\frac{2+\beta_1}{1+\beta_1}} \ml{L}_1\bigg( \frac{\ell_1^+\Qcuta^{\frac{1}{1+\beta_1}}}{\mu^{\frac{2+\beta_1}{1+\beta_1}}} \bigg) \Bigg) \,, \nn\\
\hat D_{C_j}\big( \ell_2^+\Qcutb^{\frac{1}{1+\beta_2}} ,\beta_2,\mu \big) &=& \delta\big( \ell_2^+\Qcutb^{\frac{1}{1+\beta_2}}\big) - \frac{\alpha_s C_j}{\pi}\Bigg( \delta\big( \ell_2^+\Qcutb^{\frac{1}{1+\beta_2}} \big) \frac{2+\beta_2}{1+\beta_2}\Big( -\frac{1}{2\epsilon^2} + \frac{\pi^2}{24}\Big) \nn \\
&+& \frac{1}{\epsilon}\mu^{-\frac{2+\beta_2}{1+\beta_2}} \ml{L}_0\bigg( \frac{\ell_2^+\Qcutb^{\frac{1}{1+\beta_2}}}{\mu^{\frac{2+\beta_2}{1+\beta_2}}} \bigg) - \frac{2(1+\beta_2)}{2+\beta_2} \mu^{-\frac{2+\beta_2}{1+\beta_2}} \ml{L}_1\bigg( \frac{\ell_2^+\Qcutb^{\frac{1}{1+\beta_2}}}{\mu^{\frac{2+\beta_2}{1+\beta_2}}} \bigg) \nn
\Bigg) \,,
\eea
where ${\cal L}_n(x)$ is a plus-function that integrates to zero on the region $x\in [0,1]$:
\be
\ml{L}_n = \Big(\theta(x)\frac{\ln^n(x)}{x} \Big)_+ \,.
\ee
The convolution (\ref{eq:convol}) can be factorized by applying the Laplace transform
\be
\widetilde{f}(y) = \int_0^\infty \diff \Delta m^2\, \exp\big(-ye^{-\gamma_E} \Dms \big) f(\Dms) \,.
\ee
In Laplace space we have
\be
\hat{\widetilde{P}}_j{}^{\!\!\rm CD} (y, Q, \tilde{z}_{{\rm cut}\,i},\beta_i,\mu) = \hat{\widetilde{S}}_{C_j}\pig(yQ\Qcuta^{\frac{-1}{1+\beta_1}},\beta_1,\mu \pig)
\hat{\widetilde{D}}_{C_j}\pig(yQ\Qcutb^{\frac{-1}{1+\beta_2}},\beta_2,\mu \pig) \,,
\ee
where $\hat{\widetilde{S}}_{C_j}$'s first variable is the Laplace conjugate to the first variable in $\hat S_{C_j}$, and likewise for $\hat{\widetilde{D}}_{C_j}$ and $\hat D_{C_j}$. 
The renormalized one-loop CS functions in Laplace space are given by 
\bea
\label{eq:cs_1loop1}
\hat{\widetilde{S}}_{C_j}\pig(yQ\Qcuta^{\frac{-1}{1+\beta_1}},\beta_1,\mu \pig) &=& 1 + \frac{\alpha_s C_j}{\pi} \frac{2+\beta_1}{1+\beta_1} \bigg( -\ln^2\frac{\mu \,y^{\frac{1+\beta_1}{2+\beta_1}} Q^{\frac{1+\beta_1}{2+\beta_1}}}{Q_{{\rm cut}1}^{\frac{1}{2+\beta_1}}} + \frac{\pi^2}{24} \bigg) \,, \\
\label{eq:cs_1loop2}
\hat{\widetilde{D}}_{{C}_j}\pig(yQ\Qcutb^{\frac{-1}{1+\beta_2}},\beta_2,\mu \pig) &=& 1 - \frac{\alpha_s C_j}{\pi} \frac{2+\beta_2}{1+\beta_2} \bigg( -\ln^2\frac{\mu \,y^{\frac{1+\beta_2}{2+\beta_2}} Q^{\frac{1+\beta_2}{2+\beta_2}}}{Q_{{\rm cut}2}^{\frac{1}{2+\beta_2}}} + \frac{\pi^2}{24} \bigg)  
\,. \nonumber
\eea
We note that the argument $yQQ_{{\rm cut}\,i}^{\frac{-1}{1+\beta_i}}$ of the collinear-soft function appears only in terms of logarithms in Eq.~(\ref{eq:cs_1loop1}). They satisfy the following RG equations that are generally valid at higher loops as well
\bea
\frac{\diff}{\diff\ln\mu} \ln\hat{\widetilde{S}}_{C_j}\pig(yQ\Qcuta^{\frac{-1}{1+\beta_1}},\beta_1,\mu \pig)
 &=&  2C_j \Gamma_{\ma{cusp}}[\alpha_s]\,
\ln\frac{Q_{\ma{cut}1}^{\frac{1}{1+\beta_1}}}{\mu^{\frac{2+\beta_1}{1+\beta_1}}Qy} + \gamma_{S_{C_j}}[\alpha_s]  
\,, \\
\frac{\diff}{\diff\ln\mu} \ln\hat{\widetilde{D}}_{{C}_j}
 \pig(yQ\Qcutb^{\frac{-1}{1+\beta_2}},\beta_2,\mu \pig) &=& -  2C_j \Gamma_{\ma{cusp}}[\alpha_s]\,
\ln\frac{Q_{\ma{cut}2}^{\frac{1}{1+\beta_2}}}{\mu^{\frac{2+\beta_2}{1+\beta_2}}Qy} + \gamma_{{D}_{{C}_j}}[\alpha_s] \,.
\nn
\eea
The solutions to the RG equations are given by
\bea
\label{eq:sol_rg_cs1}
\hat{\widetilde{S}}_{C_j} \pig( yQQ_{\text{cut}1}^{\frac{-1}{1+\beta_1}}, \beta_1, \mu \pig) &=& \hat{\widetilde{S}}_{C_j} \pig( yQQ_{\text{cut}1}^{\frac{-1}{1+\beta_1}}, \beta_1, \mu_{{\rm cs}1} \pig) 
\Bigg( \frac{Q_{\text{cut}1}^{\frac{1}{1+\beta_1}}}{yQ \mu_{{\rm cs}1}^{\frac{2+\beta_1}{1+\beta_1}}} \Bigg)^{2C_j\omega(\mu_{{\rm cs}1}, \mu)}  \nn\\
&\times& \exp\bigg( -2C_j\frac{2+\beta_1}{1+\beta_1} K(\mu_{{\rm cs}1}, \mu) + \omega_{S_{C_j}}(\mu_{{\rm cs}1}, \mu) \bigg) 
 \,, \nn \\
\label{eq:sol_rg_cs2}
\hat{\widetilde{D}}_{{C}_j} \pig(
yQQ_{\text{cut}2}^{\frac{-1}{1+\beta_2}}, \beta_2, \mu \pig) &=& \hat{\widetilde{D}}_{{C}_j} \pig( yQQ_{\text{cut}2}^{\frac{-1}{1+\beta_2}}, \beta_2, \mu_{{\rm cs}2} \pig) \Bigg( \frac{Q_{\text{cut}2}^{\frac{1}{1+\beta_2}}}{yQ \mu_{{\rm cs}2}^{\frac{2+\beta_2}{1+\beta_2}}} \Bigg)^{-2C_j\omega(\mu_{{\rm cs}2}, \mu)}  \nn \\
&\times& \exp\bigg( 2C_j\frac{2+\beta_2}{1+\beta_2} K(\mu_{{\rm cs}2}, \mu) + \omega_{D_{{C}_j}}(\mu_{{\rm cs}2}, \mu) \bigg) 
\,. 
\eea
At one loop, $\gamma_{S_{C_j}} = \gamma_{{D}_{{C}_j}} = 0$, thus we can set $\omega_{S_{C_j}} = \omega_{D_{{C}_j}}=0$ for LL and NLL accuracy. To minimize the logarithmic terms in the boundary terms of the RG equations, we choose the scales of evaluation as $\mu_{{\rm cs}1}\simeq Q_{{\rm cs}1}$ and $\mu_{{\rm cs}2}\simeq Q_{{\rm cs}2}$ for $\hat{\widetilde{S}}_{C_j}$ and $\hat{\widetilde{D}}_{{C}_j}$ respectively. The inverse Laplace transform is given by
\be
f(\Dms) = \frac{1}{2\pi i} \int_{c-i\infty}^{c+i\infty} \diff y\: e^{-\gamma_E} \,\exp\big( ye^{-\gamma_E} \Dms \big) \widetilde{f}(y) \,,
\ee
where $c$ is chosen such that the integration contour is on the right of all the poles of $\widetilde{f}(y)$. For the function $\widetilde{f}(y) = 1/y^\eta$, the inverse Laplace transform gives
\be
f(\Dms) = \ml{L}^{-1}\Big(\frac{1}{y^\eta}\Big) = \frac{(e^{-\gamma_E}\Delta m^2)^\eta}{\Delta m^2\Gamma(\eta)} \Theta(\Delta m^2) \,,
\ee
where $\Theta(x)$ is the theta function. Applying the inverse Laplace transform
leads to
\begin{align}
\label{eq:Pcd_final}
\hat P_j^{\rm CD} & (\Dms, Q, \tilde{z}_{{\rm cut}\,i},\beta_i,\mu) 
\nn\\
&= \exp\biggl( -2C_j\frac{2+\beta_1}{1+\beta_1} K(\mu_{{\rm cs}1}, \mu) + \omega_{S_{C_j}}(\mu_{{\rm cs}1}, \mu) +  2C_j\frac{2+\beta_2}{1+\beta_2} K(\mu_{{\rm cs}2}, \mu) + \omega_{D_{{C}_j}}(\mu_{{\rm cs}2}, \mu) \biggr) 
\nn\\
&\times
\hat{\widetilde{S}}_{C_j} \pig( QQ_{\text{cut}1}^{\frac{-1}{1+\beta_1}}e^{-\frac{\partial}{\partial\eta}}, \beta_1, \mu_{{\rm cs}1} \pig)
\hat{\widetilde{D}}_{{C}_j} \pig( QQ_{\text{cut}2}^{\frac{-1}{1+\beta_2}}e^{-\frac{\partial}{\partial\eta}}, \beta_2, \mu_{{\rm cs}2} \pig) \frac{(e^{-\gamma_E}\Dms)^{\eta}}{\Dms\Gamma(\eta)} \bigg|_{\eta = 2C_j\omega(\mu_{{\rm cs}1}, \mu_{{\rm cs}2})} 
\nn\\
&\times \Bigg( \frac{Q_{\text{cut}1}^{\frac{1}{1+\beta_1}}}{Q \mu_{{\rm cs}1}^{\frac{2+\beta_1}{1+\beta_1}}} \Bigg)^{2C_j\omega(\mu_{{\rm cs}1}, \mu)}
\Bigg( \frac{Q_{\text{cut}2}^{\frac{1}{1+\beta_2}}}{Q \mu_{{\rm cs}2}^{\frac{2+\beta_2}{1+\beta_2}}} \Bigg)^{-2C_j\omega(\mu_{{\rm cs}2}, \mu)} \,,
\end{align}
where we have suppressed the overall $\Theta(\Dms)$.

Finally we want to review how to fix the scales for $\mu_{{\rm gs}\,i}$ and $\mu_{{\rm cs}\,i}$ in different regions of the collinear drop spectrum (see~\cite{Chien:2019osu} for further details). The natural choice for these scales is obtained by minimizing the logarithm in the boundary terms of the RG running, which gives $\mu_{{\rm gs}\,i}\simeq Q_{{\rm gs}\,i}$ and $\mu_{{\rm cs}\,i} \simeq Q_{{\rm cs}\,i}$. Here the $\simeq$ indicates that we fix the scale near this value, which in practice means that central results are obtained at this value while variations in a region parametrically nearby are used to estimate perturbative uncertainties. Special treatments are needed when $\mu_{{\rm gs}\,i}=\mu_{{\rm cs}\,i}$, which happens when the $i$-th GS and CS modes merge and the $i$-th soft drop in the collinear drop grooming becomes ineffective. The $i$-th soft drop starts to become ineffective when the CD jet mass satisfies
\be
\Delta m^2 \geq \Delta m_{{\rm cut}\,i}^2 =  (p_TR)^2 z_{{\rm gs}\,i}\,,
\ee
which is obtained by solving $Q_{{\rm gs}\,i}=Q_{{\rm cs}\,i}$ for $\Delta m^2$. When $\Delta m_{{\rm cut}1}^2 \leq \Delta m^2 \leq \Delta m_{{\rm cut}2}^2$, the factorization formulas that have been discussed above still apply if we fix the scales as $\mu_{{\rm gs}1}=\mu_{{\rm cs}1} = \frac{\Delta m^2}{p_TR}$. When $\Delta m^2 \geq \Delta m_{{\rm cut}2}^2$, the second soft drop becomes ineffective and there is no more phase space for radiation that passes collinear drop, and this therefore corresponds to the endpoint of the collinear drop jet mass spectrum. Thus the differential cross section goes to zero when $\Delta m^2 \geq \Delta m_{{\rm cut}2}^2$ and the cumulative cross section that will be discussed in the next section becomes constant.

\subsection{Cumulative Jet Mass}
\label{sec:cumulative}

The cumulative cross section of the collinear drop jet mass can be obtained from integrating the differential cross section (\ref{eq:differential_sigma}) over $\Dms$:
\be
\Sigma(\Delta m_c) = \frac{1}{\sigma} \int_0^{\Dms_c} \diff \Dms \,\frac{\diff \sigma}{\diff \Dms} \,.
\ee
The inclusion of the $1/\sigma$ ensures that $\Sigma(\Delta m_c^2)\to 1$ when $\Delta m_c^2$ gets to the endpoint of the spectrum. 
The $\Dms$-dependent part of the differential cross section can be seen from Eq.~(\ref{eq:Pcd_final}) and after the integration it leads to
\be
\int_0^{\Dms_c}\diff \Dms \,\frac{e^{-\gamma_E \eta}}{\Gamma(\eta)} \big(\Dms\big)^{\eta-1} = \frac{\big( e^{-\gamma_E} \Dms_c \big)^\eta}{\Gamma(1+\eta)}  \,.
\ee
For the purely perturbative result for the cumulative cross section we then obtain
\begin{align}
\label{eq:Sigfull}
\hat \Sigma &(\Delta m_c) = \sum_{j=q,g} f_j\: \hat \Sigma_{j}(\Delta m_c) 
\,,
\end{align}
where 
\begin{align}  \label{eq:fj_full}
f_j &= \frac{ H_j(p_T,\eta_J, R)
  \otimes_\Omega S_{G_j}(Q_{\rm gs1},R,\beta_1,\mu_{\rm gs1})
  \otimes_\Omega S_{\overline G_j}(Q_{\rm gs2},R,\beta_2,\mu_{\rm gs2}) }
  {S_{G_j}^{ee}(Q_{\rm gs1},\beta_1,\mu_{\rm gs1})
   S_{\overline G_j}^{ee}(Q_{\rm gs2},\beta_2,\mu_{\rm gs2})}
  \: \frac{\sigma_j}{\sigma} 
\end{align} 
are the fractions of quark and gluon jets with $\sigma$ the full jet cross section. Here we include the ratio between the full global-soft functions $S_{G_j}$, $S_{\overline G_j}$  that are tied to $H_j$ through the calculation of non-global logarithms beyond NLL order, and the global-soft functions $S_{G_j}^{ee}$, $S_{\overline G_j}^{ee}$ for hemisphere jets in $e^+e^-$ collisions whose structure is fully global. For the latter we (artificially) employ the same global-soft scales as in \eqn{mu_gs}. At one loop the results for $S_{G_j}^{ee}$, $S_{\overline G_j}^{ee}$ are identical to those in \eqn{SG-oneloop}. This is convenient since logarithms of $\mu_{{\rm gs}\,i}$ in these global-soft functions fully suffice to cancel $\mu_{{\rm gs}\,i}$ dependence in the renormalization group improved calculation of $\hat\Sigma_j$ order-by-order, and furthermore the ratios in \eqn{fj_full} are also 
$\mu_{{\rm gs}\,i}$ independent order-by-order. In \eqn{fj_full} the $\sigma_j=1+\ldots $ is a fixed order series which provides the proper normalization for $f_j$ at higher orders, and is defined to ensure $\hat \Sigma_j\to 1$ at the upper limit of $\Delta m_c^2$. 
Using the all-orders form of the perturbative factorization theorem in \Eq{eq:Pcd_final} gives the all-orders result
\begin{align} \label{eq:Sigma_allorder}
\hat \Sigma_{j} &= \!\frac{1}{\sigma_j}
\exp\biggl[\!\frac{2C_j}{1\!+\!\beta_1} K(\mu_{{\rm gs}1},\mu) - \frac{2C_j}{1\!+\!\beta_2} K(\mu_{{\rm gs}2},\mu)  -2C_j\frac{2\!+\!\beta_1}{1\!+\!\beta_1} K(\mu_{\text{cs}1}, \mu) \!+\! 2C_j\frac{2\!+\!\beta_2}{1\!+\!\beta_2} K(\mu_{\text{cs}2}, \mu)\! \biggr]
\nn\\ 
&\times \Big( \frac{\mu_{{\rm gs}1}}{Q_{{\rm gs}1}} \Big)^{\frac{2C_j}{1+\beta_1}\omega(\mu_{{\rm gs}1},\mu)} 
\Big( \frac{\mu_{{\rm gs}2}}{Q_{{\rm gs}2}} \Big)^{\frac{-2C_j}{1+\beta_2}\omega(\mu_{{\rm gs}2},\mu)}
\Bigg( \frac{Q_{\text{cut}1}^{\frac{1}{1+\beta_1}}}{Q \mu^{\frac{2+\beta_1}{1+\beta_1}}_{\text{cs}1}} \Bigg)^{2C_j\omega(\mu_{\text{cs}1}, \mu)}
\Bigg( \frac{Q_{\text{cut}2}^{\frac{1}{1+\beta_2}}}{Q \mu^{\frac{2+\beta_2}{1+\beta_2}}_{\text{cs}2}} \Bigg)^{-2C_j\omega(\mu_{\text{cs}2}, \mu)} 
\nn\\
& \times 
\exp\bigl[  \omega_{S_{C_j}}(\mu_{{\rm cs}1}, \mu) 
+ \omega_{D_{C_j}}(\mu_{{\rm cs}2}, \mu) 
+ \omega_{S_{G_j}}(\mu_{\text{gs}1},\mu)
+ \omega_{S_{{\overline G}_j}}(\mu_{\text{gs}2},\mu) \bigr]
\nn\\
&\times
S_{G_j}^{ee}(Q_{\rm gs1},\beta_1,\mu_{\rm gs1})
   S_{\overline G_j}^{ee}(Q_{\rm gs2},\beta_2,\mu_{\rm gs2})
\nn\\
&\times
\hat{\widetilde{S}}_{C_j} \pig( QQ_{\text{cut}1}^{\frac{-1}{1+\beta_1}}e^{-\frac{\partial}{\partial\eta}}, \beta_1, \mu_{{\rm cs}1} \pig)
\hat{\widetilde{D}}_{{C}_j} \pig( QQ_{\text{cut}2}^{\frac{-1}{1+\beta_2}}e^{-\frac{\partial}{\partial\eta}}, \beta_2, \mu_{{\rm cs}2} \pig)
\frac{( e^{-\gamma_E} \Dms_c)^\eta}{\Gamma(1+\eta)}
\bigg|_{\eta = 2C_j \omega(\mu_{\text{cs}1}, \mu_{\text{cs}2})} \,.
\end{align}
To make predictions for this cumulative cross section with resummed large logarithms we should fix the mass dependent scales in terms of $\Delta m_c^2$, so the canonical values are $\mu_{{\rm cs}\,i} = Q_{{\rm cs}\,i}(\Delta m^2 \to \Delta m_c^2)$.

At LL and NLL accuracy, the boundary terms in the solutions of the RG equations can be set to be the tree-level values, which are unity for both the GS functions in momentum space and the CS functions in the Laplace space. Also $\sigma_j=1$, and non-global effects are neglected. Furthermore, the non-cusp anomalous dimension vanishes at one loop for both GS and CS functions, so we can set $\omega_{S_{G_j}} = \omega_{S_{\overline{G}_j}} = \omega_{S_{C_j}} = \omega_{D_{{C}_j}} =0$. For the purely perturbative result for the cumulative cross section we then obtain
\begin{align}
\label{eq:ll_nll}
\hat \Sigma^{{\rm NLL}}&(\Delta m_c) = \sum_{j=q,g} f_j\: \hat \Sigma^{\rm NLL}_{j}(\Delta m_c) 
 \,,
\end{align}
where here the fractions of quark and gluon jets are given by $f_j= H_j(p_T,\eta_J, R)/\sigma_0$ with $\sigma_0$ the tree-level jet cross section, and
\begin{align} \label{eq:SigNLL}
\hat \Sigma^{\rm NLL}_{j} &=
\exp\biggl[\!\frac{2C_j}{1\!+\!\beta_1} K(\mu_{{\rm gs}1},\mu) - \frac{2C_j}{1\!+\!\beta_2} K(\mu_{{\rm gs}2},\mu)  -2C_j\frac{2\!+\!\beta_1}{1\!+\!\beta_1} K(\mu_{\text{cs}1}, \mu) \!+\! 2C_j\frac{2\!+\!\beta_2}{1\!+\!\beta_2} K(\mu_{\text{cs}2}, \mu)\! \biggr]
\nn\\ 
&\times \Big( \frac{\mu_{{\rm gs}1}}{Q_{{\rm gs}1}} \Big)^{\frac{2C_j}{1+\beta_1}\omega(\mu_{{\rm gs}1},\mu)} 
\Big( \frac{\mu_{{\rm gs}2}}{Q_{{\rm gs}2}} \Big)^{\frac{-2C_j}{1+\beta_2}\omega(\mu_{{\rm gs}2},\mu)}
\Bigg( \frac{Q_{\text{cut}1}^{\frac{1}{1+\beta_1}}}{Q \mu^{\frac{2+\beta_1}{1+\beta_1}}_{\text{cs}1}} \Bigg)^{2C_j\omega(\mu_{\text{cs}1}, \mu)}
\Bigg( \frac{Q_{\text{cut}2}^{\frac{1}{1+\beta_2}}}{Q \mu^{\frac{2+\beta_2}{1+\beta_2}}_{\text{cs}2}} \Bigg)^{-2C_j\omega(\mu_{\text{cs}2}, \mu)} 
\nn\\
&\times
\frac{( e^{-\gamma_E} \Dms_c)^\eta}{\Gamma(1+\eta)}
\bigg|_{\eta = 2C_j \omega(\mu_{\text{cs}1}, \mu_{\text{cs}2})} \,.
\end{align}

In the form given in \eqn{Sigma_allorder} or \eqn{SigNLL} with quark and gluon fractions $f_j$, the results for $\hat\Sigma_j$ apply equally well for a jet produced in $pp$ and $e^+e^-$ collisions.  For our analysis we will commonly refer to scales that are appropriate for the currently more phenomenologically relevant $pp$ case.
To evaluate the perturbative result in \eqn{SigNLL}, we need to specify the scales at which the perturbative RG evolution is stopped. Naturally, 
we choose the canonical scales
\bea \label{eq:gscs-scales}
\mu_{{\rm gs}\,i} &=& Q_{{\rm gs}\,i} = p_T R z_{{\rm gs}\,i} 
  \,, \\
\mu_{{\rm cs}\,i} &=& Q_{{\rm cs}\,i} = \Big( \frac{\Dms_c}{Q} \Big)^{\frac{1+\beta_i}{2+\beta_i}} Q_{{\rm cut}\,i}^{\frac{1}{2+\beta_i}} 
 \,. \nn
\eea
The CS scales chosen $\mu_{{\rm cs}\,i}$ depend on the CD cumulative jet mass. When we apply the factorization formula to the small CD jet mass region, the CS scales can become nonperturbative and the perturbative RG running has to be frozen. We will explain how to incorporate the nonperturbative effects using shape functions in the next section. At the moment, we just focus on the purely perturbative calculations and freeze out the CS scales to the value $\mu_0$ continuously and smoothly by taking
\be \label{eq:musctomu0}
\mu_{{\rm cs}\,i} = \begin{cases}
	\mu_{{\rm cs}\,i}\,, &  \mu_{{\rm cs}\,i} \geq 2\mu_0  \\
	\mu_0\Big(1+\frac{\mu_{{\rm cs}\,i}^2}{4\mu_0^2}\Big)\,, & \mu_{{\rm cs}\,i} < 2\mu_0 
\end{cases} \,.
\ee
The perturbative results depend on the choice of $\mu_0$. However, the $\mu_0$ dependence will be canceled by the $\mu_0$ dependence in shape functions, which will be discussed in the next section.

\begin{figure}[t]
	\begin{subfigure}[t]{0.49\textwidth}
		\centering
		\includegraphics[height=2.2in]{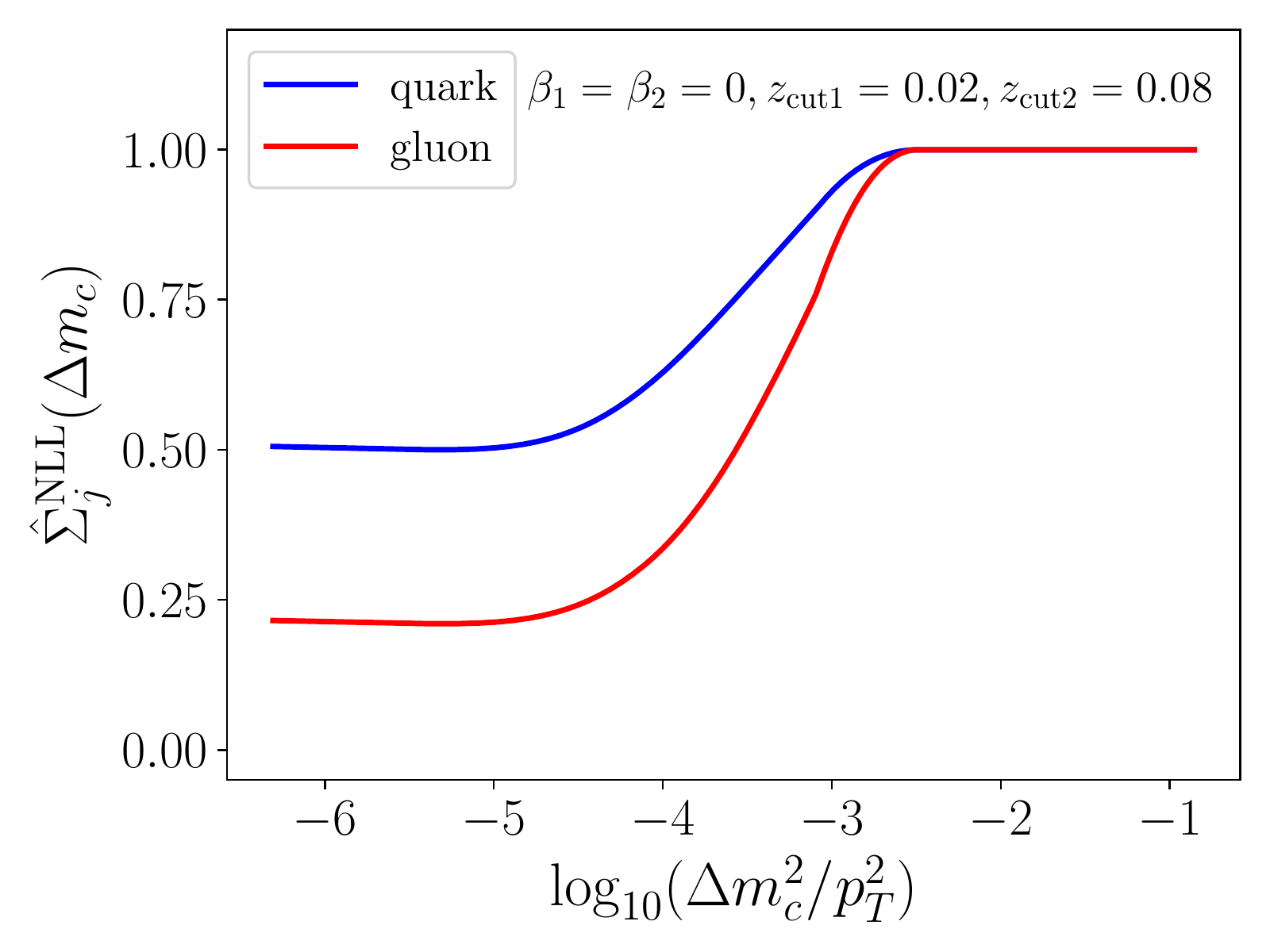}
		\caption{$\beta_1=\beta_2=0,z_{{\rm cut}1}=0.02,z_{{\rm cut}2}=0.08$.}
		\label{fig:Sigma1}
	\end{subfigure}%
	~
	\begin{subfigure}[t]{0.49\textwidth}
		\centering
		\includegraphics[height=2.2in]{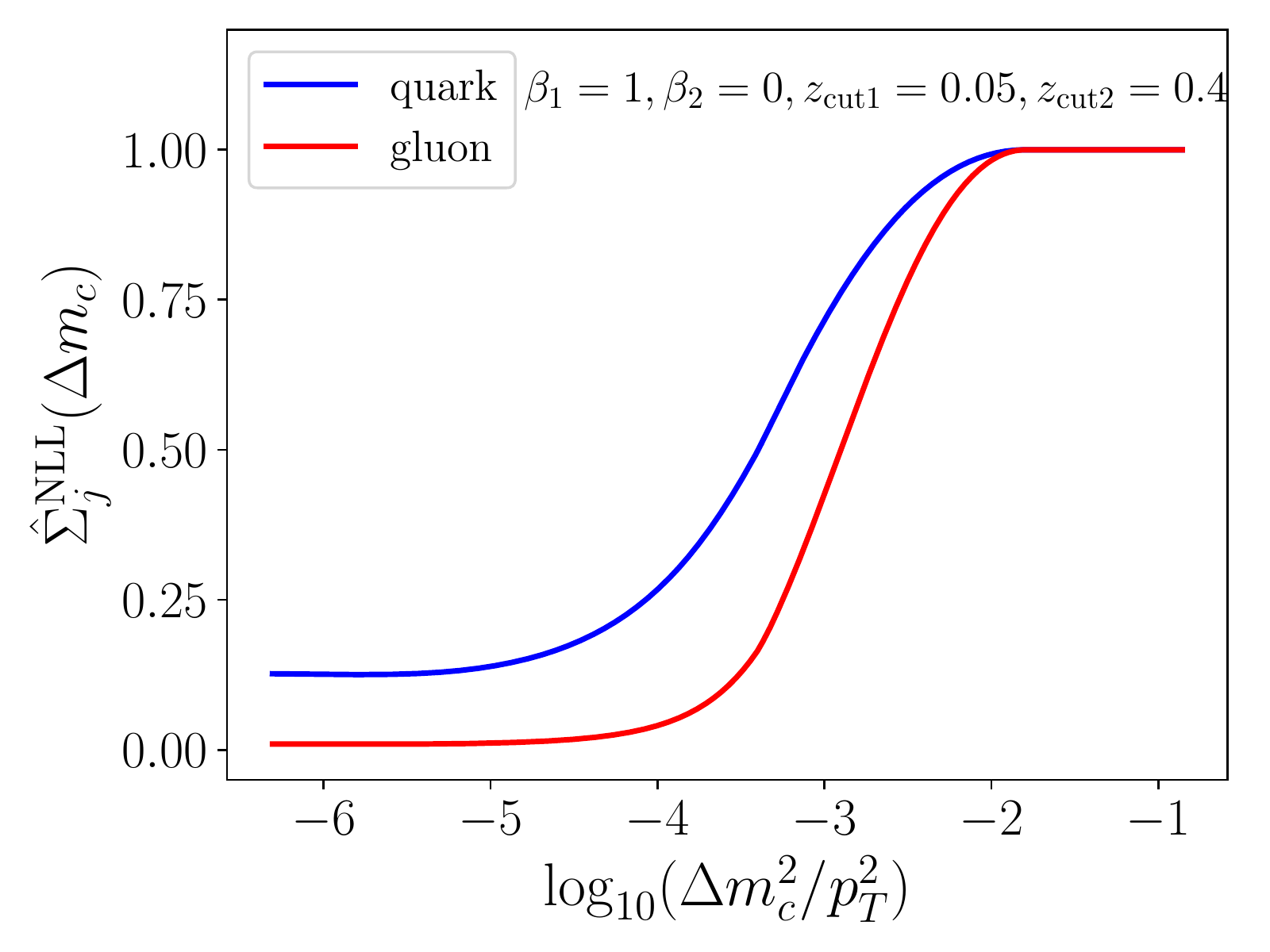}
		\caption{$\beta_1=1,\beta_2=0,z_{{\rm cut}1}=0.05,z_{{\rm cut}2}=0.4$.}
		\label{fig:Sigma2}
	\end{subfigure}%
	\caption{Perturbative results of the cumulative jet mass in collinear drop for quark and gluon jets. The constant feature in the limit $\Delta m_c^2\to0$ depends on both the jet content and the collinear drop parameters.}
	\label{fig:pert_results}
\end{figure}

In Fig.~\ref{fig:pert_results} we plot the perturbative results of the cumulative jet mass $\hat \Sigma^{\rm NLL}_{j}$ for quarks and gluons using $p_T=800\,{\rm GeV}$, $\eta_J=0$, $R=0.2$, and $\mu_0=1$ GeV. The two panels show two different sets of CD parameters; for the left panel $\beta_1=\beta_2=0,z_{{\rm cut}1}=0.02,z_{{\rm cut}2}=0.08$, while for the right panel $\beta_1=1,\beta_2=0,z_{{\rm cut}1}=0.05,z_{{\rm cut}2}=0.4$. As discussed at the end of Section~\ref{sec:jetmass}, the cumulative jet mass spectrum becomes unity at the endpoint of the spectrum, $\Delta m_c^2 = \Delta m_{{\rm cut}2}^2$. This upper endpoint corresponds to
\begin{align} \label{eq:dmc-upper}
  \log_{10}\Big(\frac{\Delta m_c^2}{p_T^2}\Big) 
   = \log_{10}\bigg(\frac{R^{2+\beta_2}}{R_0^{\beta_2}} z_{\rm cut2} \bigg) \,,
\end{align}
which corresponds to 
$\log_{10}(\Delta m_c^2/p_T^2)= \{-2.49, -1.80 \}$
for the left and right panels of Fig.~\ref{fig:pert_results}, respectively. 
On the opposite side of the plots we see that the perturbative cumulative cross section of the jet mass in collinear drop does not go to zero as $\Delta m_c^2\to 0$, but instead approaches a constant value. This is a special feature of the collinear drop jet mass and is physically different from the soft drop jet mass (and many other observables), whose perturbative cumulative cross section vanishes in the limit of zero jet mass. In soft drop, zero jet mass means no radiation passes the soft drop grooming procedure, but since the energetic collinear core will always pass soft drop the probability of having no radiation vanishes, and thus so does the cross section.  For collider observables this behavior is induced by the ubiquitous presence of a Sudakov exponential that predicts zero probability for no radiation.   In contrast, in collinear drop the jet mass is defined from an intermediate (soft) region of phase space with cuts from two sides induced by the collinear drop procedure, which is obtained from the difference of two soft drop jet masses. Here zero collinear drop jet mass corresponds to events where the two soft drop jet masses are equal, which happens for a finite number of events.  The contribution in this bin is represented by a $\delta(\Delta m^2)$ in the collinear drop spectrum, and by the constant that appears for $\hat\Sigma(\Delta m_c^2\to 0)$. Thus the constant value in the limit $\Dms_c\to 0$ corresponds to the fraction of events without intermediate soft radiation, namely those with only collinear and soft radiation that is groomed away. This constant value is sensitive to the jet content, i.e., whether it is a quark or a gluon jet. Furthermore, as we can see by comparing the two plots in Fig.~\ref{fig:pert_results}, the constant values also depend sensitively on the choice of the collinear drop parameters.
This region of constant $\hat \Sigma_j$ begins when the cross section transitions from the perturbative to nonperturbative regimes, corresponding to the scales $\mu_{{\rm cs}\,i}$ from \eqn{gscs-scales} reaching $2\mu_0$ and transitioning to the fixed value $\mu_0$, as in \eqn{musctomu0}.  Since $\mu_{{\rm cs}2}$ reaches small values later than $\mu_{{\rm cs}1}$, we can solve for when $\mu_{{\rm cs}2}=\mu_{\rm np}$ to give
\begin{align}  \label{eq:dmc-np}
  \log_{10}\Big(\frac{\Delta m_c^2}{p_T^2}\Big)
  &= - \frac{2+\beta_2}{1+\beta_2} \log_{10}\frac{p_T}{\mu_{\rm np}} +
     \frac{1}{1+\beta_2} \log_{10}\frac{R_0^{\beta_2}}{z_{{\rm cut}2}} \,.
\end{align}
Recalling that we use $R_0=1$, the transition to the flat behavior occurs near 
$\mu_{\rm np}=2\mu_0=2\,{\rm GeV}$, which corresponds to
$\log_{10}(\Delta m_c^2/p_T^2)= \{-4.11, -4.81 \}$
for the left and right panels of Fig.~\ref{fig:pert_results}, respectively.
For the gluon channel in the right panel the cumulative falls somewhat more steeply, reaching the plateau region already at 
$\log_{10}(\Delta m_c^2/p_T^2)\simeq -4.2$.

The motivating observation for our construction of pure quark and gluon observables 
was this constant behavior in the limit $\Dms_c\to0$, and exploiting its dependence on the $z_{{\rm cut}\,i}$ parameters. In the region where the perturbative result of the cumulative cross section becomes constant, nonperturbative effects also become important, and we will discuss how to include these effects in Section~\ref{sec:nonper}.  The construction of our pure quark and gluon observables will then be given in Section~\ref{sec:obs}. 
We will see that our construction works for both the nonperturbative region where $\hat \Sigma_j$ is flat in $\Delta m_c^2$ and for the perturbative resummation region where $\hat \Sigma_j$ has non-trivial dependence on $\Delta m_c^2$.

\section{Nonperturbative Effects in the Small Jet Mass Regime}
\label{sec:nonper}

\subsection{Nonperturbative Regime}
As discussed in the previous section, nonperturbative effects will become important in the limit $\Delta m_c^2\to 0$, in particular when $\Delta m_c^2 \lesssim \Delta m_{\Lambda\,i}^2$ given in \Eq{eq:mass_in_nonpert}. 
In the fully nonperturbative regime, corresponding to Figs.~\ref{fig:modes}c and \ref{fig:modes}d, perturbative calculations of the CS functions are no longer reliable. To see this more explicitly, we can examine the boundary terms in the solutions to the RG equations of the CS functions \Eq{eq:sol_rg_cs1}. One-loop results of the boundary terms are given in \Eqs{eq:cs_1loop1}{eq:cs_1loop2}. To minimize the logarithms in the fixed-order results, we choose the scale of evaluation $\mu_{{\rm cs}\,i}$ to satisfy
\be
\frac{\mu_{{\rm cs}\,i} \,y^{\frac{1+\beta_i}{2+\beta_i}} Q^{\frac{1+\beta_i}{2+\beta_i}}}{Q_{{\rm cut}\,i}^{\frac{1}{2+\beta_i}}} \sim 1\,,
\ee
which with $y^{-1}\sim \Dms_c$ gives $\mu_{{\rm cs}\,i} \sim Q_{{\rm cs}\,i}$. 
However, when $\mu_{{\rm cs}\,i}=Q_{{\rm cs}\,i}\sim\Lambda_{\rm QCD}$, the fixed-order results of the CS functions at the CS scales are no longer reliable. This CS scale becomes nonperturbative $Q_{{\rm cs}\,i}\sim \Lambda_{\rm QCD}$ when the cumulative jet mass is smaller than
\be
\Delta m_c^2 \lesssim  \Delta m_{\Lambda\,i}^2= \Lambda_{\rm QCD}\, Q\, 
\Big( \frac{\Lambda_{\rm QCD}}{Q_{{\rm cut}\,i}} \Big)^{\frac{1}{1+\beta_i}}
 \,.
\ee

In this section, we will focus on the case where both the collinear-soft modes become nonperturbative, which corresponds to the cases shown in Figs.~\ref{fig:modes}c and \ref{fig:modes}d. 
Our construction of pure quark and gluon observables will apply for the cumulative cross section in both this nonperturbative regime and the perturbative resummation region of Fig.~\ref{fig:modes}a. 
In the case depicted in Fig.~\ref{fig:modes}b, the first CS mode is nonperturbative $\Delta m_c^2 \lesssim  \Delta m_{\Lambda1}^2$, but the second CS mode is still perturbative with $\Delta m_c^2 \gg  \Delta m_{\Lambda2}^2$, which serves as a transition region between Case~\ref{fig:modes}a and Case~\ref{fig:modes}c. We leave the discussion of this transition case to future work.

\subsection{Nonperturbative Corrections via Shape Functions}
\label{sec:NPshapefunctions}

When $\Delta m_c^2 \lesssim  \Delta m_{\Lambda\,i}^2$, nonperturbative effects can be incorporated into the CS function by introducing a nonperturbative shape function.  The procedure for doing this for the collinear-soft function in soft drop jet mass has been worked out in Ref.~\cite{Hoang:2019ceu}.  Since the collinear-soft functions appearing for collinear drop are hierarchically separated for the scenarios we consider, we can directly apply this shape function setup for each of our collinear-soft functions. Each CS function is written as a convolution of the perturbative CS function, that satisfies the RG equation in Eq.~(\ref{eq:sol_rg_cs1}), and a shape function $F_i^j(k_i,\beta_i)$ that depends on the parton species $j$ initiating the jet,
\bea
S_{C_j}\pig( \ell^+_1 Q_{\text{cut}1}^{\frac{1}{1+\beta_1}}, \beta_1, \mu \pig) &=& \int_0^{+\infty} \!\! \diff k_1 \: 
\hat S_{C_j} \pig( \ell^+_1 Q_{\text{cut}1}^{\frac{1}{1+\beta_1}} 
- k_1^{\frac{2+\beta_1}{1+\beta_1}}, \beta_1, \mu \pig) 
F_1^j(k_1,\beta_1) 
 \,, \\
D_{{C}_j}\pig(\ell^+_2 Q_{\text{cut}2}^{\frac{1}{1+\beta_2}}, \beta_2, \mu \pig) 
&=& \int_0^{+\infty} \!\! \diff k_2 \: \hat D_{{C}_j} \pig( \ell^+_2 Q_{\text{cut}2}^{\frac{1}{1+\beta_2}} 
- k_2^{\frac{2+\beta_2}{1+\beta_2}}, \beta_2, \mu \pig)\, 
F_2^j(k_2,\beta_2)
\,. \nn
\eea
It is important to note that the shape function in momentum space only depends on $\beta_i$, but not on $z_{{\rm cut}\,i}$~\cite{Hoang:2019ceu}. The nonperturbative shape functions have their dominant support in the region $k_i\sim \Lambda_{\rm QCD}$, and must fall off faster than any polynomial for $k_i \gg \Lambda_{\rm QCD}$. They smear the perturbative CS functions.

With the shape functions included, the result in \Eq{eq:convol} becomes
\begin{align}
& P_j^{\rm CD}  (\Dms, Q, \tilde{z}_{{\rm cut}\,i},\beta_i,\mu) = \Qcuta^{\frac{1}{1+\beta_1}} 
\Qcutb^{\frac{1}{1+\beta_2}} \int \diff \ell_1^+ \diff \ell_2^+ \diff k_1 \diff k_2 \,\delta \big(\Dms - Q\ell_1^+ - Q\ell_2^+ \big) \nn\\
&\ \ \times
\hat S_{C_j} \pig( \ell^+_1 Q_{\text{cut}1}^{\frac{1}{1+\beta_1}} - k_1^{\frac{2+\beta_1}{1+\beta_1}}, \beta_1, \mu \pig)
\hat D_{{C}_j} \pig( \ell^+_2 Q_{\text{cut}2}^{\frac{1}{1+\beta_2}} - k_2^{\frac{2+\beta_2}{1+\beta_2}}, \beta_2, \mu \pig)
F_1^j(k_1,\beta_1)  F_2^j(k_2,\beta_2) \,.
\end{align}
Changing variables to 
$x_1 = \ell^+_1 Q_{\text{cut}1}^{\frac{1}{1+\beta_1}}- k_1^{\frac{2+\beta_1}{1+\beta_1}}$ and $x_2=\ell^+_2 Q_{\text{cut}2}^{\frac{1}{1+\beta_2}} - k_2^{\frac{2+\beta_2}{1+\beta_2}}$
we find
\begin{align}
& P_j^{\rm CD} (\Dms, Q, \tilde{z}_{{\rm cut}\,i},\beta_i,\mu) \nn\\
&\ \ = \int \diff x_1 \diff x_2 \diff k_1 \diff k_2\, \delta \bigg(\Dms - Q Q_{{\rm cut}1}^{\frac{-1}{1+\beta_1}}x_1 - Q Q_{{\rm cut}2}^{\frac{-1}{1+\beta_2}}x_2 - Q Q_{{\rm cut}1}^{\frac{-1}{1+\beta_1}}k_1^{\frac{2+\beta_1}{1+\beta_1}} - Q Q_{{\rm cut}2}^{\frac{-1}{1+\beta_2}} k_2^{\frac{2+\beta_2}{1+\beta_2}} \bigg)  \nn\\
&\ \ \ \ \times
\hat S_{C_j}( x_1, \beta_1, \mu)
\hat D_{{C}_j}(x_2, \beta_2, \mu)
F_1^j(k_1,\beta_1)  F_2^j(k_2,\beta_2) \nn\\
&\ \ = \int\!\! \diff (\Delta \hat m^2) \diff k_1 \diff k_2\,
 \delta \bigg(\Dms - \Delta \hat m^2 - Q Q_{{\rm cut}1}^{\frac{-1}{1+\beta_1}}k_1^{\frac{2+\beta_1}{1+\beta_1}} - Q Q_{{\rm cut}2}^{\frac{-1}{1+\beta_2}} k_2^{\frac{2+\beta_2}{1+\beta_2}} \bigg) 
F_1^j(k_1,\beta_1)  F_2^j(k_2,\beta_2) 
\nn\\
&\ \ \ \ \times \hat P_j^{\rm CD} (\Delta\hat m^2, Q, \tilde{z}_{{\rm cut}\,i},\beta_i,\mu)
\,,
\end{align}
which gives the result incorporating hadronization from the shape functions as a convolution with the perturbative result for $\hat P_j^{\rm CD}$, whose form with resummation was given in \eqn{Pcd_final}.

In the deep nonperturbative regime, where both the CS scales $\mu_{{\rm cs}\,i}$ enter the nonperturbative regime, we need to stop the RG evolution of both the CS functions at some small, but perturbative scales: $\mu_{{\rm cs}\,i}=\Lambda_{{\rm cs}\,i} \sim 2$ GeV. The final results of the CD jet mass are independent of the choice of $\Lambda_{{\rm cs}\,i}$ since dependence on $\Lambda_{{\rm cs}\,i}$ in the calculations of the CS functions will be canceled by the dependence in the shape functions. In this region it is convenient to define $\overline{\rm MS}$ scheme shape functions which combine the $F_1^j(k_1,\beta_1)$ and the boundary series at the low scale $\hat S_{C_j}(x_1,\beta_1,\Lambda_{{\rm cs}1})$ into a single function $F_1^{({\overline{\rm MS}})j}(k_1,\beta_1,\Lambda_{{\rm cs}1})$, and likewise combine $F_2^j(k_2,\beta_2)$ and $\hat D_{C_j}(x_2,\beta_2,\Lambda_{{\rm cs}2})$ into a single $F_2^{({\overline{\rm MS}})j}(k_2,\beta_2,\Lambda_{{\rm cs}2})$. This is most easily done in Laplace space, and the details are left to Appendix~\ref{app:laplace}. We refer to these functions as being in the $\overline{\rm MS}$ scheme since their dependence on the scales $\Lambda_{{\rm cs}\,i}$ exactly follows the $\overline{\rm MS}$ RGE for the collinear-soft functions. 
We thus obtain
\begin{align}
& P_j^{\rm CD} (\Dms, Q, \tilde{z}_{{\rm cut}\,i},\beta_i,\mu)
 \nn\\
&= \int\!\! \diff (\Delta \hat m^2) \diff k_1 \diff k_2\, \delta \bigg(\Dms - \Delta \hat m^2- Q Q_{{\rm cut}1}^{\frac{-1}{1+\beta_1}}k_1^{\frac{2+\beta_1}{1+\beta_1}} - Q Q_{{\rm cut}2}^{\frac{-1}{1+\beta_2}} k_2^{\frac{2+\beta_2}{1+\beta_2}} \bigg) 
\nn\\
&\times  \exp\bigg( -2C_j\frac{2+\beta_1}{1+\beta_1} K(\Lambda_{{\rm cs}1}, \mu) + \omega_{S_{C_j}}(\Lambda_{{\rm cs}1}, \mu) +  2C_j\frac{2+\beta_2}{1+\beta_2} K(\Lambda_{{\rm cs}2}, \mu) + \omega_{D_{{C}_j}}(\Lambda_{{\rm cs}2}, \mu) \bigg)
 \nn\\
&\times 
\Bigg( \frac{Q_{\text{cut}1}^{\frac{1}{1+\beta_1}}}{Q \Lambda_{{\rm cs}1}^{\frac{2+\beta_1}{1+\beta_1}}} \Bigg)^{2C_j\omega(\Lambda_{{\rm cs}1}, \mu)}
\Bigg( \frac{Q_{\text{cut}2}^{\frac{1}{1+\beta_2}}}{Q \Lambda_{{\rm cs}2}^{\frac{2+\beta_2}{1+\beta_2}}} \Bigg)^{-2C_j\omega(\Lambda_{{\rm cs}2}, \mu)} \,
 \frac{(e^{-\gamma_E}\Delta \hat m^2)^{\eta}}{\Delta \hat m^2 \Gamma(\eta)} \bigg|_{\eta = 2C_j\omega(\Lambda_{{\rm cs}1}, \Lambda_{{\rm cs}2})} 
\nn\\
\label{eq:Pcd_np}
&\times 
 F_1^{({\overline{\rm MS}})j}(k_1,\beta_1,\Lambda_{{\rm cs}1}) \:
 F_2^{({\overline{\rm MS}})j}(k_2,\beta_2,\Lambda_{{\rm cs}2}) \,.
\end{align}
Integrating over $\Delta \hat m^2$ to get the cumulative distribution gives
\begin{align}
&\int_0^{\Delta m_c^2}\!\!\! \diff \Delta m^2 \int\!\! \diff(\Delta \hat m^2) \,\delta \bigg(\Dms - \Delta \hat m^2 - Q Q_{{\rm cut}1}^{\frac{-1}{1+\beta_1}}k_1^{\frac{2+\beta_1}{1+\beta_1}} - Q Q_{{\rm cut}2}^{\frac{-1}{1+\beta_2}} k_2^{\frac{2+\beta_2}{1+\beta_2}} \bigg)
\frac{(\Delta \hat m^2)^{\eta}}{\Delta \hat m^2 \Gamma(\eta)}\nn\\
&= \frac{1}{\Gamma(\eta)}  \int_0^{\Delta m_c^2}\!\!\! \diff \Delta m^2 \,
\bigg(\Delta m^2  - Q Q_{{\rm cut}1}^{\frac{-1}{1+\beta_1}}k_1^{\frac{2+\beta_1}{1+\beta_1}} - Q Q_{{\rm cut}2}^{\frac{-1}{1+\beta_2}} k_2^{\frac{2+\beta_2}{1+\beta_2}} \bigg)^{\eta-1} \nn\\
&\qquad \times
\Theta\bigg(\Delta m^2  - Q Q_{{\rm cut}1}^{\frac{-1}{1+\beta_1}}k_1^{\frac{2+\beta_1}{1+\beta_1}} - Q Q_{{\rm cut}2}^{\frac{-1}{1+\beta_2}} k_2^{\frac{2+\beta_2}{1+\beta_2}} \bigg) \nn\\
&= \frac{1}{\Gamma(1+\eta)} \bigg(\Delta m^2_c  - Q Q_{{\rm cut}1}^{\frac{-1}{1+\beta_1}}k_1^{\frac{2+\beta_1}{1+\beta_1}} - Q Q_{{\rm cut}2}^{\frac{-1}{1+\beta_2}} k_2^{\frac{2+\beta_2}{1+\beta_2}} \bigg)^{\eta} 
\nn\\
&\qquad \times
\Theta\bigg(\Delta m^2_c - Q Q_{{\rm cut}1}^{\frac{-1}{1+\beta_1}}k_1^{\frac{2+\beta_1}{1+\beta_1}} - Q Q_{{\rm cut}2}^{\frac{-1}{1+\beta_2}} k_2^{\frac{2+\beta_2}{1+\beta_2}} \bigg) \,,
\end{align}
where the $\Theta$-function originates from the fact $\Delta \hat m^2\geq0$. After the integration, we can fix the scales. We are free to pick a common scale $\Lambda_{{\rm cs}1} = \Lambda_{{\rm cs}2} \equiv \Lambda_{{\rm cs}}$, and will use $\Lambda_{\rm cs}=2\,{\rm GeV}$ as our default choice. This choice can be made without loss of generality, since the $\Lambda_{{\rm cs}\,i}$ dependence of the RG evolution factors will be exactly canceled order-by-order by the dependence in the $\overline{\rm MS}$ shape functions. Effectively this just corresponds to choosing the values of $\Lambda_{{\rm cs}\,i}$ at which the shape functions in the $\overline{\rm MS}$ scheme are defined, as discussed in Appendix~\ref{app:laplace}.  With this choice the value of $\eta = 2C_j\omega(\Lambda_{{\rm cs}1},\Lambda_{{\rm cs}2})$ becomes zero and all $\Delta m_c^2$ dependence is given by the shape functions. This is compatible with the constant values obtained for the perturbative cumulative cross sections in Section~\ref{sec:cumulative}.

Putting everything together, we find the cumulative jet mass in the deep nonperturbative regime is given by 
\begin{align} \label{eq:all-order}
\Sigma(\Delta m_c) = \sum_{j=q,g} f_j\, \hat \Sigma_j \, \ml{F}_j(\Delta m_c) 
\,,
\end{align}
where the $\Delta m_c$ independent perturbative cumulant cross sections are given by
\begin{align} \label{eq:SigAllorderNP}
\hat \Sigma_j & = \frac{1}{\sigma_j} 
  S^{ee}_{G_j}(Q_{{\rm gs}1},\beta_1,\mu_{{\rm gs}1}) 
  S^{ee}_{\overline{G}_j}(Q_{{\rm gs}2},\beta_2,\mu_{{\rm gs}2}) \nn\\
&\ \ \times 
\exp \bigg[ \frac{2C_j}{1+\beta_1} K(\mu_{{\rm gs}1},\mu) + \omega_{S_{G_j}}(\mu_{{\rm gs}1},\mu) - \frac{2C_j}{1+\beta_2} K(\mu_{{\rm gs}2},\mu) + \omega_{S_{\overline{G}_j}}(\mu_{{\rm gs}2},\mu)
\bigg] \nn\\
& \ \ \times \exp \bigg[ \frac{2 C_j(\beta_1-\beta_2)}{(1+\beta_1)(1+\beta_2)} 
    K(\Lambda_{\text{cs}}, \mu) + \omega_{S_{C_j}}(\Lambda_{{\rm cs}}, \mu)
    + \omega_{D_{{C}_j}}(\Lambda_{{\rm cs}}, \mu)
    \bigg] \nn\\
&\ \ \times 
\Big( \frac{\mu_{{\rm gs}1}}{Q_{{\rm gs}1}} \Big)^{\frac{2C_j}{1+\beta_1}\omega(\mu_{{\rm gs}1},\mu)} 
\Big( \frac{\mu_{{\rm gs}2}}{Q_{{\rm gs}2}} \Big)^{\frac{-2C_j}{1+\beta_2}\omega(\mu_{{\rm gs}2},\mu)}
\Bigg( \frac{Q_{\text{cut}1}^{\frac{1}{1+\beta_1}}}     
     {\Lambda^{\frac{1}{1+\beta_1}}_{\text{cs}}}
    \frac{\Lambda^{\frac{1}{1+\beta_2}}_{\text{cs}}}
     {Q_{\text{cut}2}^{\frac{1}{1+\beta_2}}}
 \Bigg)^{2C_j\omega(\Lambda_{\text{cs}}, \mu)} \,,
\end{align}
and now we have generalized the $\Delta m_c$ dependent nonperturbative function to
\begin{align} \label{eq:two_shapeAllorder}
\ml{F}_j(\Delta m_c) & = \int\diff k_1 \diff k_2 \, F_1^{(\overline{\rm MS})j}(k_1,\beta_1, \Lambda_{\rm cs}) F_2^{(\overline{\rm MS})j}(k_2, \beta_2, \Lambda_{\rm cs}) \nn\\
& \ \ \times
\Theta\Big(\Delta m^2_c - Q Q_{{\rm cut}1}^{\frac{-1}{1+\beta_1}}k_1^{\frac{2+\beta_1}{1+\beta_1}} - Q Q_{{\rm cut}2}^{\frac{-1}{1+\beta_2}} k_2^{\frac{2+\beta_2}{1+\beta_2}} \Big) \,.
\end{align}
The shape functions in the $\overline{\rm MS}$ scheme are defined in \Eqs{eq:Fmsbar1}{eq:LapFmsbar}, and contain both the original shape functions and the boundary terms of the CS functions.
 
This result simplifies at NLL accuracy, where various fixed order contributions can be neglected. We find the cumulative jet mass in the deep nonperturbative region can be written as
\begin{align}
\Sigma^{{\rm NLL}}(\Delta m_c) = \sum_{j=q,g} f_j\, \hat \Sigma_j^{{\rm NLL}} \, \ml{F}_j(\Delta m_c) 
\,,
\end{align}
where $f_q$ and $f_g$ are again the fractions of quark and gluons jets, the 
perturbative cumulant cross sections for quarks and gluons are
\begin{align} \label{eq:SigNLLNP}
\hat\Sigma_j^{\rm NLL} & = 
\exp \bigg[ \frac{2C_j}{1+\beta_1} K(\mu_{{\rm gs}1},\mu) - \frac{2C_j}{1+\beta_2} K(\mu_{{\rm gs}2},\mu)  
+  \frac{2 C_j(\beta_1-\beta_2)}{(1+\beta_1)(1+\beta_2)} 
    K(\Lambda_{\text{cs}}, \mu) 
 \bigg]
\nn \\[5pt]
&\ \ \times
\Big( \frac{\mu_{{\rm gs}1}}{Q_{{\rm gs}1}} \Big)^{\frac{2C_j}{1+\beta_1}\omega(\mu_{{\rm gs}1},\mu)} 
\Big( \frac{\mu_{{\rm gs}2}}{Q_{{\rm gs}2}} \Big)^{\frac{-2C_j}{1+\beta_2}\omega(\mu_{{\rm gs}2},\mu)}
\Bigg( \frac{Q_{\text{cut}1}^{\frac{1}{1+\beta_1}}}     
     {\Lambda^{\frac{1}{1+\beta_1}}_{\text{cs}}}
    \frac{\Lambda^{\frac{1}{1+\beta_2}}_{\text{cs}}}
     {Q_{\text{cut}2}^{\frac{1}{1+\beta_2}}}
 \Bigg)^{2C_j\omega(\Lambda_{\text{cs}}, \mu)}
 \,, 
\end{align}
and the shape function ${\cal F}_j(\Delta m_c)$ is a simple combination of the original $F_i^j(k_i,\beta_i)$ shape functions,
\begin{align} \label{eq:two_shape}
& \ml{F}_j(\Delta m_c) = \int\diff k_1 \diff k_2 \, F_1^j(k_1,\beta_1) F_2^j(k_2,\beta_2) \,
\Theta\Big(\Delta m^2_c - Q Q_{{\rm cut}1}^{\frac{-1}{1+\beta_1}}k_1^{\frac{2+\beta_1}{1+\beta_1}} - Q Q_{{\rm cut}2}^{\frac{-1}{1+\beta_2}} k_2^{\frac{2+\beta_2}{1+\beta_2}} \Big) \,.\nn\\
\end{align}

As can be seen from either of \eqns{SigAllorderNP}{SigNLLNP}, the cumulative jet mass cross section in this small $\Delta m_c^2$ region now depends on $\Delta m_c^2$, in contrast to the purely perturbative result~(\ref{eq:ll_nll}) where it was constant. To evaluate the cumulative jet mass, we need to include the nonperturbative shape functions, for which we discuss general models in the next subsection.

\subsection{Models for Shape Functions}
\label{sec:model_shape}

Due to confinement, any moment of the momentum space shape functions must exist, implying that they fall off at large momentum faster than any polynomial.
We consider expanding the momentum space shape function $F_i^j(k_i,\beta_i)$ in terms of some basis of functions that are integrable on $[0,+\infty)$. To make the expansion converge fast, a necessary condition is that the $n$-th moment of the basis function does not grow with $n$, since it is expected that the $n$-th moment of the momentum space shape function scales as $(\Lambda_{\rm QCD})^n$. A good basis for expanding the shape function has been constructed in Ref.~\cite{Ligeti:2008ac}, where the expansion can be written as
\bea
\label{eq:sf_expansion}
F^j_i(k_i, \beta_i) = \frac{1}{\Lambda} \bigg[ \sum_{n=0}^\infty c^j_n(\beta_i) f_n(x,p) \bigg]^2 \,,
\eea
where $i=1,2$ for the two shape functions, $j=q,g$ represents the jet content (quark or gluon), $x={k_i}/{\Lambda}$, the expansion coefficients $c_n^j(\beta_i)$ are numbers, and $\Lambda\sim \Lambda_{\rm QCD}$ is a nonperturbative scale introduced to make the mass dimension of $F^j_i(k_i, \beta_i)$ to be $-1$ and the basis functions $f_n(x,p)$ dimensionless. The orthonormal basis functions are given by
\begin{align}
f_n(x,p) &= \sqrt{(2n+1)Y(x,p)} \: P_n\big(y(x,p)\big) 
\,, 
& y(x,p) &= -1 + 2\int_0^x \diff x' \, Y(x',p) \,,\nn \\
Y(x,p) &= \frac{(p+1)^{p+1}}{\Gamma(p+1)} x^p e^{-(p+1)x} \,,
\end{align}
where $p>0$ is a parameter and $P_n$ are the standard Legendre polynomials. If we sum over all of the basis functions, the completeness of the basis functions make the final result independent of $p$. However, in practical applications the sum on $n$ is truncated after some number of terms, and the choice of $p$ affects how well this truncated series describes any given shape function model with only a finite number of terms.  For our analysis we will vary the value of $p$. Note that any choice with $p>0$ will cause the collinear drop jet mass cross section to go to zero as $\Delta m^2\to 0$, and thus makes the assumption that nonperturbative radiation always populates the collinear drop region. In contrast, the choice $p=0$ gives a non-zero probability of having no nonperturbative radiation in this region.  The basis functions are orthonormal with $\int dx f_n(x,p) f_m(x,p)=\delta_{n,m}$. This implies that a normalization condition on $F_i^j(k_i,\beta_i)$ can be implemented as a constraint that the sum of squares of the coefficients $c_n^j(\beta_i)$ is equal to one. Since the shape functions in soft drop (and thus in collinear drop) are not normalized~\cite{Hoang:2019ceu}, we do not impose any such normalization constraint in our implementation of the shape functions that are used here.

\section{Pure Quark and Gluon Observables}
\label{sec:obs}

In this section, we construct observables that are pure quark or pure gluon, by exploiting the structure of the perturbative result of the cumulative jet mass distribution in resummation and small jet mass regions and the property of the shape functions $F_i^j(k_i,\beta_i)$ that they are independent of ${z}_{{\rm cut}\,i}$.  We demonstrate that the constructed observables work equally well in the region where perturbative resummation dominates and in the nonperturbative region.

\subsection{Construction}

\subsubsection{Linear Combinations}

We take two sets of collinear drop parameters $(z^{(a)}_{{\rm cut}\,i}, \beta_i)$ and $(z^{(b)}_{{\rm cut}\,i}, \beta_i)$ where the $z_{{\rm cut}\,i}$ parameters are different while the $\beta_i$ parameters are the same between the set $(a)$ and set $(b)$. We then take linear combination of the cumulative jet mass cross sections in the two sets
\begin{align} \label{eq:QGdefns}
\ml{Q} &=  \Sigma(\Delta m_c^{(b)},p_T,\eta_J,R,z^{(b)}_{{\rm cut}\,i}, \beta_i) - \xi_g \Sigma(\Delta m_c^{(a)},p_T,\eta_J,R,z^{(a)}_{{\rm cut}\,i}, \beta_i) \,, \\
\ml{G} &=  \Sigma(\Delta m_c^{(b)},p_T,\eta_J,R,z^{(b)}_{{\rm cut}\,i}, \beta_i) - \xi_q \Sigma(\Delta m_c^{(a)},p_T,\eta_J,R,z^{(a)}_{{\rm cut}\,i}, \beta_i) \,, \nn
\end{align}
where $\xi_g$ and $\xi_q$ are coefficients of the linear combinations that we are still free to pick, and which will be fixed below. The form in \eqn{QGdefns} is suitable for use in experimental measurement once the $\xi_j$ values have been provided.
 
To fix values for the $\xi_j$ that yield pure quark and gluon observables we make use of our factorization framework. We will first calculate the $\xi_j$
at small $\Delta m_c^2$, where we are in the fully nonperturbative regime (pictured in \Fig{fig:modes}c,d). Then we will prove that the same values of the $\xi_j$ also lead to pure quark and gluon observables in the fully perturbative regime.

In the fully nonperturbative regime, our result for the cumulative distribution gave
\be
\Sigma(\Delta m_c) = \sum_{j=q,g} f_j\, \hat\Sigma_j\, \ml{F}_j(\Delta m_c) \,,
\ee
where the only $\Delta m_c$ dependence is in the nonperturbative $\ml{F}_j$, while the resummed coefficients $\hat\Sigma_j$ are constants depending on the collinear drop parameters.
This leads to
\bea 
\ml{Q} &=& \sum_{j=q,g} f_j\, \Big( \hat \Sigma_j^{(b)} \, \ml{F}_j^{(b)} - \xi_g \hat\Sigma_j^{(a)} \, \ml{F}_j^{(a)} \Big) \,,\\
\ml{G} &=& \sum_{j=q,g} f_j\, \Big( \hat\Sigma_j^{(b)} \, \ml{F}_j^{(b)} - \xi_q \hat\Sigma_j^{(a)} \, \ml{F}_j^{(a)}\Big) \,,\nn
\eea
where the superscripts $(a)$ and $(b)$ indicate that the parameters used in $\hat\Sigma_j$ and ${\cal F}_j$ differ for the two sets.
Recall that the hard processes do not know anything about the grooming and the quark and gluon fractions depend on the jet kinematics: $p_T$, $\eta_J$ and $R$ and are insensitive to the grooming parameters.\footnote{The $f_j$ are explicitly independent of the grooming parameters at NLL, and beyond NLL our expressions for $f_j$ could potentially have small dependence on the grooming parameters due to non-global contributions, which should however be canceled order-by-order in perturbation theory, due to the $\hat \Sigma_j(\Delta m_c^{\rm max})=1$ constraints. Hence it is reasonable to assume they are independent for our construction. } To remove the $j=g$ contribution from the ${\cal Q}$ observable, and the $j=q$ contribution from the ${\cal G}$ observable, the coefficients of the linear combinations should be chosen such that
\bea \label{eq:xiqgF}
\hat \Sigma_g^{(b)} \, \ml{F}_g^{(b)} - \xi_g \hat \Sigma_g^{(a)} \, \ml{F}_g^{(a)} =0 \,,
\qquad\qquad
\hat \Sigma_q^{(b)} \, \ml{F}_q^{(b)} - \xi_q \hat \Sigma_q^{(a)} \, \ml{F}_q^{(a)} = 0 \,.
\eea
However, immediately we see a problem for this construction so far. The solutions to $\xi_{g}$ and $\xi_q$ in \eqn{xiqgF} depend on the nonperturbative shape functions, which are not known. This means we cannot use these results to predict the linear combination coefficients needed to define the pure quark and gluon observables. We can overcome this difficulty by exploiting our ability to use different bins $\Delta m_c^{(a)}$ and $\Delta m_c^{(b)}$ for the two jet masses associated with the two CD parameters, as we will see next.

\subsubsection{Binning Jet Masses}
\label{sec:binnedmJ}

Let us have a closer look at the nonperturbative shape functions appearing in the cumulative jet mass distributions in the nonperturbative regime, which are given by Eq.~(\ref{eq:two_shape}):
\bea
\ml{F}_j\big( \Delta m_c^{(a)} \big) &=& \int\diff k_1 \diff k_2 \, F_1^{(\overline{\rm MS})j}(k_1,\beta_1,\Lambda_{\rm cs}) F_2^{(\overline{\rm MS})j}(k_2,\beta_2,\Lambda_{\rm cs}) \,
\nn\\
&\times& \Theta\Big(1 - \frac{Q (Q_{{\rm cut}1}^{(a)})^{\frac{-1}{1+\beta_1}}}{(\Delta m_c^{(a)})^2} k_1^{\frac{2+\beta_1}{1+\beta_1}} - \frac{Q (Q_{{\rm cut}1}^{(a)})^{\frac{-1}{1+\beta_2}}}{(\Delta m_c^{(a)})^2} k_2^{\frac{2+\beta_2}{1+\beta_2}} \Big) 
\,,\nn \\
\ml{F}_j\big( \Delta m_c^{(b)} \big) &=& \int\diff k_1 \diff k_2 \, F_1^{(\overline{\rm MS})j}(k_1,\beta_1,\Lambda_{\rm cs}) F_2^{(\overline{\rm MS})j}(k_2,\beta_2,\Lambda_{\rm cs})
 \nn\\
&\times& \Theta\Big(1 - \frac{Q (Q_{{\rm cut}1}^{(b)})^{\frac{-1}{1+\beta_1}}}{(\Delta m_c^{(b)})^2} k_1^{\frac{2+\beta_1}{1+\beta_1}} - \frac{Q (Q_{{\rm cut}1}^{(b)})^{\frac{-1}{1+\beta_2}}}{(\Delta m_c^{(b)})^2} k_2^{\frac{2+\beta_2}{1+\beta_2}} \Big) \,.
\eea
Since the CD parameters $\beta_i$ are the same in the set $(a)$ and set $(b)$, we find that we can make the shape functions $\ml{F}_j^{(a)}$ and $\ml{F}_j^{(b)}$ the same by choosing the jet masses and the CD parameters $z_{{\rm cut}\,i}^{(a,b)}$ such that
\bea
\label{eq:mass_ratio1}
(Q^{(a)}_{{\rm cut}1})^{\frac{1}{1+\beta_1}} (\Delta m_c^{(a)})^2 &=& (Q^{(b)}_{{\rm cut}1})^{\frac{1}{1+\beta_1}}(\Delta m_c^{(b)})^2
 \,, \\
(Q^{(a)}_{{\rm cut}2})^{\frac{1}{1+\beta_2}}(\Delta m_c^{(a)})^2 &=& (Q^{(b)}_{{\rm cut}2})^{\frac{1}{1+\beta_2}}(\Delta m_c^{(b)})^2 
 \nn \,.
\eea
These can be rewritten as
\begin{align} \label{eq:QGchoice}
  \frac{(\Delta m_c^{(a)})^2}{(\Delta m_c^{(b)})^2}
 &= \bigg( \frac{Q^{(b)}_{{\rm cut}1}}{Q^{(a)}_{{\rm cut}1}}
    \bigg)^{\frac{1}{1+\beta_1}}
  = \bigg( \frac{Q^{(b)}_{{\rm cut}2}}{Q^{(a)}_{{\rm cut}2}}
    \bigg)^{\frac{1}{1+\beta_2}} 
  \nn\\
 &= \bigg( \frac{ z^{(b)}_{{\rm cut}1} }{z^{(a)}_{{\rm cut}1}}
    \bigg)^{\frac{1}{1+\beta_1}}
  = \bigg( \frac{ z^{(b)}_{{\rm cut}2} }{z^{(a)}_{{\rm cut}2}}
    \bigg)^{\frac{1}{1+\beta_2}} 
 \,,
\end{align}
where in the second line we have used $ Q_{{\rm cut}\,i}^{(a,b)} =  z_{{\rm cut}\,i}^{(a,b)} Q \big( 2\cosh\eta_J/R_0 \big)^{\beta_i}$. In practice we solve these equations by specifying $z_{{\rm cut}2}^{(b)}$ and $(\Delta m_c^{(b)})^2$ in terms of the other variables:
\begin{align} \label{eq:z_ratio}
 (\Delta m_c^{(b)})^2 &= (\Delta m_c^{(a)})^2
  \bigg( \frac{ z^{(a)}_{{\rm cut}1} }{z^{(b)}_{{\rm cut}1}}
  \bigg)^{\frac{1}{1+\beta_1}}
   \,,
& z_{{\rm cut}2}^{(b)} & = z_{{\rm cut}2}^{(a)} \bigg( \frac{z_{{\rm cut}1}^{(b)}}{z_{{\rm cut}1}^{(a)}} \bigg)^{\frac{1+\beta_2}{1+\beta_1}} \,.
\end{align}
Recall that we considered soft drop parameters that satisfy the constraint $z_{{\rm cut}1}\le z_{{\rm cut}2}$ to ensure that the SD$_2$ grooming is stronger than that of SD$_1$. The solution for $z_{{\rm cut}2}^{(b)}$  in \Eq{eq:z_ratio} is always compatible with this constraint for $\beta_1=\beta_2$, while for $\beta_1>\beta_2$ it is always compatible as long as $z_{{\rm cut}1}^{(b)}\le z_{{\rm cut}1}^{(a)}$.  If we have $z_{{\rm cut}1}^{(b)}> z_{{\rm cut}1}^{(a)}$ then it may still be compatible, but we must confirm this for each set of parameters considered.

Using \Eq{eq:z_ratio}, we have by construction a common nonperturbative shape function for the $(a)$ and $(b)$ sets,
\be
\ml{F}_j^{(a)} = \ml{F}_j^{(b)} \equiv \ml{F}_j\,.
\ee
With this setup the pure quark and gluon observables in \eqn{xiqgF} now become
\bea \label{eq:nonpertQG}
\ml{Q} &=& \sum_{j=q,g} f_j\, \ml{F}_j \Big( \hat \Sigma_j^{(b)} 
    - \xi_g  \hat \Sigma_j^{(a)}  \Big)  
\,,  \\
\ml{G} &=& \sum_{j=q,g} f_j\, \ml{F}_j \Big( \hat \Sigma_j^{(b)} - \xi_q \hat \Sigma_j^{(a)} \Big) 
\,. \nn
\eea
Now the desired linear combination coefficients in \eqn{xiqgF} are fixed by purely perturbative functions, and can be solved to give
\be  \label{eq:xisolns}
\xi_g = \frac{\hat\Sigma_g^{(b)}}{\hat\Sigma_g^{(a)}} \,,
\qquad\qquad
 \xi_q = \frac{\hat\Sigma_q^{(b)}}{\hat\Sigma_q^{(a)}} \,.
\ee
This entire construction, including these perturbative expressions, can be used even beyond NLL order as long as non-global dependence on collinear drop parameters is confirmed to be small in the $f_j$. We note that due to \eqn{QGchoice} all dependence on the scale $\Lambda_{\rm cs}$ fully cancels out in the ratios in \eqn{xisolns}, so that calculations of $\xi_{g,q}$ are only sensitive to perturbative results at and above the global-soft scale.
With these choices we have the final results for our pure quark and pure gluon observables, so far in the nonperturbative regime, namely
\bea \label{eq:QG}
\ml{Q} &=& f_q\, \ml{F}_q \Big( \hat \Sigma_q^{(b)}
     - \xi_g \hat \Sigma_q^{(a)}  \Big)  
\,,  \\
\ml{G} &=&  f_g\, \ml{F}_g \Big( \hat \Sigma_g^{(b)}
     - \xi_q \hat \Sigma_g^{(a)} \Big) 
\,. \nn
\eea
This demonstrates that with the above choices these observables are predicted to be entirely quark or gluon dominated as desired. Here the entire contributions in brackets are perturbative. Furthermore, the above construction leaves $\beta_1$, $\beta_2$, $z_{{\rm cut}1}^{(a)}$, $z_{{\rm cut}2}^{(a)}$ and $z_{{\rm cut}1}^{(b)}$ as free variables that can be varied, thus giving a number of possible variables which have the pure quark or pure gluon property for a range of choices for the kinematic mass variable $\Delta m_c^{(a)}$.

Recall that for an experimental measurement of the ${\cal Q}$ or ${\cal G}$ observable we use \eqn{QGdefns}, which only requires perturbative input to determine the values to use for $\xi_q$ and $\xi_g$. Using the NLL result in \eqn{SigNLLNP} as input for \eqn{xisolns} we obtain
\begin{align}
\label{eq:xipre}
\xi_j &= \exp\bigg\{ \frac{2C_j}{1+\beta_1} \Big[ K(\mu_{{\rm gs}1}^{(b)},\mu) - K(\mu_{{\rm gs}1}^{(a)},\mu) \Big] - \frac{2C_j}{1+\beta_2} \Big[ K(\mu_{{\rm gs}2}^{(b)},\mu) - K(\mu_{{\rm gs}2}^{(a)},\mu) 
\Big]\bigg\} 
 \\
&\ \ \times 
\biggl( \frac{\mu^{(b)}_{{\rm gs}1}}{Q^{(b)}_{{\rm gs}1}} \biggr)^{\frac{2C_j}{1+\beta_1}\omega(\mu^{(b)}_{{\rm gs}1},\mu)} 
\biggl( \frac{\mu^{(b)}_{{\rm gs}2}}{Q^{(b)}_{{\rm gs}2}} \biggr)^{\frac{-2C_j}{1+\beta_2}\omega(\mu^{(b)}_{{\rm gs}2},\mu)} 
\biggl( \frac{\mu^{(a)}_{{\rm gs}1}}{Q^{(a)}_{{\rm gs}1}} \biggr)^{\frac{-2C_j}{1+\beta_1}\omega(\mu^{(a)}_{{\rm gs}1},\mu)} 
\biggl( \frac{\mu^{(a)}_{{\rm gs}2}}{Q^{(a)}_{{\rm gs}2}} \biggr)^{\frac{2C_j}{1+\beta_2}\omega(\mu^{(a)}_{{\rm gs}2},\mu)} \,.
 \nn
\end{align}
Though the above expression of $\xi_j$ seems to depend on the scale $\mu$, it can be simplified to a form that is $\mu$-independent.
First we use the definition of $K(\mu_1,\mu_2)$ and $\omega(\mu_1,\mu_2)$ in Eq.~(\ref{eq:K}), and $\diff \ln\mu = \diff \alpha_s/\beta(\alpha_s)$ to obtain the relation
\begin{align} \label{eq:mu_indep}
 K(\mu_1,\mu) - K(\mu_2,\mu) 
&= K(\mu_1,\mu_2) + \omega(\mu_2,\mu)\: \ln\frac{\mu_2}{\mu_1} \,.
\end{align}
Using this in \eqn{xipre} gives
\begin{align}
\xi_j  &=\exp\bigg\{
 \frac{2C_j}{1+\beta_1} K(\mu_{{\rm gs}1}^{(b)}, \mu_{{\rm gs}1}^{(a)})
 - \frac{2C_j}{1+\beta_2} K(\mu_{\rm{gs}2}^{(b)}, \mu_{\rm{gs}2}^{(a)})
 \bigg\}  \\
& \ \ \times
\biggl( \frac{\mu^{(b)}_{{\rm gs}1}}{Q^{(b)}_{{\rm gs}1}} \biggr)^{\frac{2C_j}{1+\beta_1}\omega(\mu^{(b)}_{{\rm gs}1},\mu)} 
\biggl( \frac{\mu^{(b)}_{{\rm gs}2}}{Q^{(b)}_{{\rm gs}2}} \biggr)^{\frac{-2C_j}{1+\beta_2}\omega(\mu^{(b)}_{{\rm gs}2},\mu)} 
\biggl( \frac{\mu^{(b)}_{{\rm gs}1}}{Q^{(a)}_{{\rm gs}1}} \biggr)^{\frac{-2C_j}{1+\beta_1}\omega(\mu^{(a)}_{{\rm gs}1},\mu)} 
\biggl( \frac{\mu^{(b)}_{{\rm gs}2}}{Q^{(a)}_{{\rm gs}2}} \biggr)^{\frac{2C_j}{1+\beta_2}\omega(\mu^{(a)}_{{\rm gs}2},\mu)} 
  \,. \nn
\end{align}
Inserting $1=Q_{{\rm gs}1}^{(b)}/Q_{{\rm gs}1}^{(b)}$ and $1=Q_{{\rm gs}2}^{(b)}/Q_{{\rm gs}2}^{(b)}$ in the last two ratios, recollecting common fractions, and using the relation in \eqn{QGchoice} then gives our final NLL result
\begin{align} \label{eq:finalxij}
\xi_j  &= \exp\bigg[ 
  \frac{2C_j}{1+\beta_1}
  K\bigl(\mu_{{\rm gs}1}^{(b)}, \mu_{{\rm gs}1}^{(a)}\bigr) 
  - \frac{2C_j}{1+\beta_2} 
  K\bigl(\mu_{\rm{gs}2}^{(b)}, \mu_{\rm{gs}2}^{(a)}\bigr) 
  + \frac{2C_j}{1+\beta_1}
  \omega\bigl(\mu_{\rm{gs}1}^{(a)},\mu_{\rm{gs}2}^{(a)}\bigr)
  \ln\frac{z_{{\rm cut}1}^{(a)}}{z_{\rm{cut}1}^{(b)}} 
  \bigg]
 \nn\\
&\quad\times \biggl( \frac{\mu^{(b)}_{{\rm gs}1}}{Q^{(b)}_{{\rm gs}1}} \biggr)^{\frac{2C_j}{1+\beta_1}\omega\bigl(\mu^{(b)}_{{\rm gs}1},\,\mu^{(a)}_{{\rm gs}1}\bigr)} 
\biggl( \frac{\mu^{(b)}_{{\rm gs}2}}{Q^{(b)}_{{\rm gs}2}} \biggr)^{\frac{-2C_j}{1+\beta_2}\omega\bigl(\mu^{(b)}_{{\rm gs}2},\,\mu^{(a)}_{{\rm gs}2}\bigr)} 
 \,,
\end{align}
which is now manifestly $\mu$-independent. We see here explicitly that the $\xi_j$ are sensitive to perturbative contributions at and above the global-soft scales. Later we exploit the dependence on the global-soft scales $\mu_{{\rm gs}\,i}^{(a,b)}$ in the evaluations of these coefficients to investigate the perturbative uncertainty in the NLL determination of $\xi_{q,g}$, and hence the definition of the ${\cal Q}$ and ${\cal G}$ observables.  

It is worth emphasizing that although we have quoted explicit results for $\xi_j$ by working with NLL expressions, the general construction still applies at higher orders in the resummed perturbation theory. (The only potential caveat that must be checked at higher orders is that non-global corrections have small enough dependence on the collinear drop parameters in the $f_j$, that they can continue to be pulled out as common factors.)

\subsubsection{Pure Quark and Gluon Observables in the Perturbative Region}

Next we consider the pure quark and gluon observables in the $\Delta m_c^2$ region that is dominated by perturbative contributions in the collinear drop resummation region, corresponding to that illustrated in \Fig{fig:modes}a. In this region the nonperturbative corrections are power suppressed and \eqn{QGdefns} gives
\begin{align}  \label{eq:QGpertstart}
\ml{Q} &= \sum_{j=q,g} f_j\, \Big[ 
  \hat \Sigma_j^{(b)}\big(\Delta m_c^{(b)}\big)  
   - \xi_g \hat\Sigma_j^{(a)}\big(\Delta m_c^{(a)}\big) \Big] 
  \times \Bigg[1 + 
 {\cal O}\Bigg(\frac{Q\: \Lambda_{\rm QCD}^{\frac{2+\beta_1}{1+\beta_1}} }
 {\big(\Delta m_c^{(b)}\big)^2\, (Q^{(b)}_{{\rm cut}1})^{\frac{1}{1+\beta_1}} }
 \Bigg) \Bigg]
 \,,\\
\ml{G} &= \sum_{j=q,g} f_j\, \Big[ 
  \hat \Sigma_j^{(b)}\big(\Delta m_c^{(b)}\big)  
   - \xi_q \hat\Sigma_j^{(a)}\big(\Delta m_c^{(a)}\big) \Big] 
 \times \Bigg[1 + 
 {\cal O}\Bigg(\frac{Q\: \Lambda_{\rm QCD}^{\frac{2+\beta_1}{1+\beta_1}} }
 {\big(\Delta m_c^{(b)}\big)^2\, (Q^{(b)}_{{\rm cut}1})^{\frac{1}{1+\beta_1}} }
 \Bigg) \Bigg]
  \,,\nn
\end{align}
where the scaling for the dominant power suppressed terms is shown in the $+\,{\cal O}(\cdots)$.  In this region it turns out that the same choice for $\xi_g$ and $\xi_q$ given by \eqn{xisolns}, again leads to pure quark and gluon observables.  This occurs because the weights that are needed to obtain quark and gluon observables in this region, $\xi_g=\hat\Sigma_g^{(b)}(\Delta m_c^{(b)})/\hat\Sigma_g^{(a)}(\Delta m_c^{(a)})$ and $\xi_q=\hat\Sigma_q^{(b)}(\Delta m_c^{(b)})/\hat\Sigma_q^{(a)}(\Delta m_c^{(a)})$, are independent of $\Delta m_c$.

To demonstrate this independence of $\xi_j$ to $\Delta m_c$ we first recall that \eqn{z_ratio} fixes $\Delta m_c^{(b)}$ in terms of $\Delta m_c^{(a)}$.  Examining the all-orders perturbative cumulative cross section in \eqn{Sigma_allorder} we see that there are two places that $\Delta m_c$ dependence arises, through the explicit $(\Delta m_c^2)^\eta$ and through the canonical collinear-soft scales which are functions of the collinear drop jet mass, $\mu_{{\rm cs}\,i}(\Delta m_c)=Q_{{\rm cs}\,i}$ given in \eqn{gscs-scales}. The conditions used to define the pure quark and pure gluon observables in \eqn{QGchoice} imply that these canonical scales are actually related by
\begin{align} \label{eq:musc-relation}
\mu_{{\rm cs}\,i}^{(a)}\big(\Delta m_c^{(a)}\big) 
  &= \mu_{{\rm cs}\,i}^{(b)}\big(\Delta m_c^{(b)}\big) \,.
\end{align}
This implies that in the ratios $\hat\Sigma_j^{(b)}(\Delta m_c^{(b)})/\hat\Sigma_j^{(a)}(\Delta m_c^{(a)})$ all dependence on factors like $K(\mu_{{\rm cs}\,i},\mu)$, $\omega_{S_{C_j}}(\mu_{{\rm cs}1})$, and $\omega_{S_{D_j}}(\mu_{{\rm cs}2})$ immediately cancels. The remaining $\mu_{{\rm cs}\,i}$ dependent factors on the second and fifth lines of \eqn{Sigma_allorder} can be assembled into the form
\begin{align} \label{eq:resDelmc}
 \big(\Delta m_c^2\big)^{-\eta} 
 \hat{\widetilde{S}}_{C_j}\! \pig( QQ_{\text{cut}1}^{\frac{-1}{1+\beta_1}}e^{-\frac{\partial}{\partial\eta}}, \beta_1, \mu_{{\rm cs}1} \pig) 
 \hat{\widetilde{D}}_{{C}_j}\! \pig( QQ_{\text{cut}2}^{\frac{-1}{1+\beta_2}}e^{-\frac{\partial}{\partial\eta}}, \beta_2, \mu_{{\rm cs}2} \pig)
\frac{( e^{-\gamma_E} \Dms_c)^\eta}{\Gamma(1+\eta)}
\bigg|_{\eta = 2C_j \omega(\mu_{\text{cs}1}, \mu_{\text{cs}2})}
 .
\end{align}
Thus the $\xi_j$ involve one factor of \eqn{resDelmc} in the numerator and denominator for the $(b)$ and $(a)$ sets respectively. 
Here the action of the derivative operators in $\hat{\widetilde{S}}_{C_j}$ and $\hat{\widetilde{D}}_{{C}_j}$ is to induce in the perturbative series various numerical factors plus logarithms of the form 
\begin{align}
  \ln\Bigg( \frac{ Q \mu_{{\rm cs}\,i}^{\frac{2+\beta_i}{1+\beta_i}} }
    { Q_{{\rm cut}\,i}^{\frac{1}{1+\beta_i}} \Delta m_c^2 } \Bigg) \,.
\end{align}
These logarithms vanish for the canonical scale choice in \eqn{gscs-scales}, while residual dependence through $\alpha_s(\mu_{{\rm cs}\,i})$ is always systematically canceled out order-by-order in the resummed perturbation theory.
The explicit $(\Delta m_c^2)^{-\eta}(\Delta m_c^2)^\eta$ terms then cancel separately in the numerator and denominator. Finally, the remaining dependence on $\eta$ cancels between the numerator and denominator of the $\xi_j$ ratios due to the relation in \eqn{musc-relation}.  Thus the same values for $\xi_j$ that were determined for the nonperturbative region in \sec{binnedmJ} also work equally well for the perturbative resummation region.

Thus for the perturbative resummation region we again have  pure quark and pure gluon observables 
\begin{align}  \label{eq:QGpert}
\ml{Q} &= f_q\, \Big( \hat \Sigma_q^{(b)}  - \xi_g \hat\Sigma_q^{(a)}  \Big)  
\,,\\
\ml{G} &= f_g\, \Big( \hat\Sigma_g^{(b)}   - \xi_q \hat\Sigma_g^{(a)} \Big) 
 \,,\nn
\end{align}
where for simplicity we have not indicated the continued presence of the same dominant power corrections that were indicated above in \eqn{QGpertstart}. 

Our analysis so far has thus obtained pure quark and pure gluon observables in two regions of phase space, for the smallest values of $\Delta m_c^2$ where nonperturbative corrections are important, and for an intermediate range of small $\Delta m_c^2$ where perturbative resummed contributions dominate. In between these two we have the region illustrated by \Fig{fig:modes}b, where 
one collinear-soft mode becomes nonperturbative while the other is still perturbative. Although we have not treated this region explicitly in our analysis, by continuity we fully expect that the same values of $\xi_j$ will work equally well in this region too.  Taken together, this yields a significant region of phase space over which we obtain pure quark or pure gluon observables, thus yielding the desired result.

\subsection{Optimizing the Parameter Choice}
\label{sec:optimizing}

With the expression of $\xi_j$, we have constructed a class of pure quark and gluon observables $\ml{Q}$ and $\ml{G}$, that depend on $\beta_1$, $\beta_2$, $z_{{\rm cut}1}^{(a)}$, $z_{{\rm cut}2}^{(a)}$ and $z_{{\rm cut}1}^{(b)}$. In practice there will be both theoretical and experimental uncertainties, and thus it is worthwhile to exploit these five independent variables in order to maximize the ability of the constructed observables to distinguish between quarks and gluons. 

Our construction of quark and gluon observables ${\cal Q}$ and ${\cal G}$ leaves $\beta_1$, $\beta_2$, $z_{{\rm cut}1}^{(a)}$, $z_{{\rm cut}2}^{(a)}$ and $z_{{\rm cut}1}^{(b)}$ as variables that can be varied for optimization.  Another important variable choice is the jet radius $R$ for the initial jet,  on which we apply collinear drop grooming, and then measure the $\Delta m_c^{(a)}$ spectrum.
There are both theoretical and practical considerations for the parameter optimization, which include:
\begin{enumerate}
\item Perturbative global-soft scales $Q_{{\rm gs}\,i}^{(a,b)}=p_T R\, z^{(a,b)}_{{\rm cut}\,i} (R/R_0)^{\beta_i} \gg \Lambda_{\rm QCD}$. This constraint is necessary to ensure that the parameters $\xi_q$ and $\xi_g$ used to specify the observables are perturbatively calculable. 

\item For discrimination power of the constructed observables, we want the values of $\xi_q$ and $\xi_g$ to be widely separated. If these parameters are too close it indicates that the cancellations needed to get pure quark and pure gluon observables are delicate, and may be spoiled by uncertainties in the determination of $\xi_{q,g}$, or experimental uncertainties.  We will see this requires $z^{(a)}_{{\rm cut}1} \ll z^{(a)}_{{\rm cut}2}$ and $z^{(a)}_{{\rm cut}1}$ and $z^{(b)}_{{\rm cut}1}$ widely separated.

\item Removing contamination from external soft radiation not associated with the jet itself, such as initial state radiation (ISR) and underlying event/multiparton interactions (MPI). This requires either i) small enough $R$, or ii) $z^{(a,b)}_{{\rm cut}1} \gtrsim 0.15$ for larger $R\sim 1$. 

\item In order to benchmark and test our proposal using the collinear drop jet mass factorization formula on which it was based, we require $z_{{\rm cut}\,i} \ll 1$. In practice, this constrains the parameters to be $z^{(a,b)}_{{\rm cut}\,i} \lesssim 0.3$. 

\end{enumerate}
To get an idea on the impact of the first constraint, taking $R_0=1$ and a jet with $p_T=800\,{\rm GeV}$ gives lower bounds of $z^{(a,b)}_{{\rm cut}\,i}\gtrsim \{ 0.004, 0.008\}$ for $R=0.5$ and $\beta=0,1$ respectively, and
$z^{(a,b)}_{{\rm cut}\,i}\gtrsim \{ 0.01, 0.05\}$ for $R=0.2$ and $\beta=0,1$ respectively.  In contrast, the fourth constraint provides an upper bound on $z^{(a,b)}_{{\rm cut}\,i}$, while, as we shall see, the second constraint pushes for these $z_{\rm cut}$ parameters to be well separated.  Thus there is a natural tension which narrows down the potential choices for the $z^{(a,b)}_{{\rm cut}\,i}$ parameters.

The third constraint turns out to be very restrictive. For large $R$ jets removing soft contamination implies a much stronger lower limit on the two parameters $z^{(a,b)}_{{\rm cut}1}$ than the first constraint. In general it is natural to take $z^{(a,b)}_{{\rm cut}2}\ge z^{(a,b)}_{{\rm cut}1}$ in order to ensure a non-zero phase space region for the collinear drop radiations, in which case this bounds all $z_{\rm cut}$ parameters. When considered together with the fourth and second constraints, the three turn out  to be impossible to simultaneously satisfy.
Due to this issue we focus on small $R$ jets to satisfy the third constraint. 

In \sec{dispower} we investigate the precise nature of the second constraint in more detail, and then in \sec{ISReffects} we study the third constraint with parton shower Monte Carlo generators. 

\begin{figure}[t]
    \begin{subfigure}[t]{0.49\textwidth}
        \centering
        \includegraphics[height=2.2in]{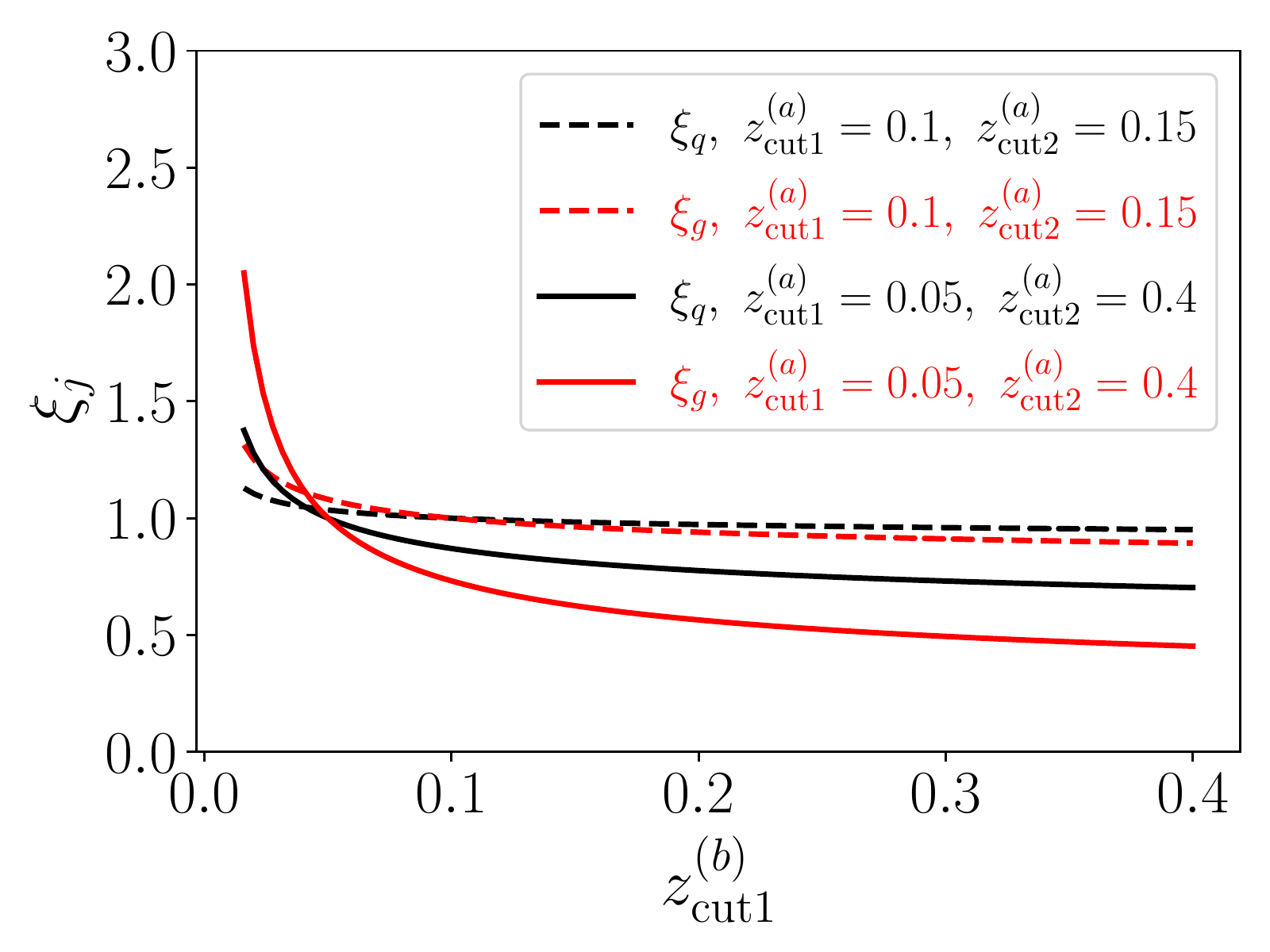}
        \caption{$\beta_1=\beta_2=1$.}
        \label{fig:beta1=beta2=1}
    \end{subfigure}%
    ~
    \begin{subfigure}[t]{0.49\textwidth}
        \centering
        \includegraphics[height=2.2in]{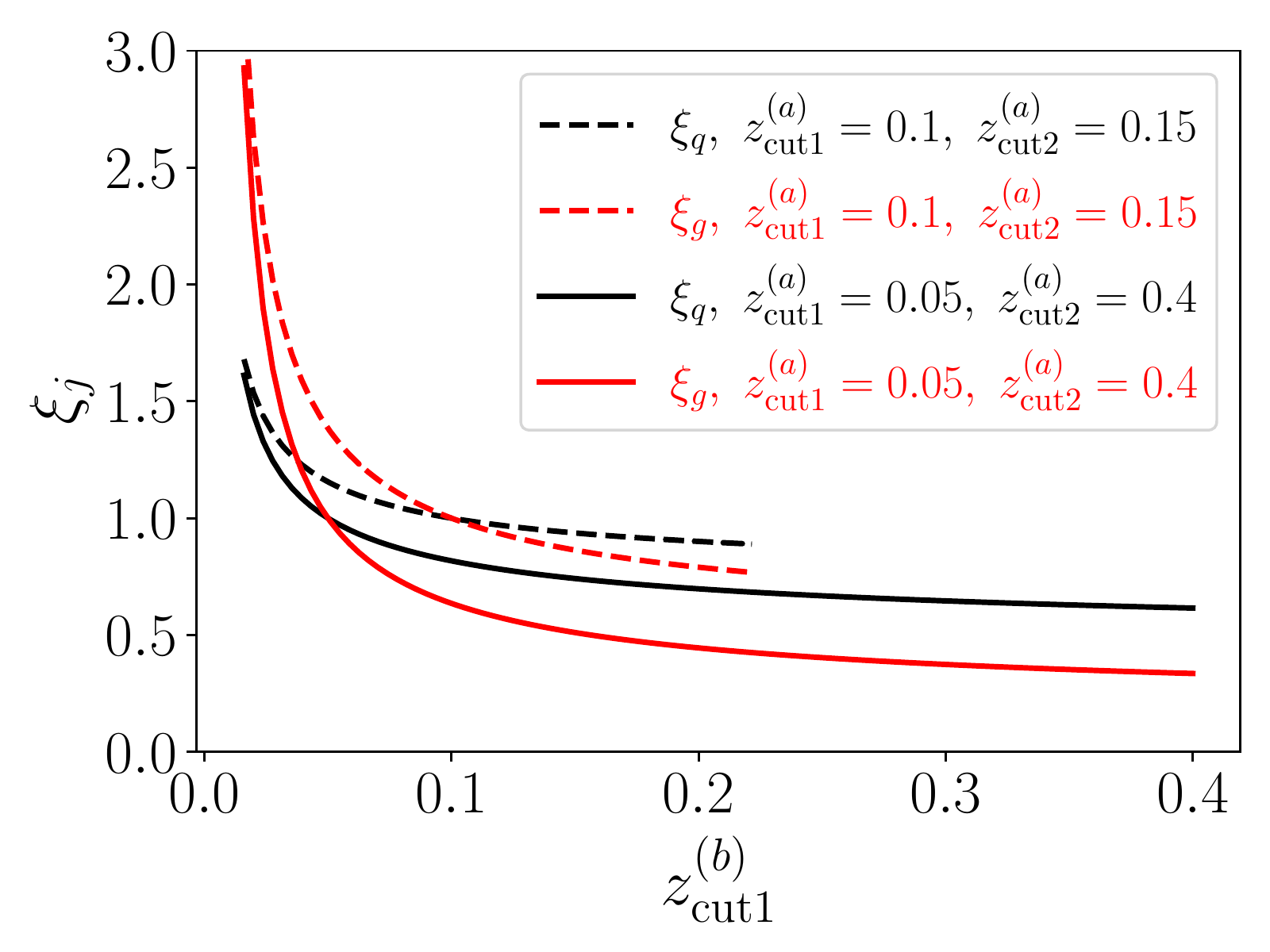}
        \caption{$\beta_1=1,\beta_2=0$.}
        \label{fig:beta1=1beta2=0}
    \end{subfigure}%
\begin{center}    
    \begin{subfigure}[t]{0.49\textwidth}
        \centering
        \includegraphics[height=2.2in]{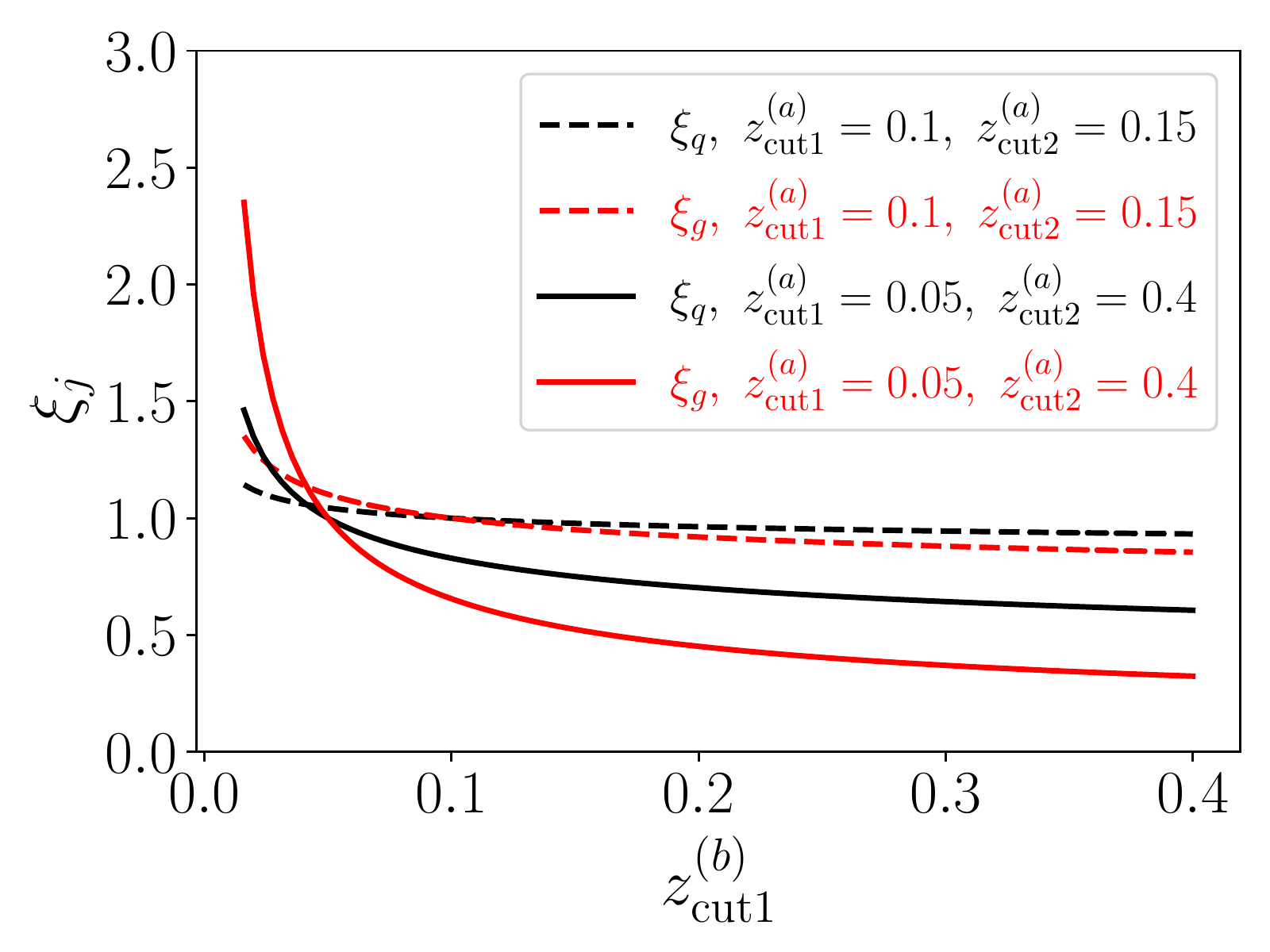}
        \caption{$\beta_1=\beta_2=0$.}
        \label{fig:beta1=beta2=0}
    \end{subfigure}%
\end{center}
\caption{Values of $\xi_j$ as functions of $z_{{\rm cut}1}^{(b)}$ for two choices of $z_{{\rm cut}\,i}^{(a)}$ for a jet with $p_T=800$ GeV, $\eta_J=0$ and $R=0.2$.
The three different panels show three choices for $\beta_1$ and $\beta_2$. For the case with $\beta_1=1$, $\beta_2=0$, $z_{{\rm cut}1}=0.1$ and $z_{{\rm cut}2}=0.15$, the curves are truncated when the condition $z_{{\rm cut}2}^{(b)} \geq z_{{\rm cut}1}^{(b)}$ is violated.}
\label{fig:xi_j}
\end{figure}

\subsubsection{Maximize Disentangling Power}
\label{sec:dispower}

An important criterion for distinguishability of the pure quark and gluon observables, ${\cal Q}$ and ${\cal G}$, is how big the numerical difference is between $\xi_g$ and $\xi_q$. These parameters are inputs provided by a theoretical calculation which has perturbative uncertainties, and hence must be distinguishable given those uncertainties. We also want $\xi_g$ and $\xi_q$ to be well separated to avoid relying on fine cancellations when taking the linear combination of experimental data.

To illustrate this,  in Fig.~\ref{fig:xi_j}  we plot the values of $\xi_j$ as functions of $z_{{\rm cut}1}^{(b)}$ for two choices of $z_{{\rm cut}\,i}^{(a)}$ and three choices of $\beta_i$. We use a jet with $p_T=800$ GeV, $\eta_J=0$, $R=0.2$, grooming parameter $R_0=1$, and choose the GS scales to be the canonical ones: $\mu_{{\rm gs}\,i}^{(a,b)} = Q_{{\rm gs}\,i}^{(a,b)}$.  This plot does not include perturbative uncertainties from the calculation of $\xi_j$, which at NLL are estimated to be $\lesssim 6\%$, and are left for discussion in \sec{analytic}.
We see that for all three choices of $\beta_i$ shown in the three panels of Fig.~\ref{fig:xi_j}, the values of $\xi_q$ (dashed black) and $\xi_g$ (dashed red) in the case with $z_{{\rm cut}1}^{(a)}=0.1$ and $z_{{\rm cut}2}^{(a)}=0.15$ are very close for most of the regions of $z_{{\rm cut}1}^{(b)}$, except for the small $z_{{\rm cut}1}^{(b)}$ region for the case with $\beta_1=1$ and $\beta_2=0$.
These close values for $\xi_j$ are problematic for distinguishability.
On the other hand, by instead taking $z_{{\rm cut}1}^{(a)}=0.05$ and $z_{{\rm cut}2}^{(a)}=0.4$, the difference between $\xi_q$ (solid black) and $\xi_g$ (solid red) becomes larger, especially when $z_{{\rm cut}1}^{(b)} < 0.03$.

We will use Fig.~\ref{fig:xi_j} as a guidance when we choose the CD parameters to enhance the disentangling power of the pure quark and gluon observables.

\begin{figure}[t!]
    \begin{subfigure}[t]{0.49\textwidth}
        \centering
        \includegraphics[height=2.2in]{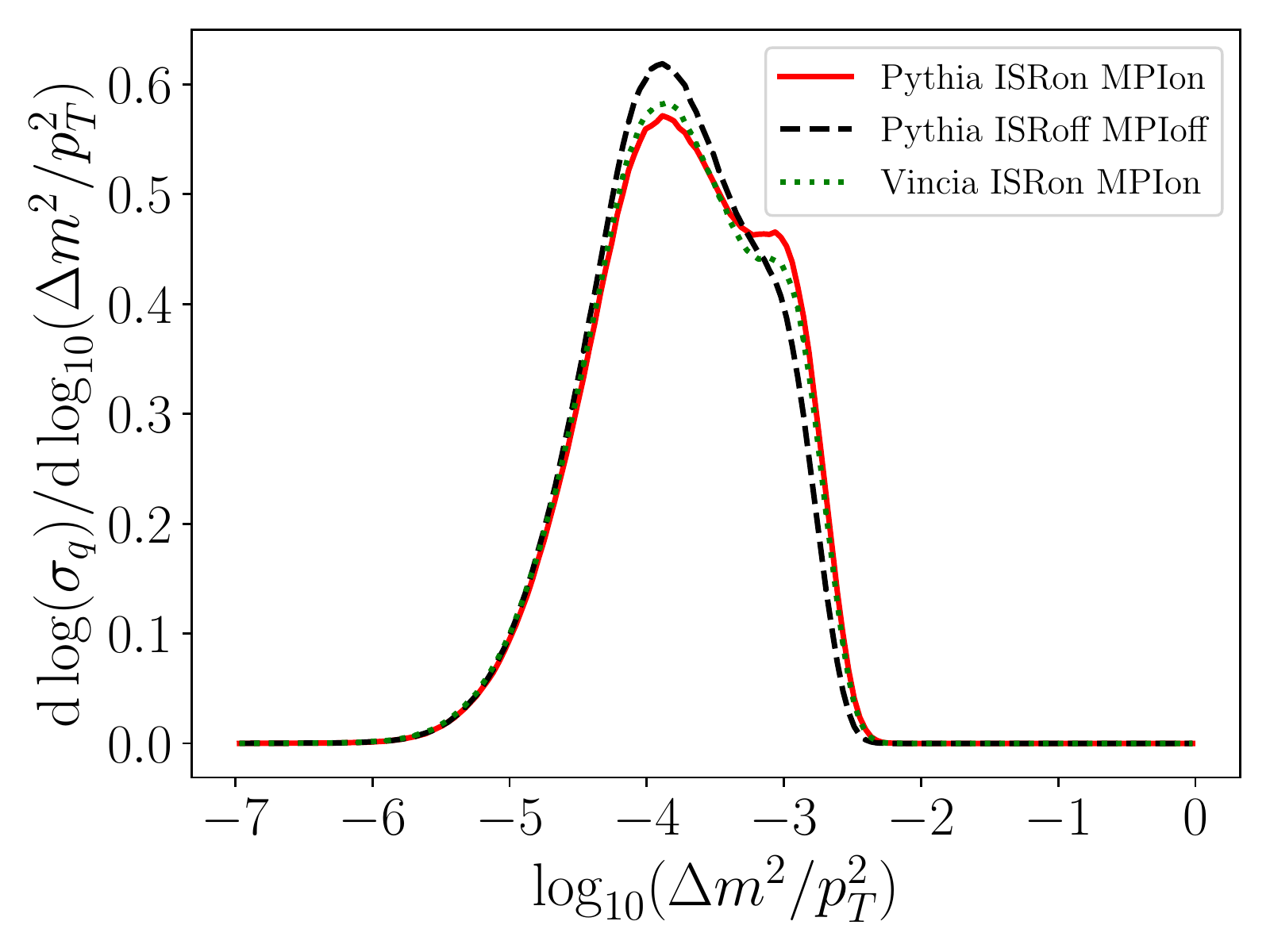}
        \caption{$\diff \sigma_q$ with $z_{{\rm cut}1}=0.02$ and $R=0.2$.}
        \label{fig:quark0.05R2}
    \end{subfigure}%
    ~
    \begin{subfigure}[t]{0.49\textwidth}
        \centering
        \includegraphics[height=2.2in]{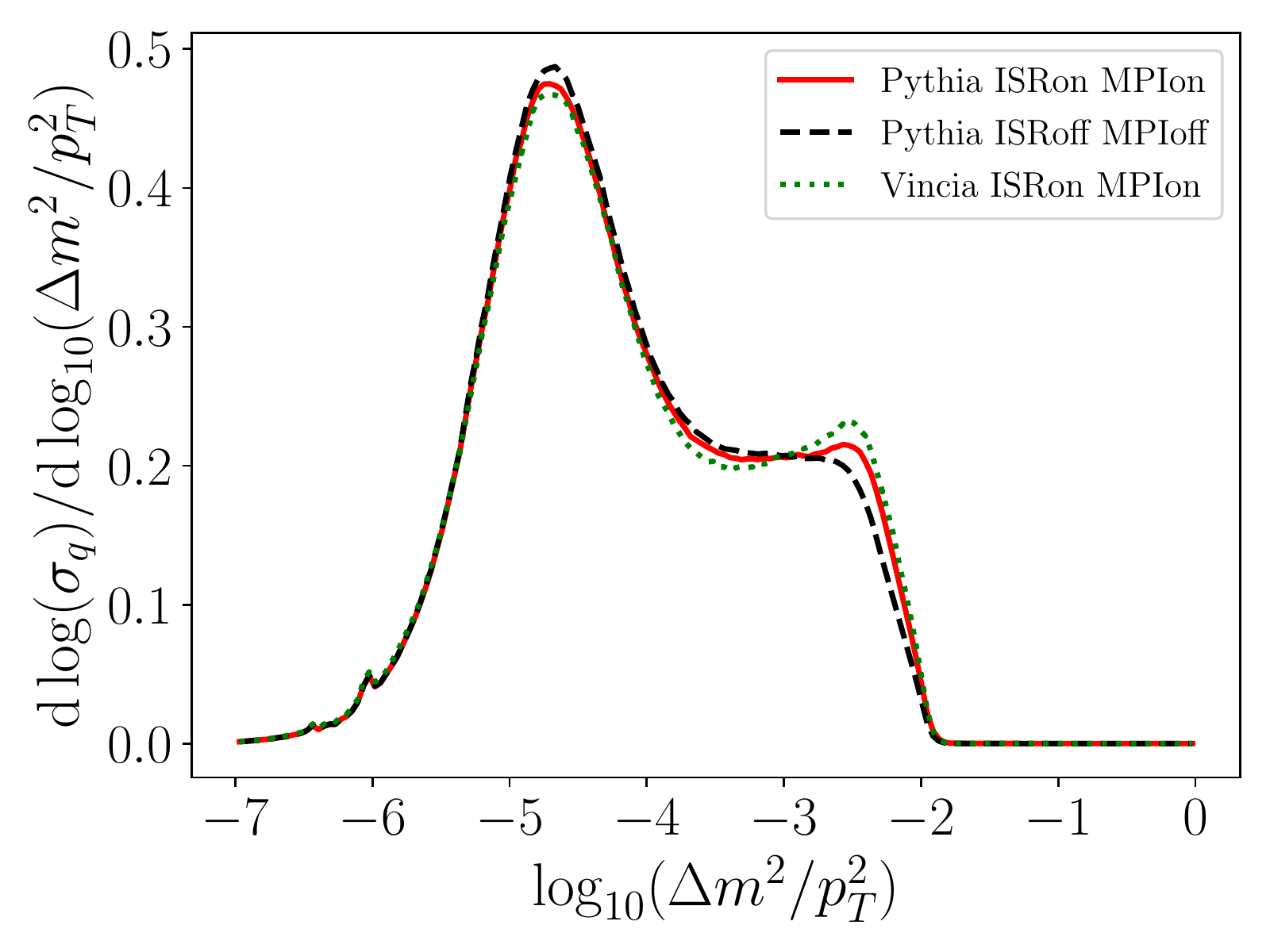}
        \caption{$\diff \sigma_q$ with $z_{{\rm cut}1}=0.1$ and $R=0.2$.}
        \label{fig:quark0.1R2}
    \end{subfigure}%
    
    \begin{subfigure}[t]{0.49\textwidth}
        \centering
        \includegraphics[height=2.2in]{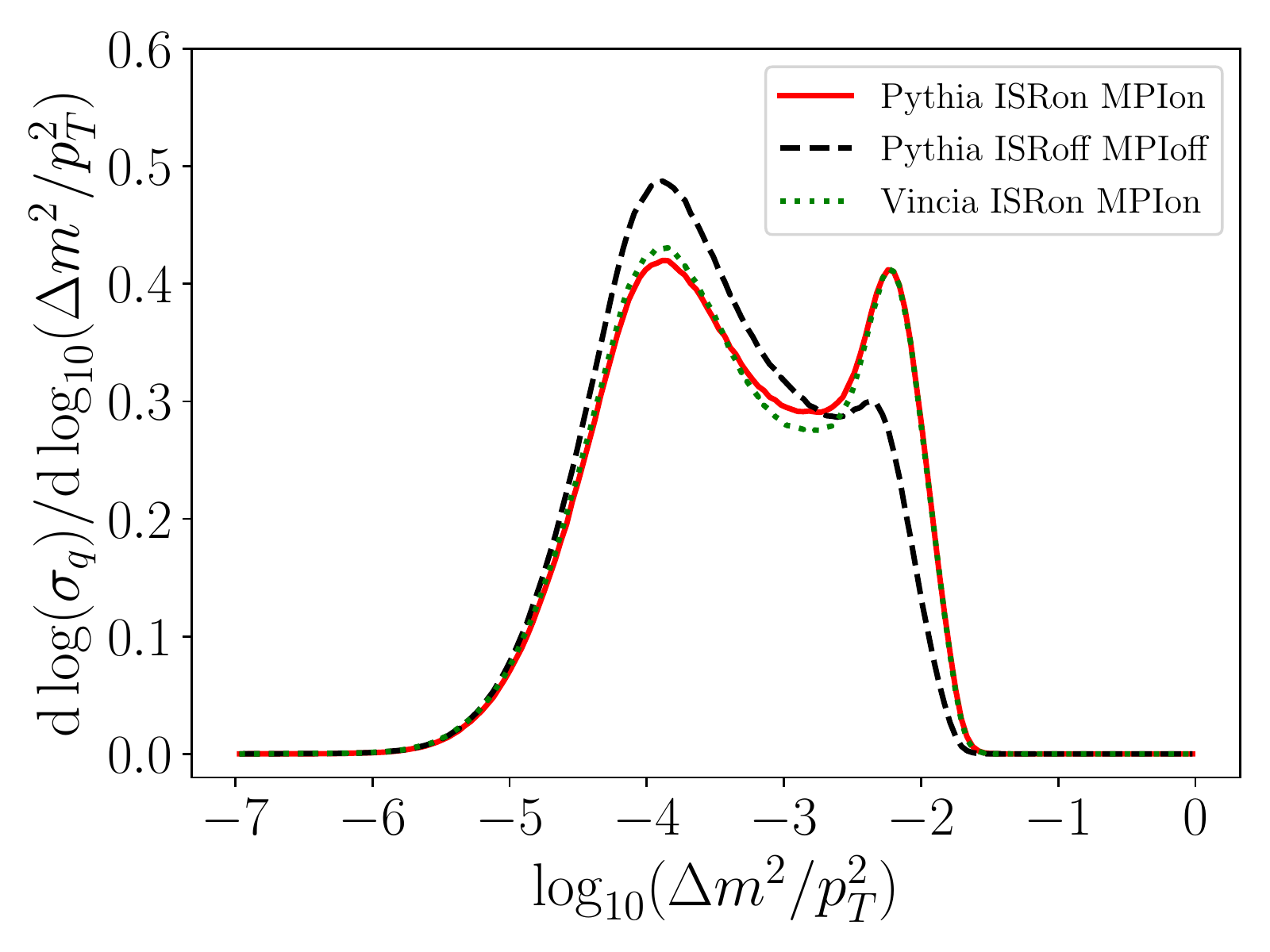}
        \caption{$\diff \sigma_q$ with $z_{{\rm cut}1}=0.02$ and $R=0.5$.}
        \label{fig:quark0.05R5}
    \end{subfigure}%
    ~
    \begin{subfigure}[t]{0.49\textwidth}
        \centering
        \includegraphics[height=2.2in]{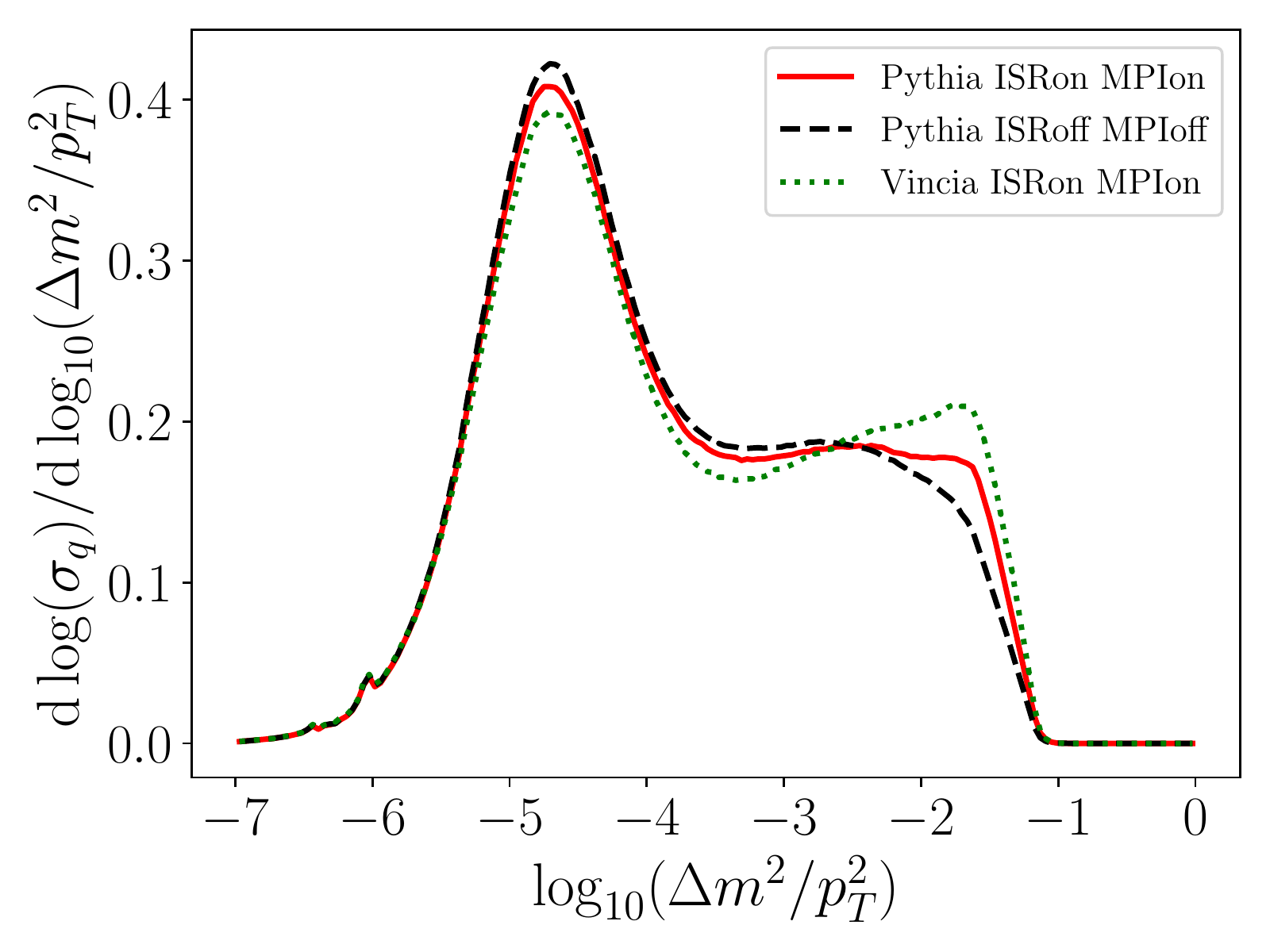}
        \caption{$\diff \sigma_q$ with $z_{{\rm cut}1}=0.1$ and $R=0.5$.}
        \label{fig:quark0.1R5}
    \end{subfigure}%
\caption{Monte Carlo results of the differential jet mass cross sections of
quark jets in collinear drop, with $\beta_1=\beta_2=0$. The top panels have $R=0.2$ and the bottom panels have $R=0.5$, while the left panels have $z_{{\rm cut}1}=0.02$ and the right panels have $z_{{\rm cut}1}=0.1$. The red solid and green dotted lines  include ISR and MPI effects and are results from \Pythia and \Vincia, respectively. The black dashed line is the \Pythia result without ISR and MPI effects.
}
\label{fig:quarkR_vs_ISR}
\end{figure}

\begin{figure}[t!]
    \begin{subfigure}[t]{0.49\textwidth}
        \centering
        \includegraphics[height=2.2in]{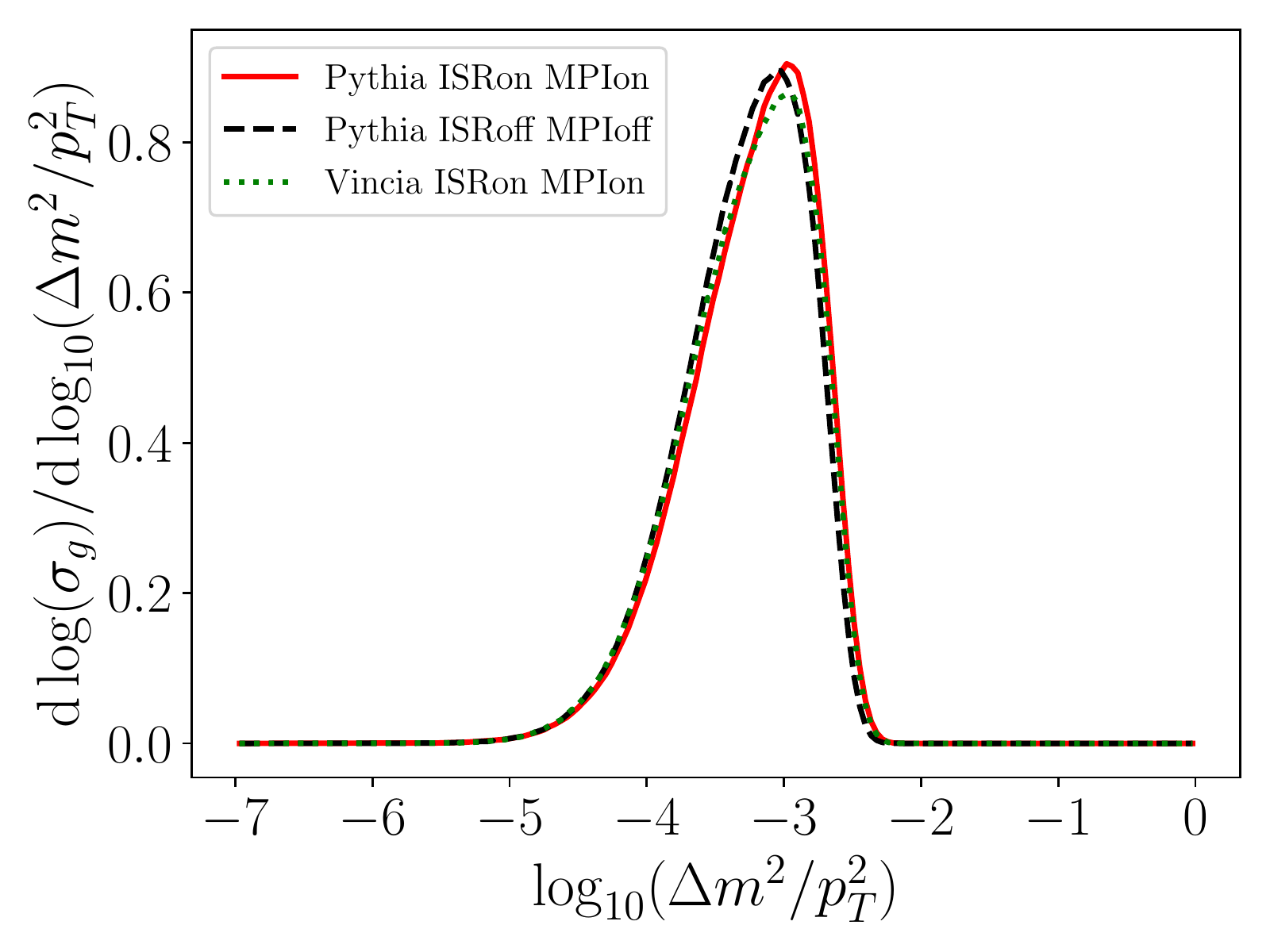}
        \caption{$\diff \sigma_g$ with $z_{{\rm cut}1}=0.02$ and $R=0.2$.}
        \label{fig:gluon0.05R2}
    \end{subfigure}%
    ~
    \begin{subfigure}[t]{0.49\textwidth}
        \centering
        \includegraphics[height=2.2in]{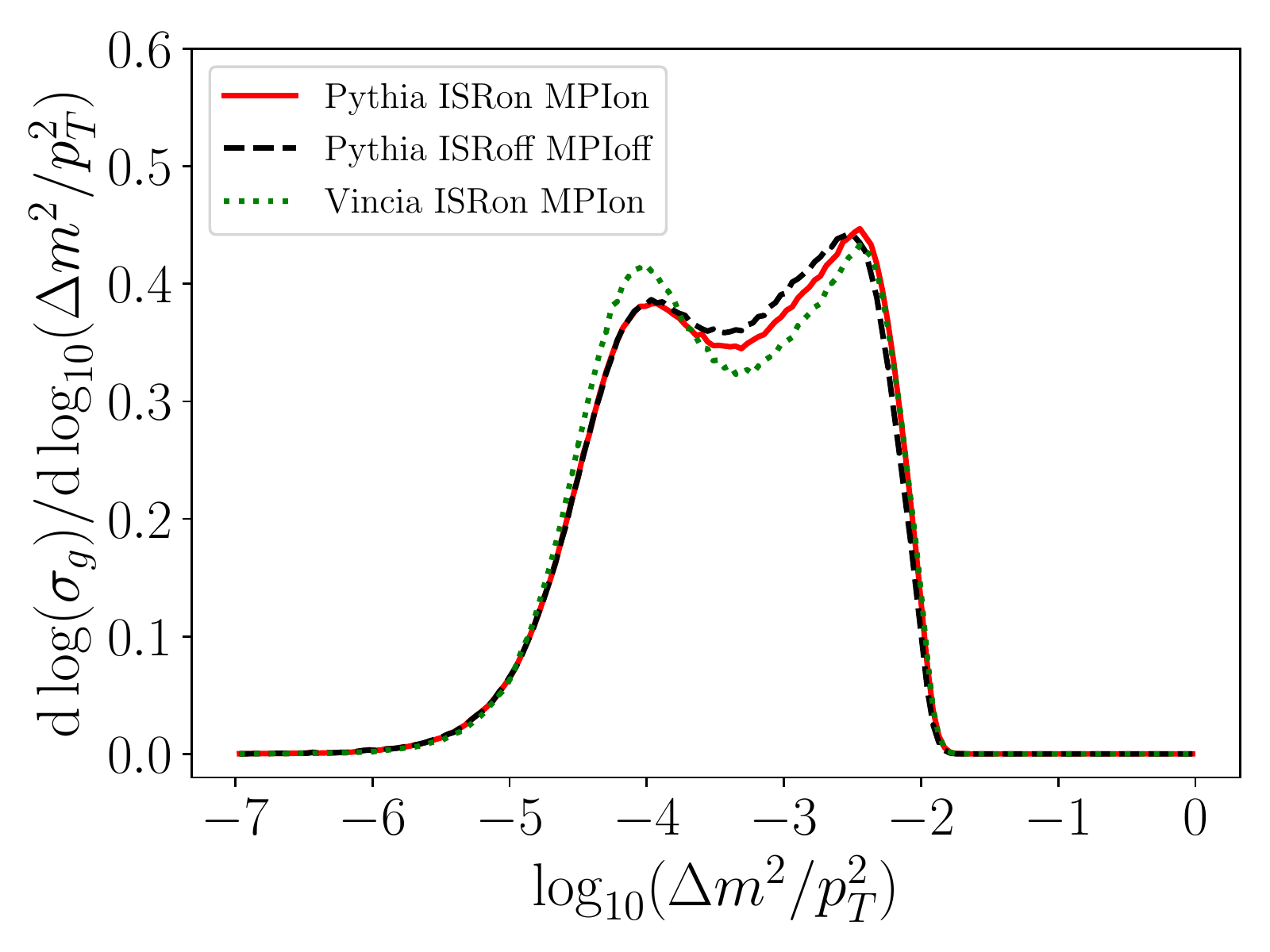}
        \caption{$\diff \sigma_g$ with $z_{{\rm cut}1}=0.1$ and $R=0.2$.}
        \label{fig:gluon0.1R2}
    \end{subfigure}%
    
    \begin{subfigure}[t]{0.49\textwidth}
        \centering
        \includegraphics[height=2.2in]{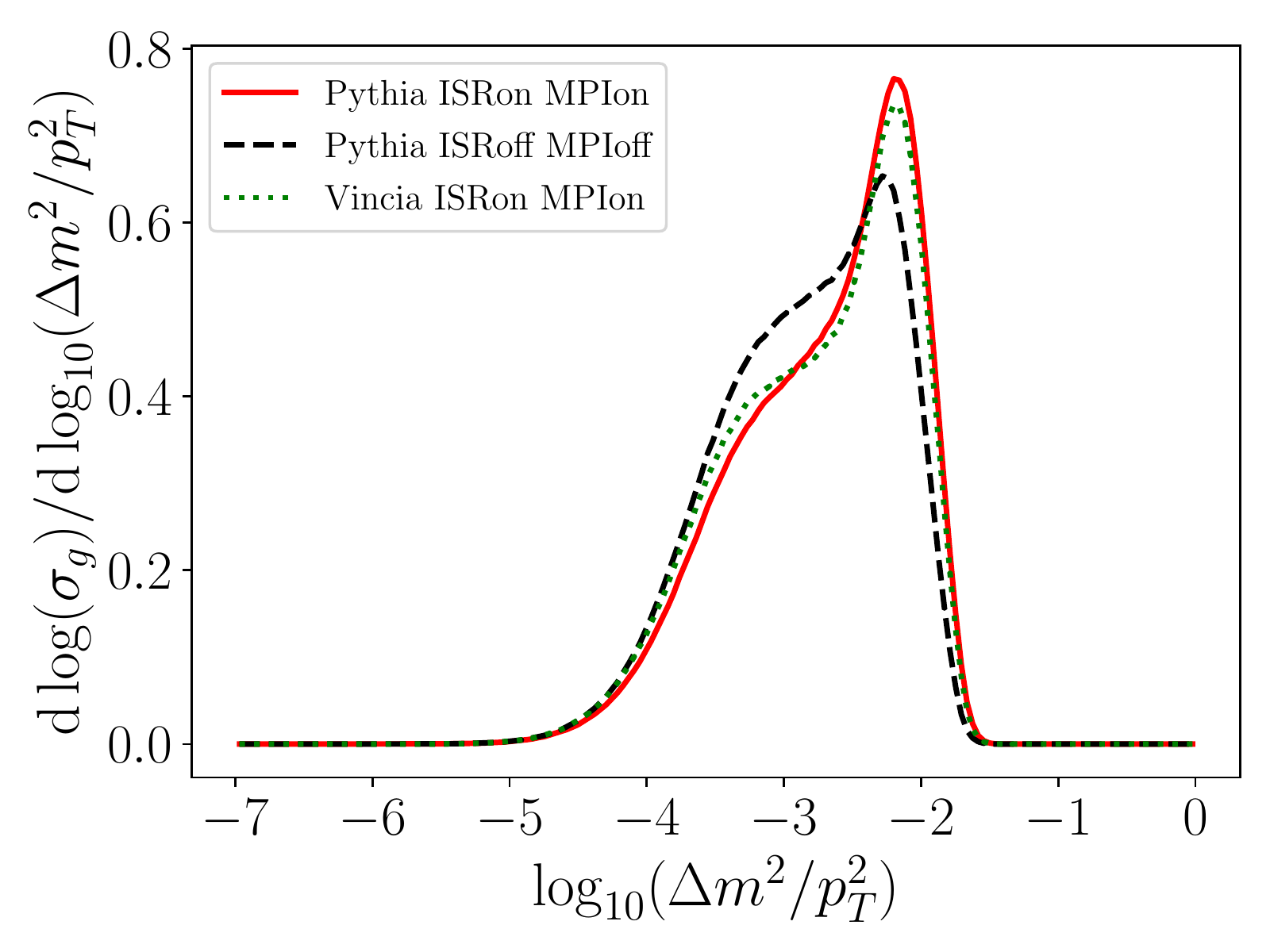}
        \caption{$\diff \sigma_g$ with $z_{{\rm cut}1}=0.02$ and $R=0.5$.}
        \label{fig:gluon0.05R5}
    \end{subfigure}%
    ~
    \begin{subfigure}[t]{0.49\textwidth}
        \centering
        \includegraphics[height=2.2in]{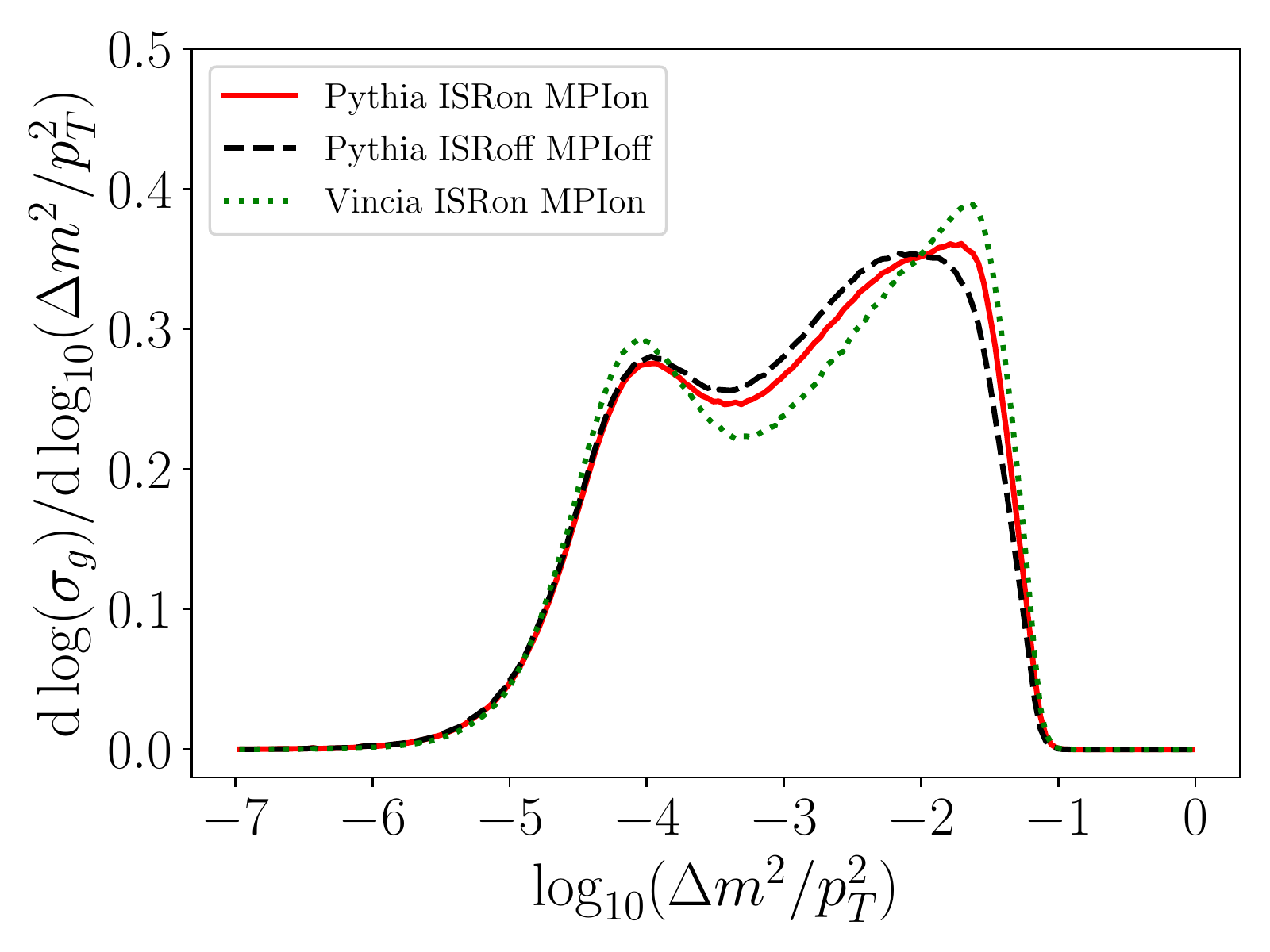}
        \caption{$\diff \sigma_g$ with $z_{{\rm cut}1}=0.1$ and $R=0.5$.}
        \label{fig:gluon0.1R5}
    \end{subfigure}%
\caption{Monte Carlo results of the differential jet mass cross sections of gluon jets in collinear drop, with $\beta_1=\beta_2=0$. The top panels have $R=0.2$ and the bottom panels have $R=0.5$, while the left panels have $z_{{\rm cut}1}=0.02$ and the right panels have $z_{{\rm cut}1}=0.1$. The red solid and green dotted lines  include ISR and MPI effects and are results from \Pythia and \Vincia, respectively. The black dashed line is the \Pythia result without ISR and MPI effects.
}
\label{fig:gluonR_vs_ISR}
\end{figure}

\subsubsection{ISR and MPI effects}
\label{sec:ISReffects}

In proton-proton collisions, the effects of ISR and underlying event (modeled by MPI) cannot be neglected, and can contaminate the construction of the pure quark and gluon observables. In practice to minimize the impact of these effects either a small $R\ll 1$ should be used, or a larger $R\sim 1$ with larger values of the soft drop grooming parameters $z_{{\rm cut}1}^{(a,b)}$. 

To demonstrate this in a quantitative manner we carry out Monte Carlo studies of the differential cross sections that enter into our pure quark and gluon observables. We use \Pythia 8~\cite{Sjostrand:2014zea} and \Vincia~\cite{Fischer:2016vfv} to generate quark and gluon jets in $13$ TeV proton-proton collisions and always consider fully hadronized events.
The jets are reconstructed by using the anti-$k_T$ algorithm, which is implemented in FastJet~\cite{Cacciari:2011ma}. Jets in the transverse momentum region $780~{\rm GeV} \le p_T \le 820~{\rm GeV}$ and the rapidity region $0 \le y_J \le 1$ are selected and passed to the collinear drop grooming procedure, implemented in JETlib~\cite{jetlib}. 
In Figs.~\ref{fig:quarkR_vs_ISR} and \ref{fig:gluonR_vs_ISR} we 
show the impact of ISR and MPI for the quark and gluon contributions to the differential cross sections, respectively.  
Comparing the \Pythia and \Vincia curves with both ISR and MPI turned on, we see that there are some noticeable difference, likely reflecting the fact that in some cases the Monte Carlos have trouble predicting the spectrum of soft radiation that dominates collinear drop observables.
Comparing only the \Pythia curves with and without ISR+MPI, we see that  
the smallest impact of ISR and MPI occurs in the top-right (b) panels when both the jet grooming and jet radius are chosen to reduce the effect ($z_{\rm cut}=0.1$ and $R=0.2$). We also see from the top-left (a) and bottom-right (d) panels that the impact of these contributions is still fairly small when only a small jet radius or more substantial jet grooming are used to mitigate these effects. In the bottom-left (c) panels we see that effects are fairly substantial if neither method is applied ($z_{\rm cut}=0.02$ and $R=0.5$). 

However, the ability to exploit smaller values of $z_{{\rm cut}1}^{(a,b)}$ is useful in order to obtain more distinct values of $\xi_q$ and $\xi_g$, as shown in Fig.~\ref{fig:xi_j}, and thus obtain stronger discrimination power of the constructed observables. This favors using small $R$ jets for the pure quark and gluon observable construction, and we will use $R=0.2$ henceforth. 

For larger $R$ jets it is possible that other procedures could be used to mitigate the impact of ISR and MPI, while still having a smaller $z_{{\rm cut}1}^{(a,b)}$.  We investigate one such possibility in Appendix~\ref{app:ISRlargeR}.

\subsection{Analytic Results for ${\cal Q}$ and ${\cal G}$}
\label{sec:analytic}

Having fixed reasonable parameter ranges to use for our analysis,
we now give results of the pure quark and gluon observables based on our factorization theorem with NLL resummation, and with and without the contributions of the shape functions. For both the quark and gluon shape function models we truncate the series at $n=0$ and take $p=1$, $c_0=1$, and $\Lambda=300$ MeV in Eq.~(\ref{eq:sf_expansion}) as our default parameters used in the cumulative distribution of the jet mass. Variations about this choice will be considered as an uncertainty.
We also choose the jet kinematics to be $p_T=800$ GeV, $\eta_J=0$ and $R=0.2$. The grooming parameter $R_0$ is chosen to be $R_0=1$.

We continue to consider three choices of $\beta_i$: $\beta_1=\beta_2=1$, $\beta_1=1$ with $\beta_2=0$, and $\beta_1=\beta_2=0$.  As discussed in \Ref{Chien:2019osu}, the perturbative series for the collinear drop jet mass does not contain leading double logarithms for cases when $\beta_1=\beta_2$. Our choices of $\beta_i$ here take two examples where this is the case, and one where it is not. The $z_{{\rm cut}\,i}$ parameters for these three cases are chosen based on improving the distinguishability following \fig{xi_j}, and listed in Table~\ref{tab:1}. 
In Table~\ref{tab:1}, once we fix $z_{{\rm cut}\,i}^{(a)}$ and $z_{{\rm cut}1}^{(b)}$, the value of $z_{{\rm cut}2}^{(b)}$ and the jet mass ratio are determined from the constraints~(\ref{eq:mass_ratio1}) and~(\ref{eq:z_ratio}), which depend on the choice of jet $p_T$, jet rapidity, and jet radius. These values are also listed for our default jet kinematics and will vary with other choices for the kinematics. 
Since it will turn out that the plots of cases with $\beta_1=\beta_2$ and those with $\beta_1\neq \beta_2$ are qualitatively similar, we will suppress some plots for the $\beta_1=1, \beta_2=0$ case.

Also shown in Table~\ref{tab:1} are our NLL predictions for the $\xi_j$ parameters using \eqn{finalxij}.  Again the calculation of these values depends on the jet $p_T$, $\eta_J$, and $R$, and we have shown values for our default kinematics. 
Since these parameters are determined perturbatively, they have a perturbative uncertainty from missing higher order contributions, which affects how well we can specify the pure quark and pure gluon observables. We will refer to this as the ``observable uncertainty'', and estimate it by varying the global-soft scales $\mu_{{\rm gs}\,i}^{(a,b)} = r_{\rm gs}^i Q_{{\rm gs}\,i}^{(a,b)}$ where central values use canonical scales with $r_{\rm gs}^i=1$ and uncertainties are estimated by factor of two variations, $r_{\rm gs}^i=0.5$ and $r_{\rm gs}^i=2$. For $i=1,2$ the up/down variations are considered independently. However, since this provides an estimate for the same missing higher order terms in the (a) and (b) cumulative distributions, it makes sense to vary these scales either up or down in both (a) and (b), which is why $r_{\rm gs}^i$ does not depend on the choice of (a) or (b). The resulting uncertainties are shown by $\pm$ entries in Table~\ref{tab:1}, and are quite small due to cancellations of common uncertainties in the (a) and (b) sets used for the ratio of perturbative cumulative cross sections. We have cross checked that the ${\cal O}(\alpha_s)$ fixed order corrections to the global-soft functions (which enter at NNLL), give shifts that are well within these uncertainty estimates.
Note that unlike other sources of theoretical uncertainty, this observable uncertainty also influences experimental predictions for the pure quark and pure gluon observables using \eqn{QGdefns}, since it is an intrinsic uncertainty in how precise these observables have been defined to do what we want them to do. 

\begin{table}
\centering
\begin{tabular}{ |c|c|c|c||c|c||c|c| }
\hline
 \multicolumn{4}{|c|}{Parameter Choice} & \multicolumn{2}{c|}{Fixed} & \multicolumn{2}{c|}{NLL results}
\\
\hline
& $z_{{\rm cut}1}^{(a)}$ & $z_{{\rm cut}2}^{(a)}$ & $z_{{\rm cut}1}^{(b)}$ & $z_{{\rm cut}2}^{(b)}$ & $\frac{\Delta m_c^{(b)}}{\Delta m_c^{(a)}}$ & $\xi_q$ & $\xi_g$ \\
\hline
$\beta_1=\beta_2=1$  & 0.1 & 0.4 & 0.05 & 0.2 & 1.19 & 
$1.11_{-0.02}^{+0.01}$ & 
$1.26_{-0.05}^{+0.02}$ \\
\hline
$\beta_1=1,\beta_2=0$ & 0.4 & 0.4 & 0.05 & 0.141 & 1.68 & $1.43_{-0.05}^{+0.01}$ & 
$2.26_{-0.16}^{+0.04}$ \\
\hline
$\beta_1=\beta_2=0$ & 0.1 & 0.4 & 0.02 & 0.08 & 2.24 & 
$1.41_{-0.05}^{+0.02}$ & 
$2.17_{-0.16}^{+0.08}$ \\
\hline
\end{tabular}
\caption{Cases considered for the $\beta_i$, $z_{{\rm cut}\,i}^{(a)}$, and $z_{{\rm cut}1}^{(b)}$ parameters. For our choice of $p_T=800$ GeV, $\eta_J=0$ and $R=0.2$ the required values for $z_{{\rm cut}2}^{(b)}$ and the ratio of jet mass bins for sets $(a)$ and $(b)$, and our NLL determination of the corresponding $\xi_q$ and $\xi_g$ are also shown.}
\label{tab:1}
\end{table}

For the $\beta_1=\beta_2=1$ case, both $z_{{\rm cut}1}^{(a)}$ and $z_{{\rm cut}1}^{(b)}$ are constrained to be relatively large to ensure that the GS scales remain perturbative (constraint 1 from \sec{optimizing}). Thus their difference becomes smaller, which results in less well separated values of $\xi_{q}$ and $\xi_g$, as seen in \Tab{tab:1}.  For the $\beta_1=\beta_2=0$ case, the GS scale is still perturbative even with $z_{{\rm cut}1}^{(b)}=0.02$, which differs significantly from $z_{{\rm cut}1}^{(a)}=0.1$. Therefore the gap in the $\beta_1=\beta_2=0$ case is large, which leads to stronger distinguishing power to separate quark and gluon jets.

When making theoretical predictions for the pure quark and pure gluon observables, we also have uncertainties associated to the calculation of the $\Sigma$s in \eqn{QGdefns}. These can be separated into two sources appearing in the use of \eqn{nonpertQG} or \eqn{QG}: A perturbative uncertainty associated to calculating $\hat\Sigma_{j}^{(a,b)}$, and a nonperturbative uncertainty associated to modeling the shape functions ${\cal F}_j$. 
To estimate the perturbative uncertainty we vary the global-soft scales in the perturbative parts of Eq.~(\ref{eq:SigNLLNP}) around their canonical values by a factor of two. This uncertainty should be treated as independent from the observable uncertainty, despite the fact that our estimate for it comes from varying the same underlying parameters. The reason is that even if we consider fixed observables with definite values of $\xi_j$, there will still be a perturbative uncertainty in predicting those observables. In contrast, the observable uncertainty provides information on how well we are able to ensure that the constructed ${\cal Q}$ and ${\cal G}$ for given values of $\xi_j$ are truly pure quark and pure gluon observables.  
When estimating the perturbative uncertainty, we do not alter the nonperturbative CS scales $\mu_{{\rm cs}\,i}=\Lambda_{{\rm cs}\,i}$, since the dependence of the final results on $\Lambda_{{\rm cs}\,i}$ is supposed to be canceled between the perturbative parts and the shape functions.  The uncertainty from varying $\Lambda_{{\rm cs}\,i}$ is therefore captured by the uncertainty in the functional form of the shape function models, which we vary to estimate the nonperturbative uncertainty.

\begin{figure}[t]
    \begin{subfigure}[t]{0.49\textwidth}
        \centering
        \includegraphics[height=2.2in]{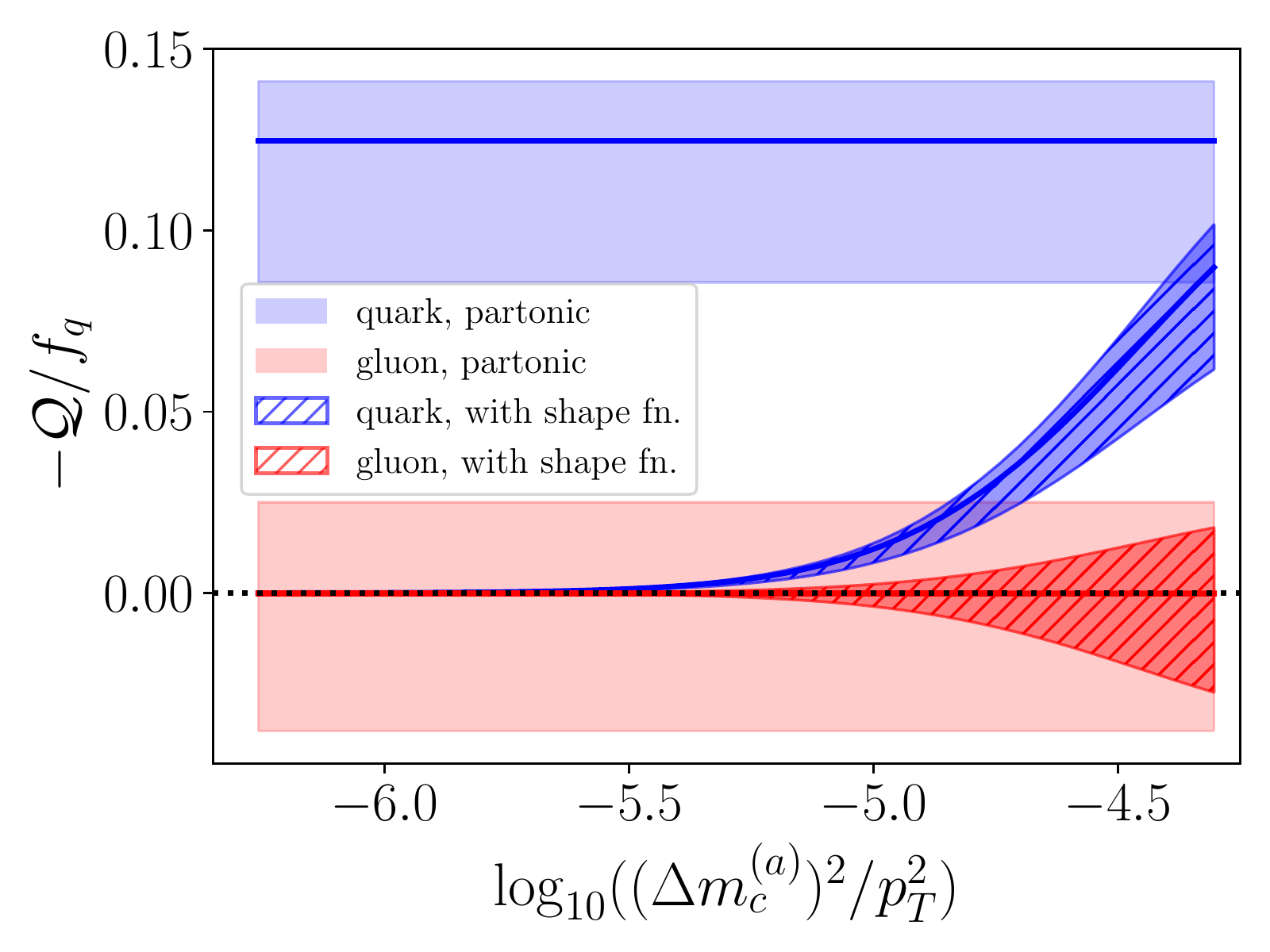}
        \caption{$-\ml{Q}$ with $\beta_1=\beta_2=1$.}
        \label{fig:Q11_shape}
    \end{subfigure}%
    ~
    \begin{subfigure}[t]{0.49\textwidth}
        \centering
        \includegraphics[height=2.2in]{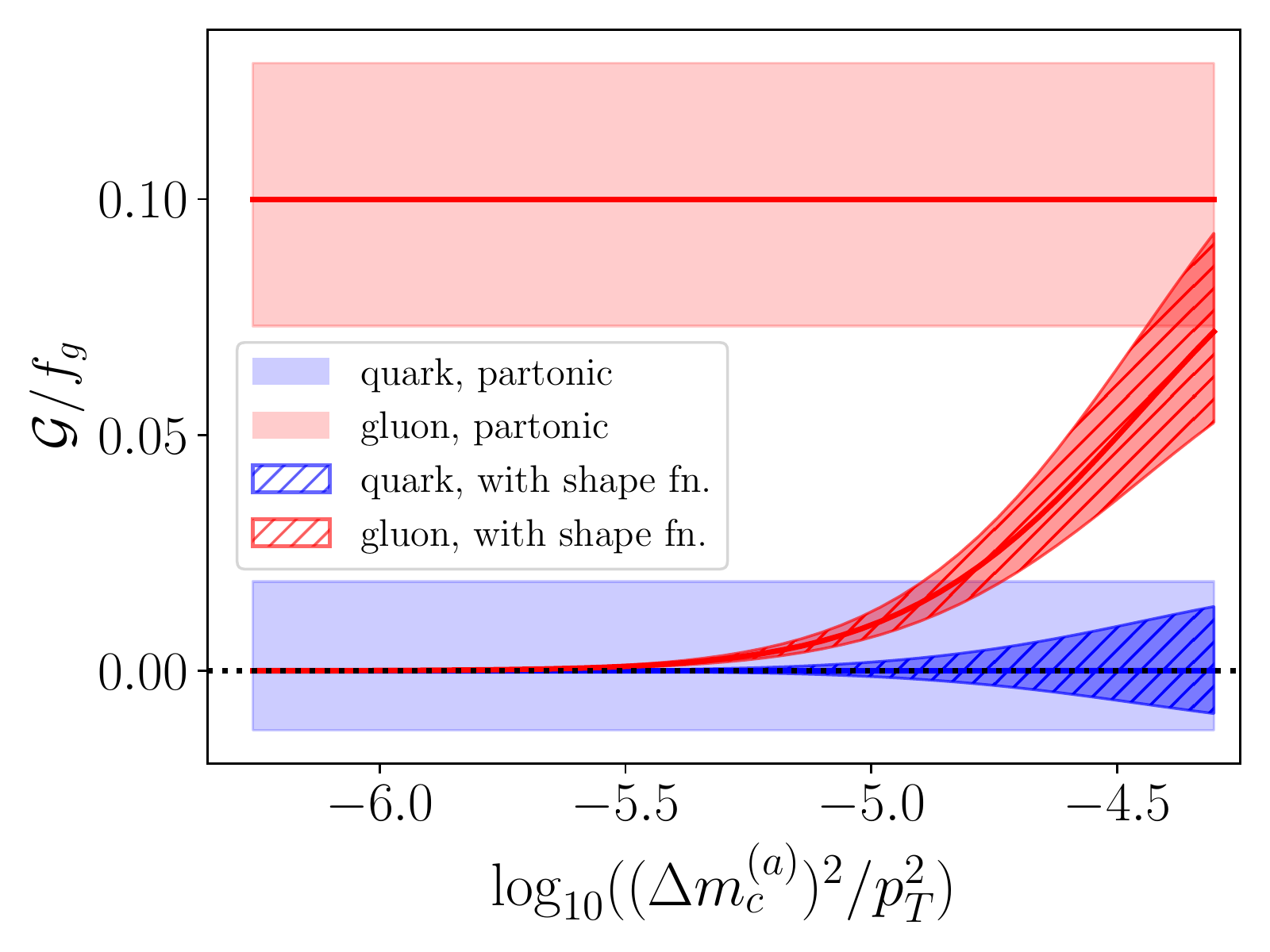}
        \caption{$\ml{G}$ with $\beta_1=\beta_2=1$.}
        \label{fig:G11_shape}
    \end{subfigure}%
    
    \begin{subfigure}[t]{0.49\textwidth}
        \centering
        \includegraphics[height=2.2in]{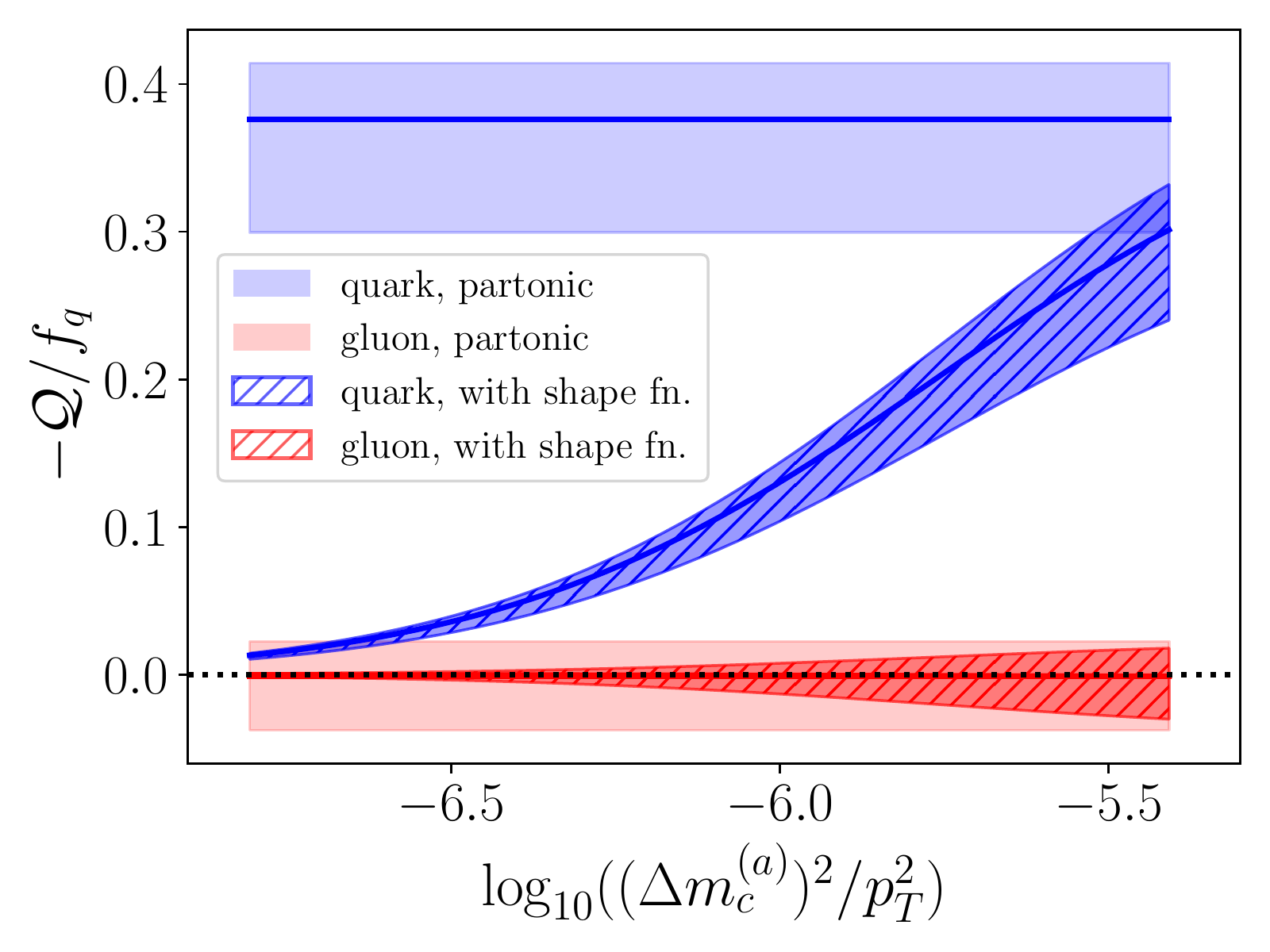}
        \caption{$-\ml{Q}$ with $\beta_1=\beta_2=0$.}
        \label{fig:Q00_shape}
    \end{subfigure}%
    ~
    \begin{subfigure}[t]{0.49\textwidth}
        \centering
        \includegraphics[height=2.2in]{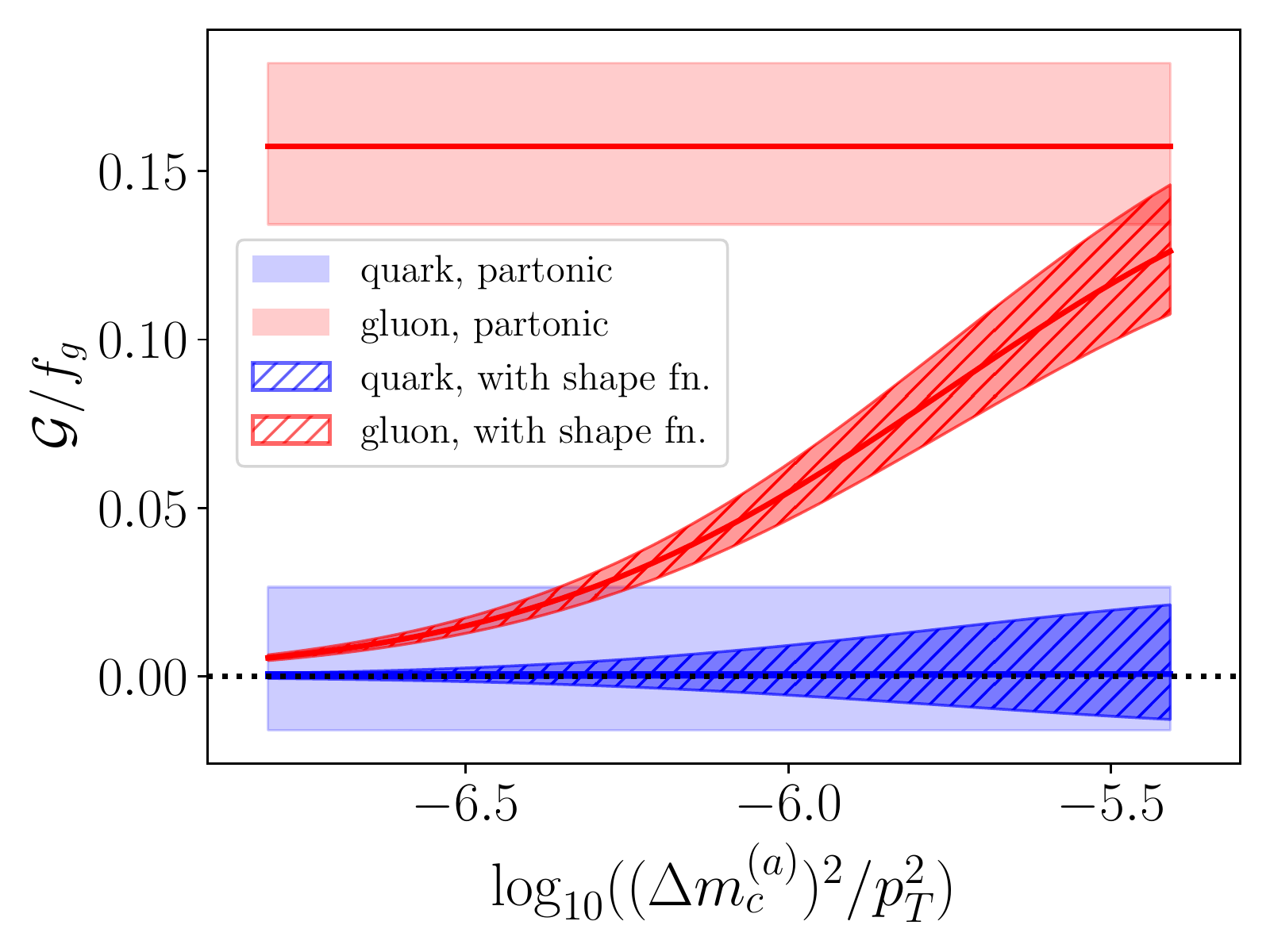}
        \caption{$\ml{G}$ with $\beta_1=\beta_2=0$.}
        \label{fig:G00_shape}
    \end{subfigure}%
\caption{Pure quark (left column) and gluon (right column) observables with and without the shape functions included for the $\beta_1=\beta_2=1$ (upper row) and $\beta_1=\beta_2=0$ (lower row) cases. The uncertainty bands add contributions from the observable uncertainty and perturbative uncertainty in quadrature. The curves with the light shading for uncertainties do not include the shape function, whereas shape functions are included for the curves with the darker hatched uncertainties.
}
\label{fig:Obs_shape}
\end{figure}

\subsubsection{Results in the Nonperturbative $\Delta m_c^2$ Region}

We start by examining the NLL results of the pure quark and gluon observables in the nonperturbative regime, with the smallest values of $\Delta m_c^2$. Results are shown in Fig.~\ref{fig:Obs_shape} for the two cases with $\beta_1=\beta_2=1$ and $\beta_1=\beta_2=0$, and with and without the shape functions to make clear how they shape the curves. For illustration purpose, we assume an equal contribution from quarks and gluons in the observable sample ($f_q=f_g=1/2$), and plot $-{\cal Q}/f_q$ and ${\cal G}/f_g$. This normalization makes the displayed non-zero signal contributions independent of the assumed quark and gluon fractions. For the pure quark observable ${\cal Q}$, the gluon contribution vanishes independent of the assumed $f_j$ fractions, but a small non-zero result will still be obtained once we account for uncertainties (and similarly for ${\cal G}$). The smaller contribution displayed for how gluons contribute to ${\cal Q}/f_q$ does depend on the chosen quark and gluon fractions, and can be scaled directly proportional to the input value of $f_g/f_q$ (and likewise for ${\cal G}/f_g$ where the quark contribution can be scaled by $f_q/f_g$).  Due to the larger values of $\xi_g$ for the pure quark observables we consider, the linear combination for ${\cal Q}$ is negative for the signal, and we choose to plot $-{\cal Q}/f_q$ so that the plots have a more uniform appearance. 

As explained in Section~\ref{sec:cumulative}, the perturbative results of the observables without the shape functions, shown in Fig.~\ref{fig:Obs_shape}, become constant in the small jet mass region, which is closely related to the fraction of events with no radiation in the phase space that is kept after the grooming. Once we include the nonperturbative shape functions, we see that the observables go to zero in the limit $\Delta m_c^2\to0$. This can be understood mathematically by examining Eq.~(\ref{eq:two_shape}). As $\Delta m_c^2\to0$, the integration measures for both $k_1$ and $k_2$ vanish and thus the integral vanishes. Physically, the shape functions represent contributions from nonperturbative soft radiation. In the limit $\Delta m_c^2\to0$, the phase space for this nonperturbative soft radiation vanishes and thus the observables go to zero when the shape function is included. The uncertainty bands shown include both the observable uncertainty for specifying $\xi_j$ from Table~\ref{tab:1}, and the perturbative uncertainty in predicting $\hat\Sigma_j^{(a,b)}$ at NLL from \eqn{QG}. These uncertainties are added in quadrature to obtain the bands shown. The percent uncertainty remains constant for the shape function curves, so the absolute uncertainties decrease as the shape function suppresses the cross section.

From Fig.~\ref{fig:Obs_shape} we see that the gap between the vanishing and nonvanishing components of each observable is sensitive to the collinear drop parameters chosen. Recall that for the  $\beta_1=\beta_2=1$ case there was less distinguishability between $\xi_{q}$ and $\xi_g$.  This is reflected in the fact that the gap between the vanishing and nonvanishing cross section components is smaller than for the other $\beta_i$ choices, accounting for the uncertainties. For the $\beta_1=\beta_2=0$ case, the GS scale is still perturbative even with $z_{{\rm cut}1}^{(b)}=0.02$, which differs significantly from $z_{{\rm cut}1}^{(a)}=0.1$. Therefore here the gap is larger, leading to better distinguishing power to separate quark and gluon jets.

We also see that the observable values depend on the jet content. More explicitly, the pure quark observables take larger values than the pure gluon observables in general. The reason is two-fold: First, the linear combination coefficients in the pure quark observables are bigger than those in the pure gluon observables $\xi_g > \xi_q$, since the quadratic Casimir of the gluon is bigger than that of the quark $C_A>C_F$, as shown in Eq.~(\ref{eq:finalxij}). (Recall that $\xi_g$ appears in the construction of the pure quark observable while $\xi_q$ appears in the construction of the pure gluon observable, as shown in Eq.~(\ref{eq:QG}).)  The bigger the $\xi_j$ value is, the bigger the observable is, since the observable is a difference between two cumulative jet mass cross sections, and only one of them is multiplied by the linear combination coefficient. The second reason is that the perturbative result of the cumulative jet mass cross section of a quark jet is bigger than that of a gluon jet, as shown in Fig.~\ref{fig:pert_results}. 

\begin{figure}[t]
    \begin{subfigure}[t]{0.49\textwidth}
        \centering
        \includegraphics[height=2.2in]{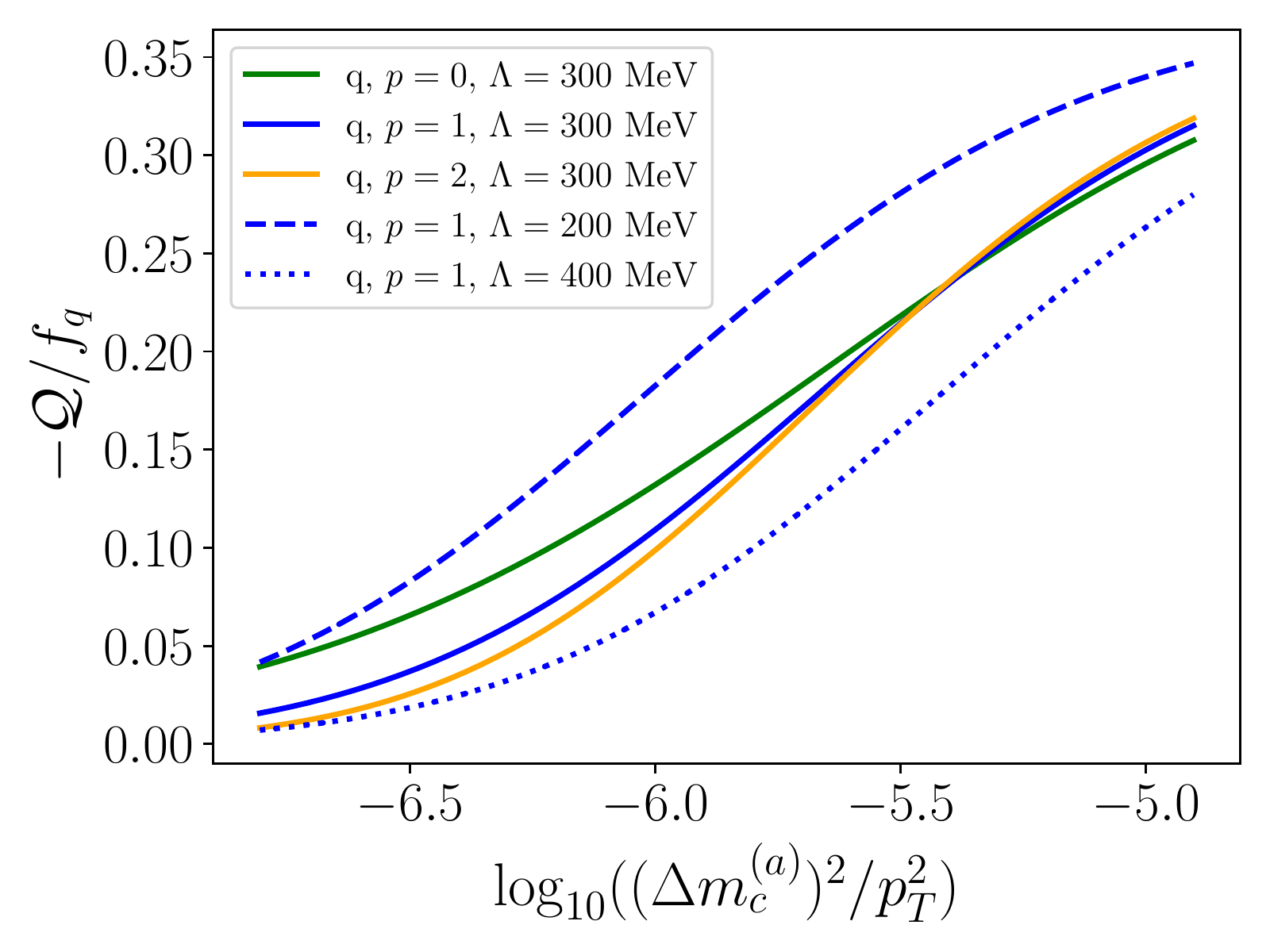}
        \caption{$-\ml{Q}$ with $\beta_1=\beta_2=0$.}
        \label{fig:Q00_varyp}
    \end{subfigure}%
    ~
    \begin{subfigure}[t]{0.49\textwidth}
        \centering
        \includegraphics[height=2.2in]{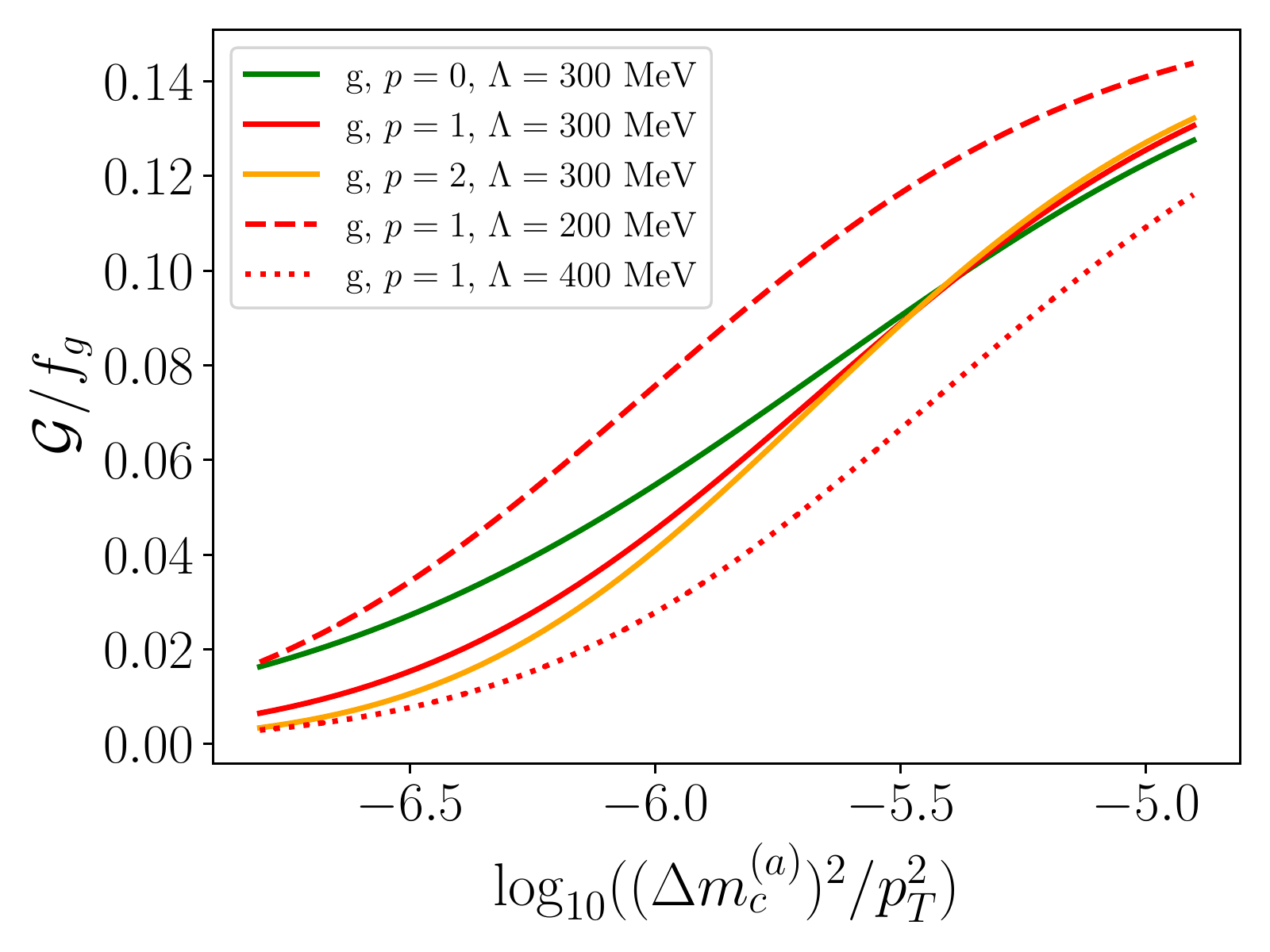}
        \caption{$\ml{G}$ with $\beta_1=\beta_2=0$.}
        \label{fig:G00_varyp}
    \end{subfigure}%
\caption{Pure quark (left) and gluon (right) observables with different shape function models included for the case with $\beta_1=\beta_2=0$. The parameters $p$ and $\Lambda$ of the shape functions are varied to show the sensitivity of the predicted values in the deep infrared region to the  nonperturbative physics these shape functions describe.}
\label{fig:vary_shape}
\end{figure}
 
Next we consider how sensitive the results are to the nonperturbative shape function models, and construct estimates for the resulting nonperturbative uncertainty in our theoretical predictions. To do this we vary the parameter $p$ and $\Lambda$ in the models of the shape functions to obtain the results shown in Fig.~\ref{fig:vary_shape}. Different values of the parameter $p$ change the jet mass dependence of the pure quark and gluon observables, but only mildly. We also see that the variation in the parameter $\Lambda$ leads to a bigger change in the results and it partially determines how fast the results approach zero. In addition to the uncertainty of the parameters $p$ and $\Lambda$, another unknown aspect of the shape function is its normalization. As discussed in Ref.~\cite{Hoang:2019ceu}, the shape functions in soft drop grooming are not normalized to be unity. So the shape function curves depicted in Fig.~\ref{fig:vary_shape} can be varied by an overall scaling factor, which would be implemented here by varying the coefficient $c_0$ in \eqn{sf_expansion}.  The scaling factor can be different for the quark and gluon jets, and it only depends on the parameter $\beta_1$ or $\beta_2$. However from \fig{vary_shape} we see that varying the overall scale with $c_0$ will be highly correlated with the result from changing $\Lambda$, and hence we only retain the latter for our nonperturbative uncertainty estimate.

\subsubsection{Including Results in the Perturbative $\Delta m_c^2$ Region}

Recall that our construction of pure quark and gluon observables is valid all the way from the nonperturbative region just considered, to the perturbative resummation region at larger $\Delta m_c^2$ values. Hence it is interesting to study the complete factorization based predictions for our pure quark and gluon observables in the full jet mass region. 

\begin{figure}[p]
    \begin{subfigure}[t]{0.49\textwidth}
        \centering
        \includegraphics[height=2.2in]{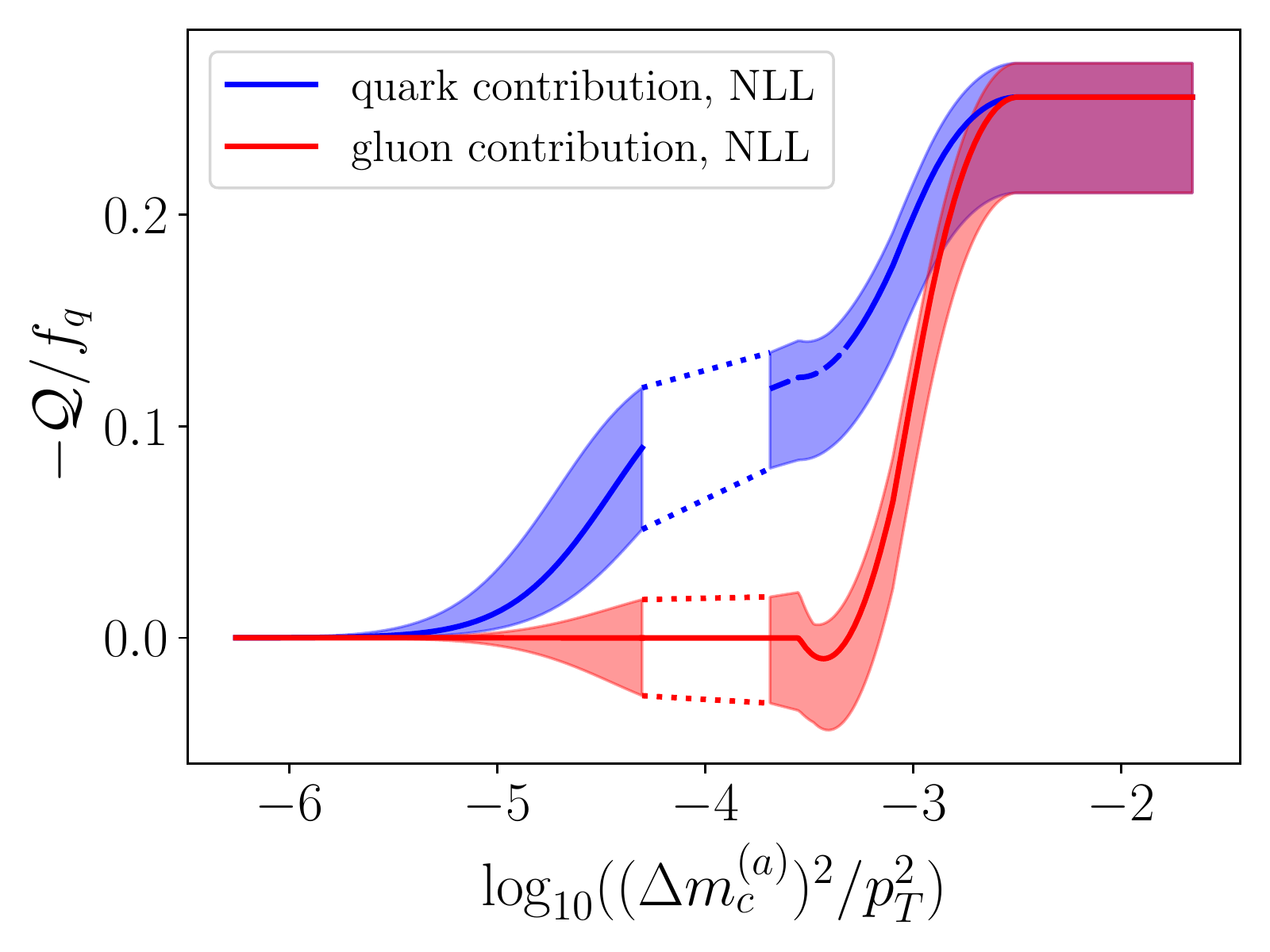}
        \caption{$-\ml{Q}$ with $\beta_1=\beta_2=1$.}
        \label{fig:Q11_full}
    \end{subfigure}%
    ~
    \begin{subfigure}[t]{0.49\textwidth}
        \centering
        \includegraphics[height=2.2in]{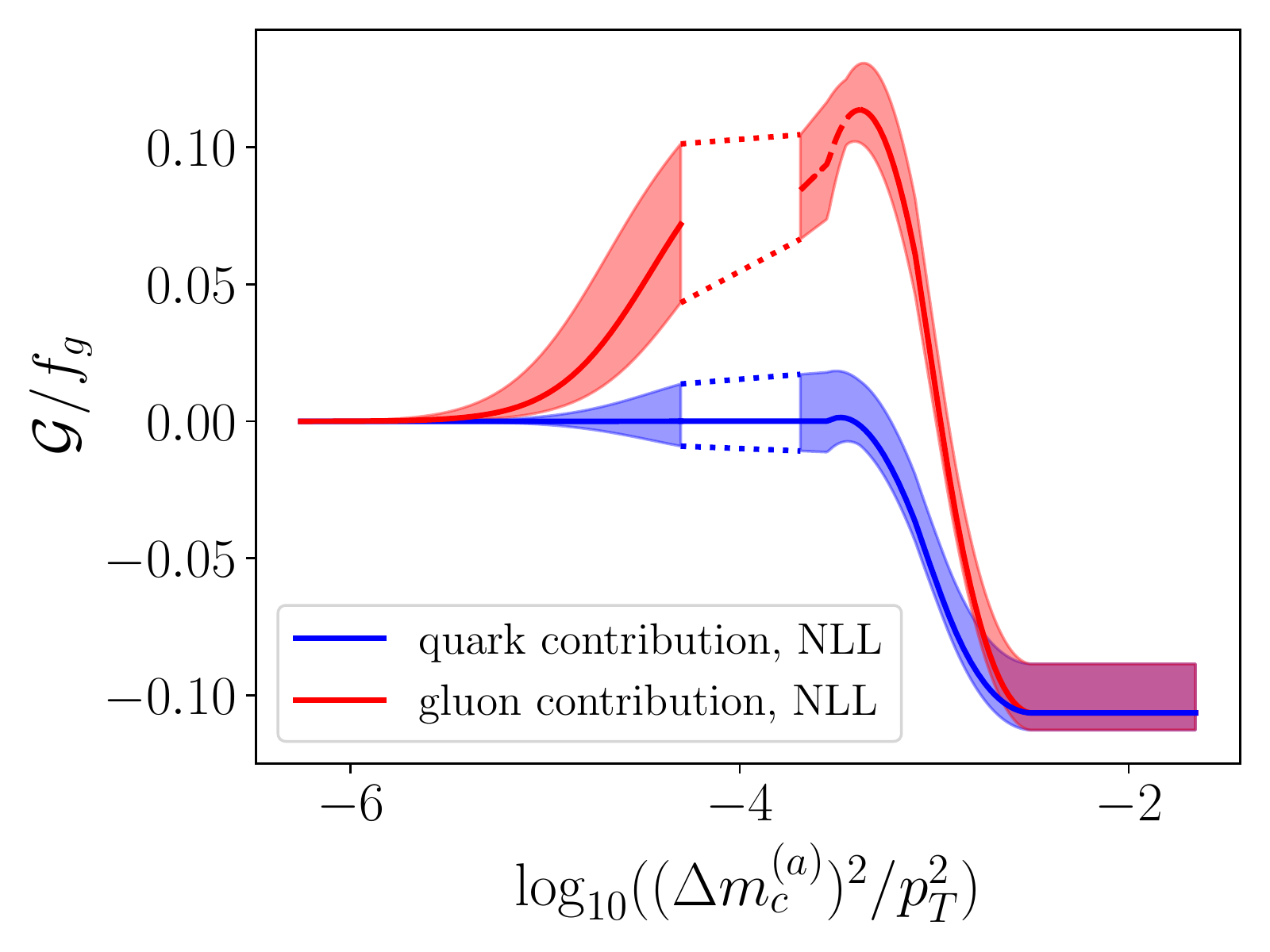}
        \caption{$\ml{G}$ with $\beta_1=\beta_2=1$.}
        \label{fig:G11_full}
    \end{subfigure}%

    \begin{subfigure}[t]{0.49\textwidth}
        \centering
        \includegraphics[height=2.2in]{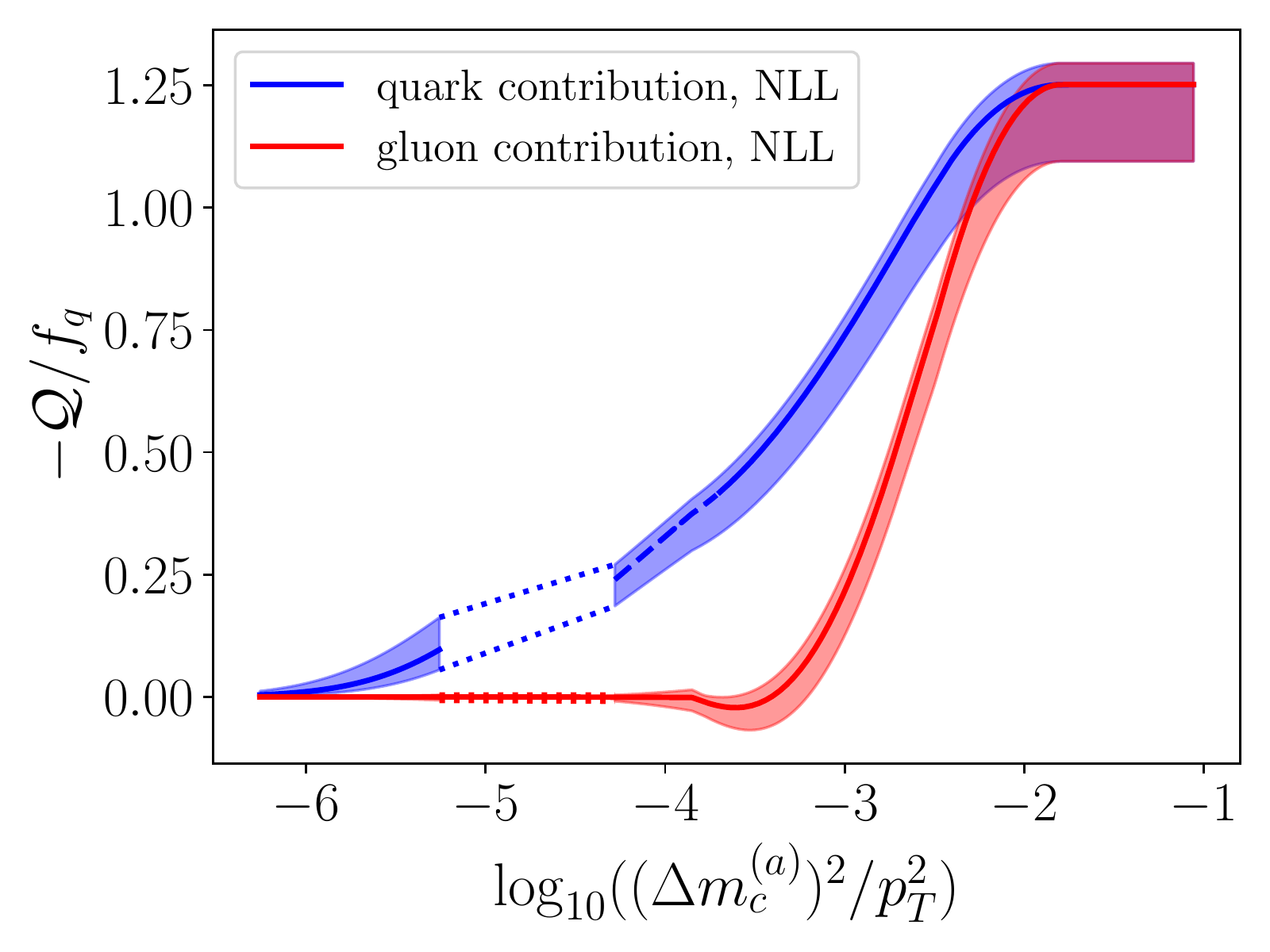}
        \caption{$-\ml{Q}$ with $\beta_1=1$, $\beta_2=0$.}
        \label{fig:Q10_full}
    \end{subfigure}%
    ~
    \begin{subfigure}[t]{0.49\textwidth}
        \centering
        \includegraphics[height=2.2in]{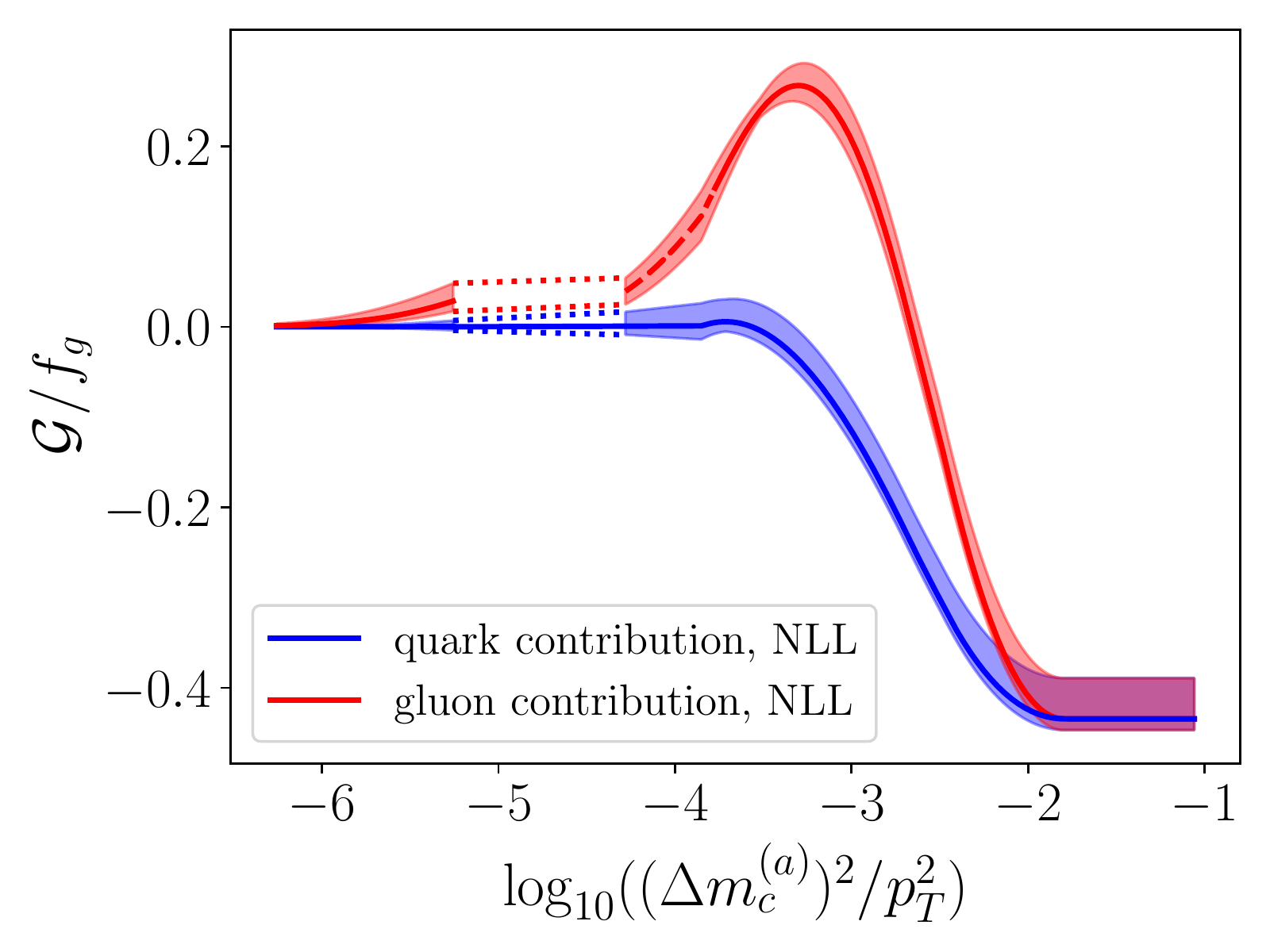}
        \caption{$\ml{G}$ with $\beta_1=1$, $\beta_2=0$.}
        \label{fig:G10_full}
    \end{subfigure}%
    
    \begin{subfigure}[t]{0.49\textwidth}
        \centering
        \includegraphics[height=2.2in]{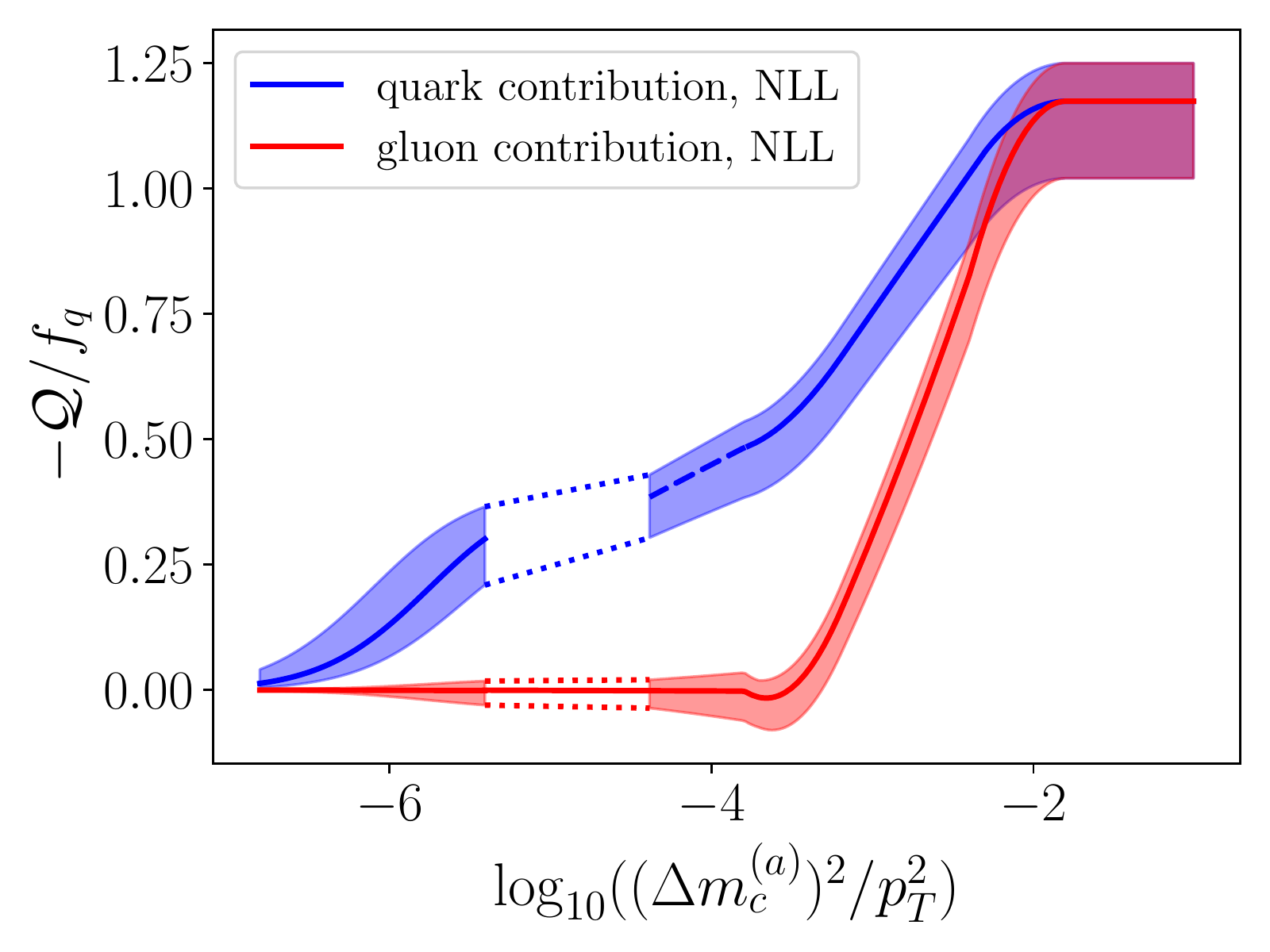}
        \caption{$-\ml{Q}$ with $\beta_1=\beta_2=0$.}
        \label{fig:Q00_full}
    \end{subfigure}%
    ~
    \begin{subfigure}[t]{0.49\textwidth}
        \centering
        \includegraphics[height=2.2in]{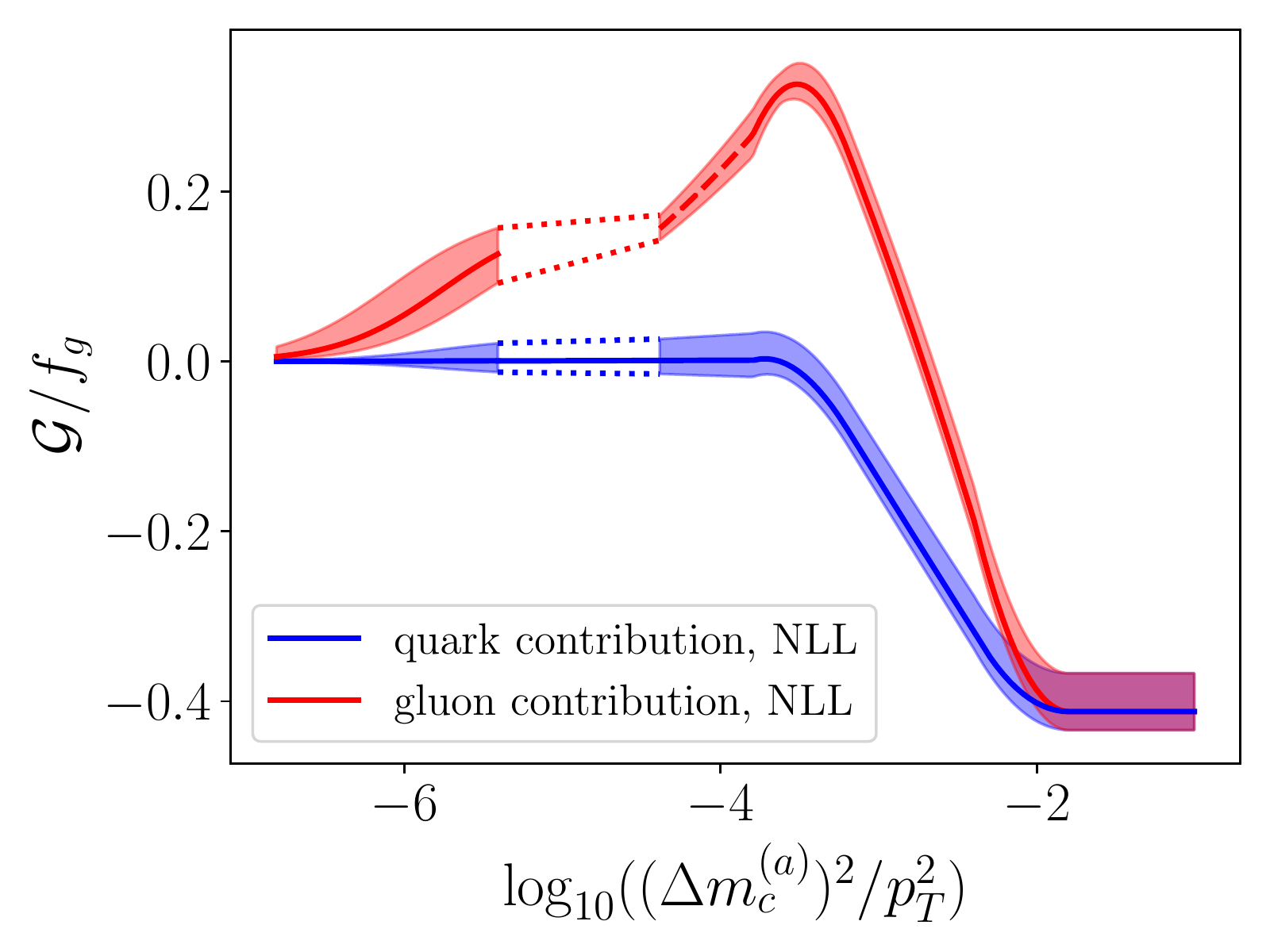}
        \caption{$\ml{G}$ with $\beta_1=\beta_2=0$.}
        \label{fig:G00_full}
    \end{subfigure}%
\caption{Pure quark (left column) and gluon (right column) observables with the shape functions included for cases with $\beta_1=\beta_2=1$ (upper row), $\beta_1=1$, $\beta_2=0$ (middle row) and $\beta_1=\beta_2=0$ (lower row) in the full jet mass region. Only the results in the deep nonperturbative regime and in the fully perturbative regime are shown. The results in the transition region are omitted here. The uncertainty bands include uncertainties from the observable definition, the perturbative calculation and the shape functions.}
\label{fig:Obs_full_equal}
\end{figure}

In Fig.~\ref{fig:Obs_full_equal} we show all the three cases with different $\beta_i$s.
Similar to Fig.~\ref{fig:Obs_shape} above we fix $f_q=f_g=1/2$, so the same discussion given there (about extending the results to other values for these quark and gluon fractions) applies here as well. Our formulas directly give results in two specific regions: i) the nonperturbative regime on the left of the plots for small $\Delta m_c^{(a)}$, where \eqn{nonpertQG} is used, and ii) the perturbative regime on the right of the plots for large $\Delta m_c^{(a)}$, where \eqn{QGpert} is used.  Directly obtaining results in between these two regimes requires a treatment of the case illustrated in \Fig{fig:modes}b, which we have not done here (it is left to future studies). Hence for these intermediate values of $\Delta m_c^{(a)}$ we simply show the expected interpolation by dotted lines in the panels of Fig.~\ref{fig:Obs_full_equal}. Since the gluon contribution to ${\cal Q}$ and quark contribution to ${\cal G}$ are predicted to vanish in the nonperturbative and perturbative regions, the interpolation in the intermediate region is also predicted to remain purely quark or purely gluon. In this intermediate region the interpolation gives an estimate for the size of the non-zero contribution to the observable as well as the uncertainties. 

The experimental application of the pure quark and pure gluon observables for quark and gluon jet separation, only relies on our prediction for the absence of the ``wrong parton'' contributions, and in particular does not require perturbative predictions for the non-zero value of the observables from the ``right parton'' (which do show more uncertainty in the interpolation in the figures).  For this reason the uses of the pure quark and pure gluon observables for quark and gluon jet tagging, are quite robust, and meet the original goal of working over a wide region of values for the phase space variable $\Delta m_c^{(a)}$.  

In Fig.~\ref{fig:Obs_full_equal} we also include uncertainty bands on the factorization based predictions in both the nonperturbative and perturbative regimes of $\Delta m_c^{(a)}$. The bands for the nonperturbative regime are determined by computing the observable uncertainty, perturbative uncertainty, and nonperturbative uncertainty using the methods described above, and then summing these in quadrature.  
In the fully perturbative region, the perturbative uncertainty is estimated by varying the $\mu_{{\rm cs}\,i}$ and $\mu_{{\rm gs}\,i}$  scales simultaneously up/down by a factor of two.  These simultaneous variations ensure that the scales never go into unphysical configurations, such as by crossing each other. These variations are done in a correlated way for the two sets $(a)$ and $(b)$, since we are again providing an estimate for the same missing higher order terms in these two sets of parameter choices from the same cumulative cross section. As can be seen, the distinguishing power remains robust in the presence of the uncertainties, in particular for the lower most panels with $\beta_1=\beta_2=0$. The observable and perturbative uncertainty can both be reduced by carrying out higher order calculations, and it is clear that this would be beneficial for the $\beta_1=\beta_2=1$ case. The gluon observable for the $\beta_1=1$, $\beta_2=0$ case may be difficult to use due to the more rapid fall off in the gluon contribution in the region where the quark contribution has become zero. 

A relevant question is the acceptance for the jets used in our analysis, i.e., what fraction of the jets are retained by restricting to the necessary region of $\Delta m_c^{(a)}$.  This can be obtained directly by considering the cumulative cross sections $\Sigma^{(a,b)}(\Delta m_c^{(a)})$ which enter ${\cal Q}$ and ${\cal G}$. Since these cross sections are normalized to one at the maximum value of $\Delta m_c^{(a)}$, the acceptance is obtained from the smaller value of $\Sigma^{(a,b)}$ at the value of $\Delta m_c^{(a)}$ below which the pure quark and gluon observables are active. For the $\beta_1=\beta_2=0$ case, varying the quark fraction in the range $0.25$--$0.75$,
we find that the acceptances for our method are in the range $37$--$52\%$. For this same range of quark fractions, we find the acceptances fall in the ranges $18$--$33\%$ and $57$--$70\%$ for the $\beta_1=1$, $\beta_2=0$ and $\beta_1=\beta_2=1$ cases respectively.

It is also interesting to examine in more detail the reason why the curves undergo several changes in slope in Fig.~\ref{fig:Obs_full_equal}.  Starting from the right side of the plots at large $\Delta m_c^{(a)}$ we are beyond the endpoint of the spectrum in \eqn{dmc-upper}, and hence have $\hat \Sigma_j^{(a)}=1$ and $\hat \Sigma_j^{(b)}=1$, so from \eqn{QGpert} the curves are flat with their constant value given by $1-\xi_j$.  Moving to smaller $\Delta m_c^{(a)}$ we reach the endpoint of the $\Delta m_c^{(a)}$ spectrum, after which $\hat \Sigma_j^{(a)}$ is changing, and we then reach the endpoint of the $\Delta m_c^{(b)}$ spectrum whereby $\hat \Sigma_j^{(b)}$ is changing too. For the three cases $\{\beta_i=1, \beta_1-1=\beta_2=0, \beta_i=0\}$ the endpoint for the $\Delta m_c^{(a)}$ spectrum occurs for 
$\log_{10}((\Delta m_c^{(a)})^2/p_T^2)=\{-2.5,-1.8,-1.8\}$
respectively, 
while the endpoint for the $\Delta m_c^{(b)}$ spectrum happens for 
$\log_{10}((\Delta m_c^{(a)})^2/p_T^2)=\{-3.0,-2.7,-3.2\}$ respectively.
Shortly after this transition we enter the perturbative resummation dominated regime and the gluon contribution to ${\cal Q}$ and quark contribution to ${\cal G}$ vanish as predicted. Next we enter the intermediate region which is indicated by dashed uncertainty bands. Then finally we hit the fully nonperturbative region at the values given by \eqn{dmc-np}, which for the three cases correspond to 
$\log_{10}( (\Delta m_c^{(a)})^2/p_T^2)=\{-4.2,-5.4,-5.4\}$
respectively.

From Fig.~\ref{fig:Obs_full_equal} we see that the case with $\beta_1=\beta_2=0$ does the best job of separating quark and gluon jets of the cases we have considered, since the non-vanishing contributions are well separated from zero in a wide kinematic region, even after the uncertainties are taken into account. This is mostly due to the larger difference between $\xi_q$ and $\xi_g$ and the larger perturbative results of the cumulative jet mass cross sections, as shown in Fig.~\ref{fig:pert_results}. The values of $\xi_j$ depend on the choice of the $z_{{\rm cut}\,i}^{(a,b)}$ parameters and larger differences in the $z_{{\rm cut}\,i}^{(a,b)}$ lead to bigger values of $\xi_j$, as in the  $\beta_1=\beta_2=0$ case.
These observations make clear the importance of scanning through the parameter space of the remaining free variables to maximize the discrimination power, and a more detailed analysis than what we have carried out here may well be warranted.

Since the pure quark and gluon observables are applicable even in the intermediate region (with dashed curves), this motivates future theoretical studies to obtain better control of the theoretical interpolation and uncertainties in this region. 

\subsection{Monte Carlo Results for ${\cal Q}$ and ${\cal G}$}
\label{sec:mc}

For realistic proton-proton collision events, the ${\cal Q}$ and ${\cal G}$  observables will have contamination from  initial state radiation (ISR) and multiparton interaction (MPI) effects. 
The size and impact of these effects are not tested by our analytic predictions for ${\cal Q}$ and ${\cal G}$ in Section~\ref{sec:analytic}, and hence we will study them here using Monte Carlo.  Here we focus on the $\beta_1=\beta_2=0,\,z_{{\rm cut}1}^{(a)}=0.1,\,z_{{\rm cut}2}^{(a)}=0.4,\,z_{{\rm cut}1}^{(b)}=0.02$ and $z_{{\rm cut}2}^{(b)}=0.08$ case.

As we have seen from the Monte Carlo studies of collinear drop jet mass cross sections in Section~\ref{sec:ISReffects}, 
smaller values of the jet radius $R$ (or larger values of $z_{{\rm cut}1}^{(a,b)}$) are favored in order to remove the dominant ISR and MPI effects.  This motivated our choice of $R=0.2$.  
Here we again use \Pythia 8~\cite{Sjostrand:2014zea} and \Vincia~\cite{Fischer:2016vfv} with the same setup described in Section~\ref{sec:ISReffects}. For convenience we again take $f_q=f_g=1/2$ by normalizing the cumulant jet mass cross sections for the quark and gluon jet events independently.  As discussed previously, results for any other values are obtained by a simple rescaling.
We will separately consider Monte Carlo results with and without ISR and MPI effects.

\begin{figure}[p]
    \begin{subfigure}[t!]{0.49\textwidth}
        \centering
        \includegraphics[height=2.2in]{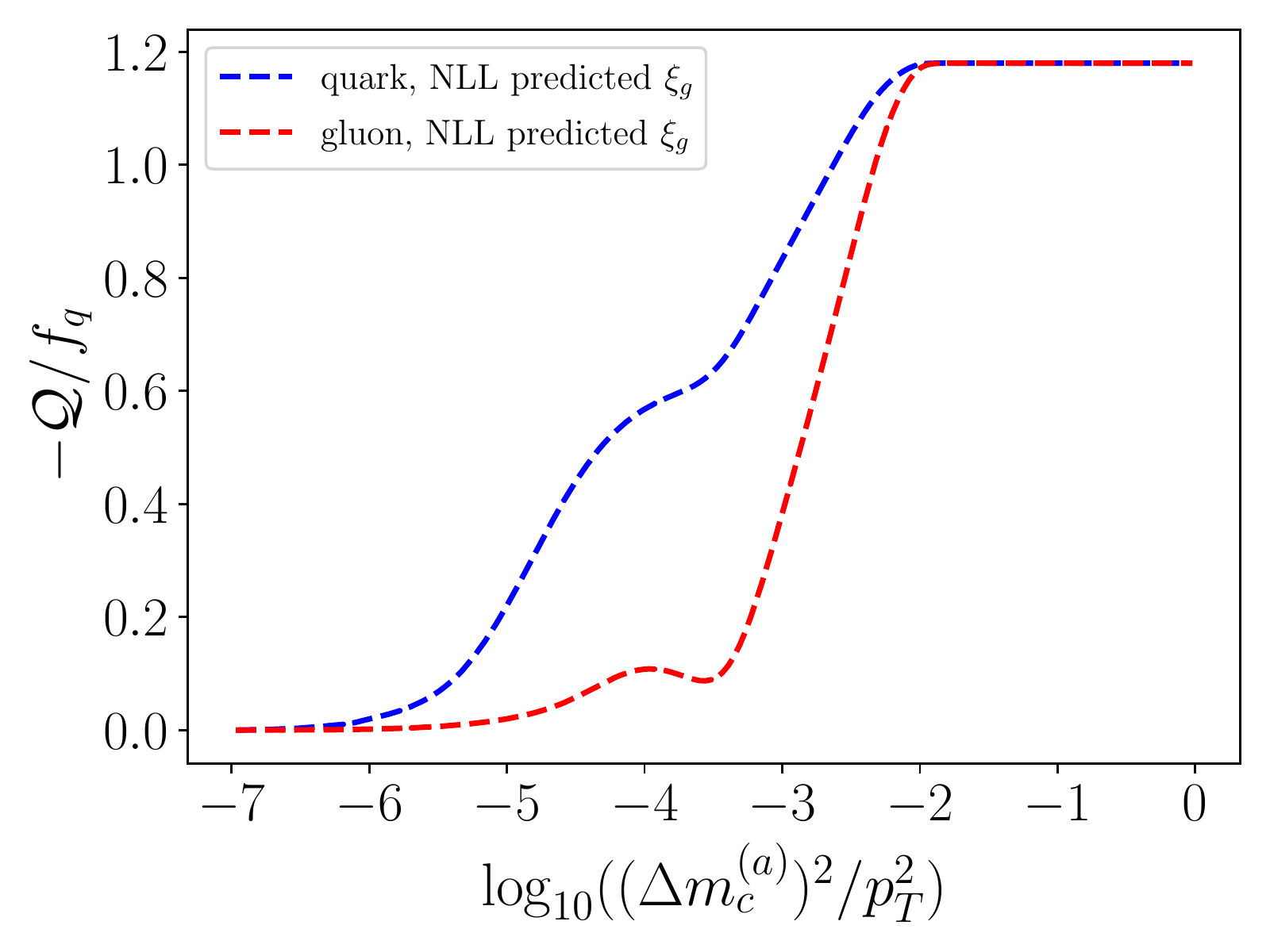}
        \caption{$-\ml{Q}$ with NLL predicted $\xi_g$.}
        \label{fig:Qpythia}
    \end{subfigure}%
    ~
     \begin{subfigure}[t!]{0.49\textwidth}
        \centering
        \includegraphics[height=2.2in]{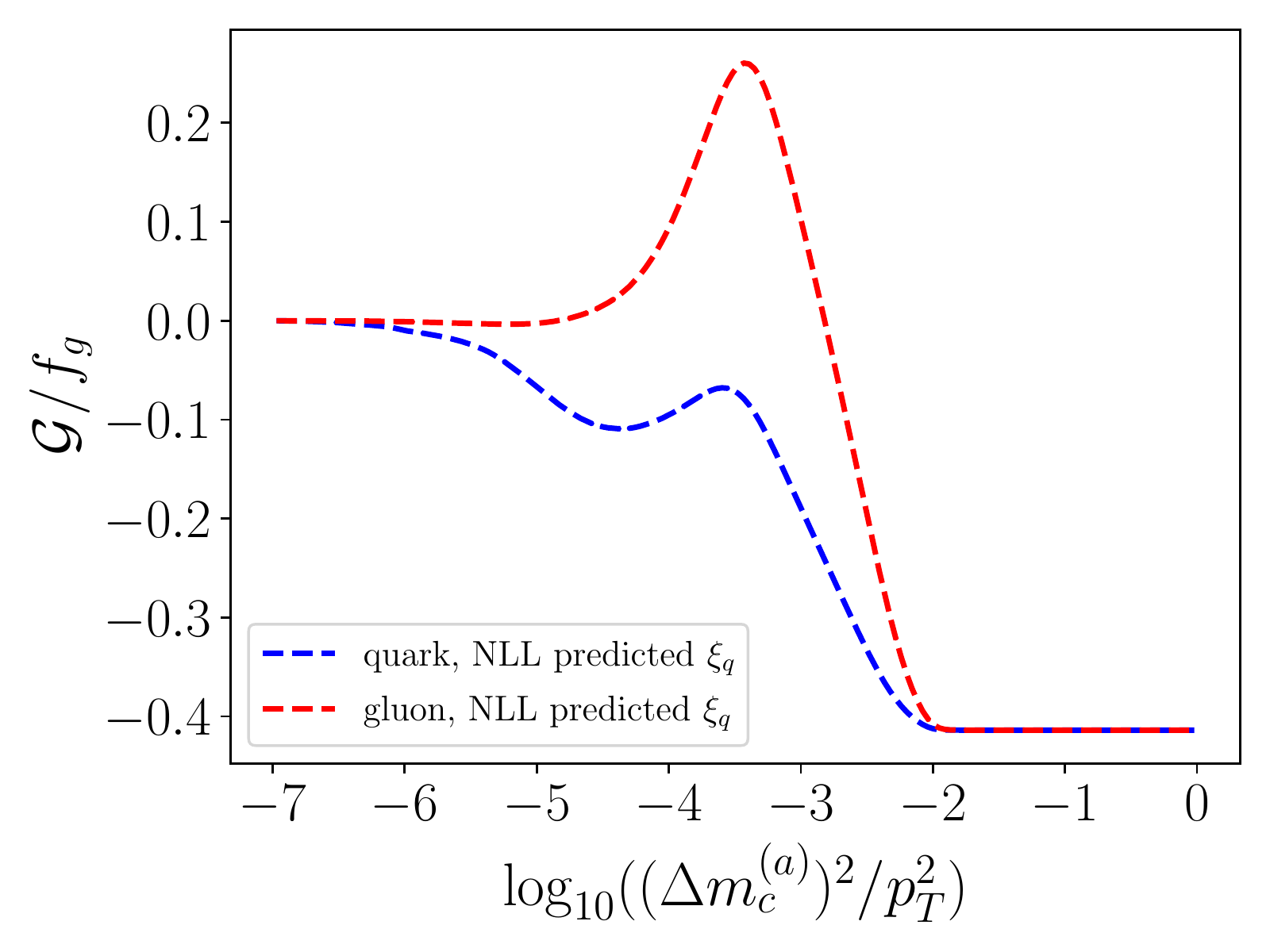}
        \caption{$\ml{G}$ with NLL predicted $\xi_q$.}
        \label{fig:Gpythia}
    \end{subfigure}%

    \begin{subfigure}[t!]{0.49\textwidth}
        \centering
        \includegraphics[height=2.2in]{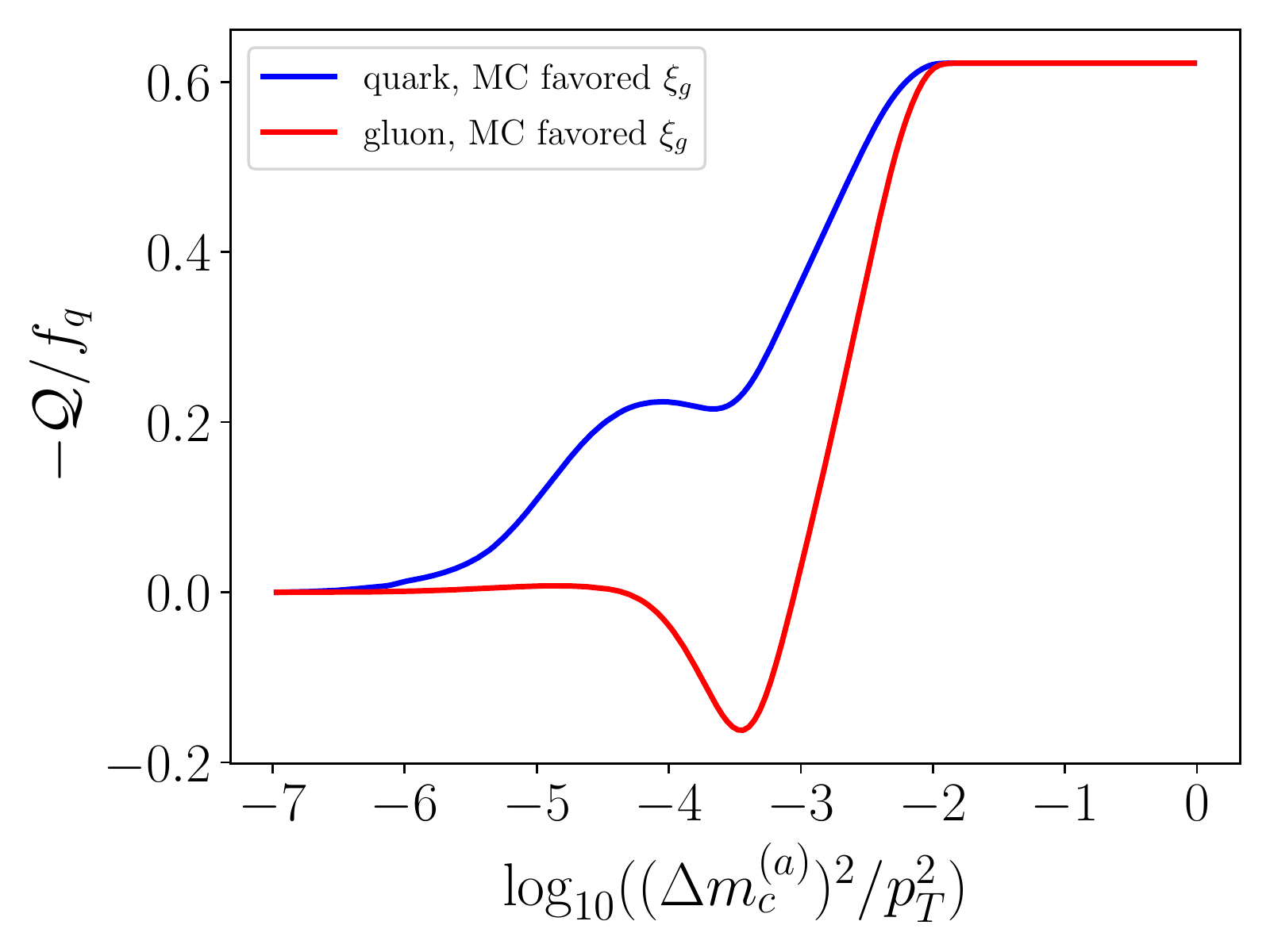}
        \caption{$-\ml{Q}$ with Monte Carlo favored $\xi_g$.}
        \label{fig:Qpythia_mc}
    \end{subfigure}%
    ~
    \begin{subfigure}[t!]{0.49\textwidth}
        \centering
        \includegraphics[height=2.2in]{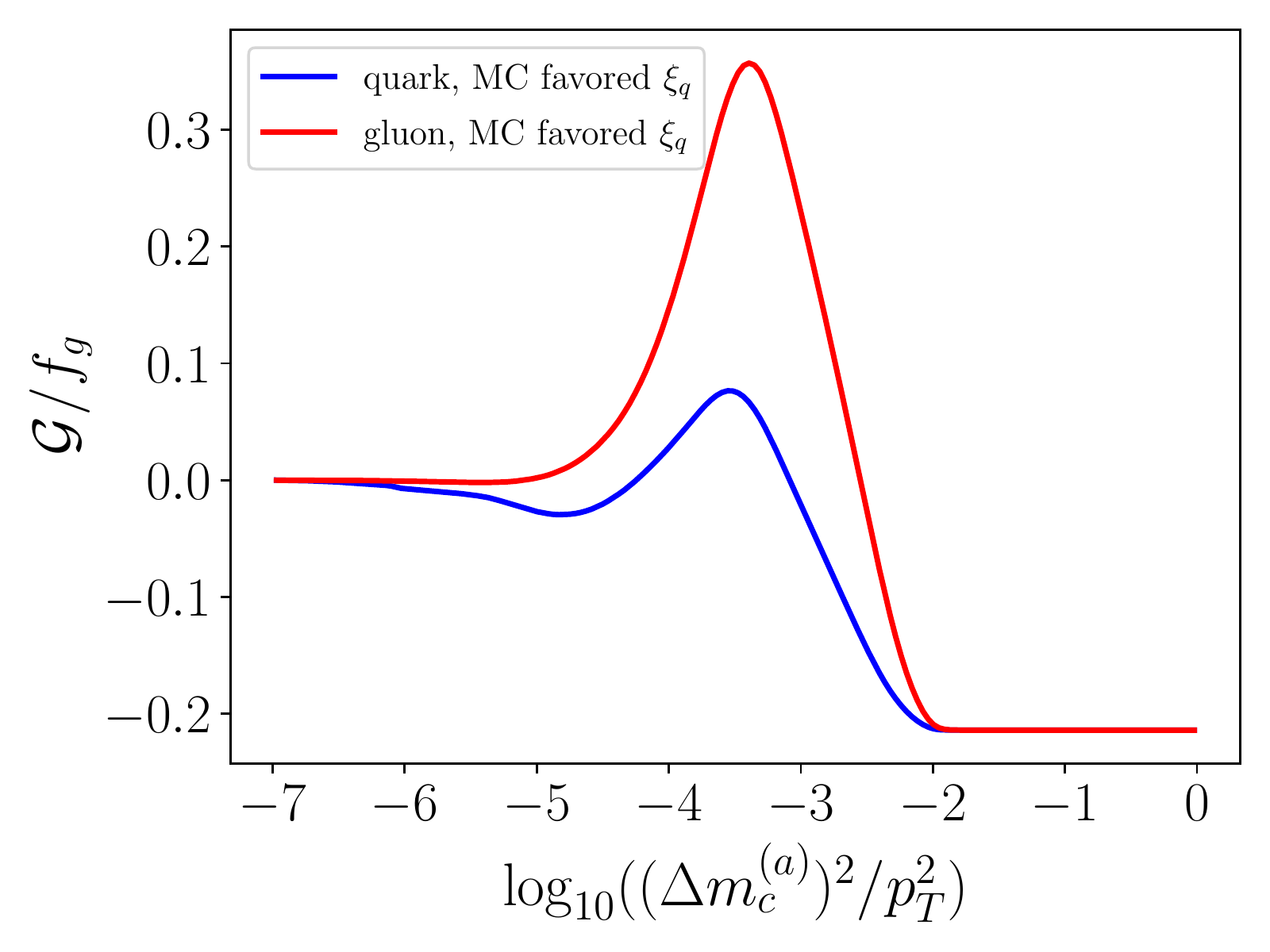}
        \caption{$\ml{G}$ with Monte Carlo favored $\xi_q$.}
        \label{fig:Gpythia_mc}
    \end{subfigure}%

    \begin{subfigure}[t!]{0.49\textwidth}
        \centering
        \includegraphics[height=2.2in]{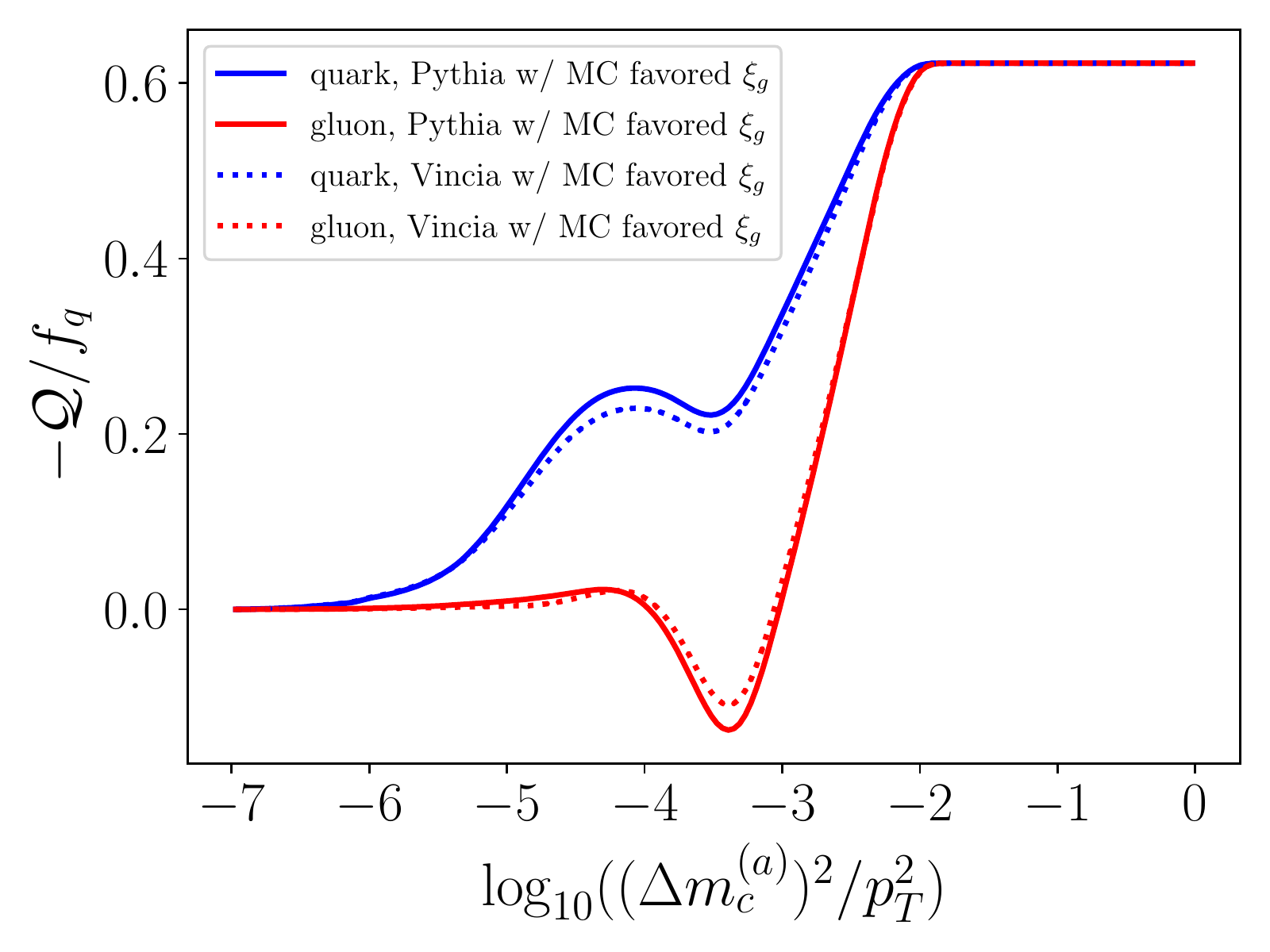}
        \caption{$-\ml{Q}$ with ISR and MPI effects.}
        \label{fig:pure_quark}
    \end{subfigure}%
    ~
    \begin{subfigure}[t!]{0.49\textwidth}
        \centering
        \includegraphics[height=2.2in]{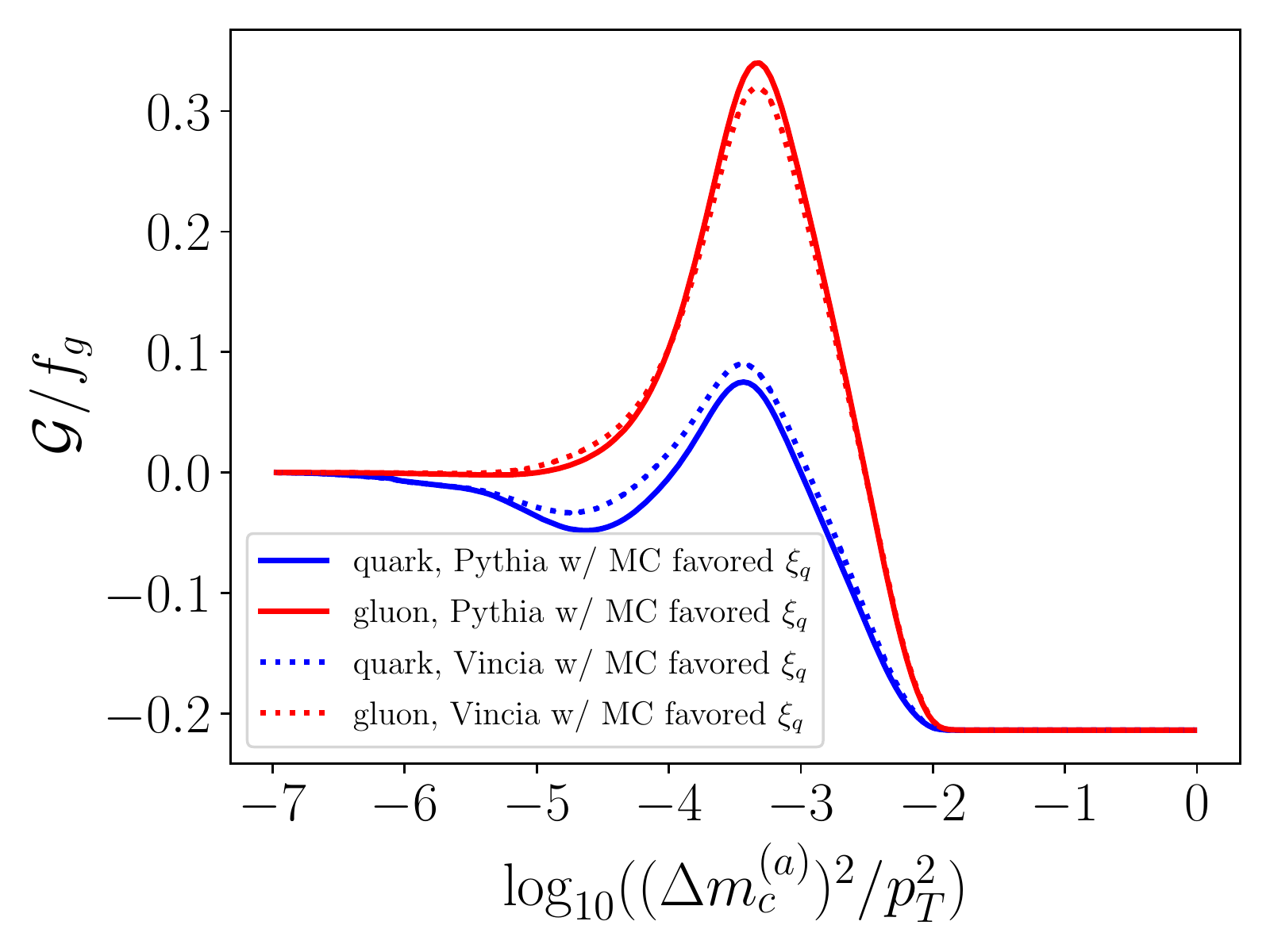}
        \caption{$\ml{G}$ with ISR and MPI effects.}
        \label{fig:pure_gluon}
    \end{subfigure}%
\caption{Monte Carlo results for the pure quark and gluon observables with $\beta_1=\beta_2=0,\,z_{{\rm cut}1}^{(a)}=0.1,\,z_{{\rm cut}2}^{(a)}=0.4$ and $z_{{\rm cut}1}^{(b)}=0.02$ from \Pythia (with results also from \Vincia in the bottom panels). The top and middle panels have ISR and MPI turned off. The dashed curves in the top panels use the $\xi_j$ predicted from our NLL  resummation. The remaining panels with solid curves use the $\xi_j$ values chosen in the text, which better purify the quark and gluon observables in \Pythia.
The bottom panels have ISR and MPI effects turned on.
}
\label{fig:pythia_noISR_ISR}
\end{figure}

We start by considering results for pure quark and gluon observables using \Pythia events that have been generated with the ISR and MPI effects turned off, which enables us to benchmark the accuracy of the Monte Carlo results against our analytic hadronic predictions.  This is important since one of the original motivations~\cite{Chien:2019osu} for constructing the collinear drop observables was to have soft sensitive observables for which predictions of Monte Carlo generators are not as reliable, and which hence could be utilized to test improvements to Monte Carlos. 

Using \eqn{QGdefns} with the linear combination coefficients $\xi_j$ calculated with NLL precision given in \Tab{tab:1}, and results for the $\Sigma$s from \Pythia without ISR or MPI, gives the results in the upper panels of Fig.~\ref{fig:pythia_noISR_ISR} (dashed curves).  These results can therefore be directly compared to those in the lower panels of \fig{Obs_full_equal}.  We observe that the curves change their slope at similar points to our analytic predictions, but in the region around $\log_{10}( (\Delta m_c^{(a)})^2 /p_T^2)=-3.7$, the \Pythia results are not purely quark or gluon, despite the fact that this is the perturbative resummation region where our NLL formulae indicate they should be. Quite possibly this result comes about due to an inability of the \Pythia shower to accurately predict the structure of the soft radiation in this region (since after all, it is designed with the goal of accurately predicting collinear radiation that is being dropped here).  Possible sources for this discrepancy in \Pythia include the use of our NLL parameter values $\xi_g=2.18$ and $\xi_q=1.41$, which encode assumptions about the resummation of radiation above and between the global-soft scales, and the soft radiation predictions for the shape of $\hat\Sigma_j(\Delta m_c)$.

We can easily test these sources by making Monte Carlo predictions using different values of the $\xi_j$ parameters. 
Taking the same values for the collinear drop parameters but using 
$\xi_g=1.62$ and $\xi_q=1.21$ to construct the ${\cal Q}$ and ${\cal G}$ observables, we obtain the results shown in the middle panels of Fig.~\ref{fig:pythia_noISR_ISR} (solid curves). We observe improved purity for the quark observable for $\log_{10}( (\Delta m_c^{(a)})^2 /p_T^2)\lesssim -4$, while there still do appear to be residual shape differences.  For the gluon observable the improvement is not very good, possibly because the gluon observable was smaller to begin with which magnifies the impact of shape deviations. 

Since our construction of the pure quark and gluon observables crucially relies on the rebinning relation between the jet masses, \eqn{QGchoice}, the Monte Carlo results indicate that only the gluon jet generated from \Pythia behaves as our analytic calculations predict after the CD grooming, in the sense of the parametric dependence of the cross section on the jet mass. As the jet mass dependence of the gluon jet cross section is consistent with our analytic study, we can make the gluon contribution vanish by modifying $\xi_g$ and obtain a mostly pure quark observable from the \Pythia Monte Carlo result. 
In contrast, the failure of the pure gluon observable in \Pythia implies a disagreement between \Pythia and our analytic prediction for the parametric dependence of the quark jet mass cross section on $\Delta m_c$. 
We have tested that these observations persist when considering ${\cal Q}$ and ${\cal G}$ observables based on other values of the $\beta_i$, and with different $p_T$. 
In the future, it will be useful to study the parametric dependence of the cross section on the jet mass in more detail and with other Monte Carlo generators to further explore this issue. 

Next we consider Monte Carlo results with both the ISR and MPI effects included. We use the same jet kinematics and CD parameters as in the studies with both ISR and MPI turned off, and again take $\xi_q = 1.21$ and $\xi_g = 1.62$ to obtain results from both \Pythia and \Vincia.  
Comparing the middle and bottom panels of \Fig{fig:pythia_noISR_ISR} we see that the ISR and MPI effects are very mild for the small $R=0.2$ jets considered here. 
Thus we find that again the pure quark observable works in the region of $\log_{10}( (\Delta m_c^{(a)})^2/p_T^2) \leq -4$, while the pure gluon observable does not work here for the same reason discussed above for the case with the ISR and MPI effects turned off.

We conclude by commenting that regardless of finding ways to improve Monte Carlo predictions for the collinear drop observables that are dominated by soft physics, if one constructs our theoretically predicted pure quark and gluon observables (e.g. from \Tab{tab:1}) in the experimental data, we can then use Eq.~(\ref{eq:fraction_ratio}) to extract the fractions of quark and gluon jets in jet samples. A cross check on the results can be obtained by carrying out the quark/gluon fraction determination with several different collinear drop observables. After determining the quark and gluon jets fractions in particular jet samples, we can obtain individual distributions of the quark and gluon jets for any observable one measures on these samples.

Note that we did not explore how the ISR and MPI effects depend on the jet kinematics such as transverse momentum $p_T$ and rapidity $\eta_J$, which might also be exploited to remove the ISR and MPI effects for larger $R$ jets. (See for example Ref.~\cite{Stewart:2014nna} for a study of these dependencies for ungroomed jets.) We have also not considered the use of fractional values of $\beta$, like $\beta=0.5$ to construct pure quark and gluon observables. We leave these investigations to future studies.

\section{Conclusions}
\label{sec:conclusions}

In this paper, we constructed a class of pure quark and gluon observables using the cumulative distributions of the jet mass with collinear drop jet grooming.  We formed linear combinations of two cumulative jet mass cross sections with the same $\beta_i$ but different $z_{{\rm cut}\,i}$ parameters in collinear drop. The linear combination coefficients $\xi_{q,g}$ are perturbatively calculable and defined such that either the quark or the gluon perturbative contribution to the cumulative cross section vanishes.  This yields pure quark and gluon observables in the perturbative region where logarithmic resummation dominates. In the small jet mass region, nonperturbative effects become important and have to be taken into account in the construction, which we did by using the theoretically predicted leading hadronization effects from nonperturbative shape functions. The choice of different jet mass bins for the two cumulative cross sections used in the linear combinations, ensures that the nonperturbative shape functions become overall factors, and thus the same purely perturbative linear combination coefficients work in this region. In this way, we constructed a class of pure quark and gluon observables, by using different collinear drop parameters. 
In practice, one can vary the remaining free parameters to maximize the disentangling power of these observables.

Analytic results of the pure quark and gluon observables for three sets of $\beta_i$s were presented. We estimated both the ``observable uncertainty'' from calculating the $\xi_j$ parameters, the standard perturbative uncertainty from scale dependence, and the nonperturbative uncertainty by varying parameters in the shape function models. 
As can be seen from \Fig{fig:Obs_full_equal}, our results in the case $\beta_1=\beta_2=0$ are most promising in the sense of experimental feasibility: The observables have a large gap between the quark and gluon contributions in a wide kinematic region, where one contribution vanishes, even if the uncertainties are accounted for. Thus, our constructed pure quark and gluon observables are promising for use with real data in the presence of experimental uncertainties. 
Our results in the $\beta_1=\beta_2=1$ case also look promising, but require predictions using perturbative results beyond NLL in order to reduce the theoretical uncertainties.

Finally we carried out Monte Carlo studies of these observables. Such studies are challenging since Monte Carlo parton showers are primarily designed to predict observables dominated by collinear radiation, and hence have some trouble predicting all the features of collinear drop spectra~\cite{Chien:2019osu}, since they are dominated by soft radiation. 
Without ISR and MPI effects the \Pythia results differ from our predictions for the pure quark and gluon observables at NLL.  By further tuning the fraction $\xi_g$ we obtained a pure quark observable in \Pythia that works over a range of phase space, while tuning $\xi_q$ did not significantly improve the \Pythia pure gluon observable.
Nevertheless we can use Monte Carlo to test the impact of initial state radiation and underlying event (MPI) effects for proton-proton collisions, on our construction, by comparing curves with and without these effects turned on. 
We found that the ISR and MPI effects 
have a negligible impact for jets with small jet radii in proton-proton collisions.

To use our pure quark and gluon observables in \eqn{QGdefns} with experimental jet samples, we recall that measurement of ${\cal Q}(\Delta m_c^{(a)})$ and ${\cal G}(\Delta m_c^{(a)})$ in two distinct jet samples (such as dijets and $Z$+jet) immediately yields a measurement of the quark and gluon jet fractions through \Eq{eq:fraction_ratio}. 
We recommend the use of the best performing parameter choice that we have identified, which gives the strongest distinguishing power taking all factors into consideration. This corresponds to  $\beta_1=\beta_2=0$, $z_{{\rm cut}1}^{(a)}=0.1$, $z_{{\rm cut}2}^{(a)}=0.4$, $z_{{\rm cut}1}^{(b)}=0.02$, which fixes $z_{{\rm cut}2}^{(b)}=0.08$ and the ratio $\Delta m_c^{(b)}/\Delta m_c^{(a)} = 2.24$. The observables are evaluated in the appropriate kinematic region of the cumulative jet mass observable $\Delta m_c^{(a)}$, shown in Figs.~\ref{fig:Obs_full_equal}e and~\ref{fig:Obs_full_equal}f. 
As a consistency check these measurements can be carried out using several different values of $\Delta m_c^{(a)}$ in the identified phase space, to make sure the same results are obtained.  
To minimize underlying event effects it is useful to consider small $R$ jets, so here we used $R=0.2$. To have valid perturbative calculations for the required $\xi_j$ linear combination parameters, it is necessary to consider sufficiently large $p_T$ jets, and for illustration we took $p_T=800\,{\rm GeV}$ at central rapidity $\eta_J=0$.  The resulting $\xi_q$ and $\xi_g$ parameters for this choice of $p_T$, $R$, and $\eta_J$ are given in \Tab{tab:1}, and can be easily recomputed at NLL for other choices of jet kinematics using \Eq{eq:finalxij}.
For this choice of parameters consideration of the appropriate region of $\Delta m_c^{(a)}$ corresponds to a rather high acceptance on the reconstructed jets, estimated to be in the range 37--52\% across a range of different jet samples.

In the future it would be worth carrying out a more substantial scan over parameters and kinematic variables to further optimize the parameter choice for the construction of pure quark and gluon observables, while satisfying various constraints.

\acknowledgments
We would like to thank Anna Ferdinand and Aditya Pathak for the use of their JETlib package to calculate collinear drop observables in Monte Carlo simulations, and Johannes Michel, Govert Nijs, and Jesse Thaler for helpful conversations.
This work is supported by the U.S. Department of Energy, Office of Science, Office of Nuclear Physics grant DE-SC0011090. I.S. was also supported in part by the Simons Foundation through the Investigator grant 327942. 

\appendix
\section{Collinear-Soft Functions and Shape Functions in Laplace Space}
\label{app:laplace}

The CS functions with the nonperturbative shape functions can be written as~\cite{Hoang:2019ceu}
\begin{align}
S_{C_j}\pig( \ell^+_1 Q_{\text{cut}1}^{\frac{1}{1+\beta_1}}, \beta_1, \mu \pig) &= \int_0^{+\infty} \diff k_1 \, \hat{S}_{C_j} \pig( \ell^+_1 Q_{\text{cut}1}^{\frac{1}{1+\beta_1}} - k_1^{\frac{2+\beta_1}{1+\beta_1}}, \beta_1, \mu \pig) F_1^j(k_1,\beta_1) 
\,, \\
D_{C_j}\pig( \ell^+_2 Q_{\text{cut}2}^{\frac{1}{1+\beta_2}}, \beta_2, \mu \pig) &= \int_0^{+\infty} \diff k_2 \, \hat{D}_{{C}_j} \pig( \ell^+_2 Q_{\text{cut}2}^{\frac{1}{1+\beta_2}} - k_2^{\frac{2+\beta_2}{1+\beta_2}}, \beta_2, \mu \pig) F_2^j(k_2,\beta_2)
 \,,\nn
\end{align}
where $k_i\sim \Lambda_{\rm QCD}$, the shape functions $F_i^j(k_i,\beta_i)$ depend on the parton species $j$ initiating the jet, and are mass dimension $-1$. These shape functions only depend on $\beta_i$, and not on $z_{{\rm cut}\,i}$. With the shape functions included, the convolution in \Eq{eq:convol} can be written as
\begin{align} \label{eq:PjSF}
 P_j^{\rm CD} (\Dms, Q, \tilde{z}_{{\rm cut}\,i},\beta_i,\mu) & = \Qcuta^{\frac{1}{1+\beta_1}} 
\Qcutb^{\frac{1}{1+\beta_2}} \int \diff \ell_1^+ \diff \ell_2^+ \diff k_1 \diff k_2 \,\delta \big(\Dms - Q\ell_1^+ - Q\ell_2^+ \big) \nn\\
&\ \ \times
\hat{S}_{C_j} \pig( \ell^+_1 Q_{\text{cut}1}^{\frac{1}{1+\beta_1}} - k_1^{\frac{2+\beta_1}{1+\beta_1}}, \beta_1, \mu \pig)
\hat{D}_{{C}_j} \pig( \ell^+_2 Q_{\text{cut}2}^{\frac{1}{1+\beta_2}} - k_2^{\frac{2+\beta_2}{1+\beta_2}}, \beta_2, \mu \pig) \nn\\[4pt]
& \ \ \times
F_1^j(k_1,\beta_1)  F_2^j(k_2,\beta_2) \,.\quad\quad
\end{align}
The Laplace transform is given by
\begin{align}
\widetilde{P}_j^{\text{CD}}(y,\mu) =  \int_0^{+\infty} \diff \Delta m^2 \, \exp\big( -y e^{-\gamma_E} \Dms \big) P_j^{\text{CD}}(\Dms,\mu) \,,
\end{align}
where for simplicity we suppress the dependence of $P_j^{\text{CD}}$ on $Q$, $z_{{\rm cut}\,i}$ and $\beta_i$. Applying this to \eqn{PjSF} we find
\begin{align}
\widetilde{P}_j^{\text{CD}}(y,\mu) 
&= Q_{\text{cut}1}^{\frac{1}{1+\beta_1}} Q_{\text{cut}2}^{\frac{1}{1+\beta_2}} \int \diff \ell^+_1 \diff \ell^+_2 \diff k_1  \diff k_2\, \exp\pig( -y e^{-\gamma_E}Q \ell_1^+ -ye^{-\gamma_E} Q \ell_2^+ \pig)  
\\
&\ \times \hat{S}_{C_j} \pig( \ell^+_1 Q_{\text{cut}1}^{\frac{1}{1+\beta_1}} - k_1^{\frac{2+\beta_1}{1+\beta_1}}, \beta_1, \mu \pig) F_1^j(k_1,\beta_1)
\hat{D}_{{C}_j} \pig( \ell^+_2 Q_{\text{cut}2}^{\frac{1}{1+\beta_2}} - k_2^{\frac{2+\beta_2}{1+\beta_2}}, \beta_2, \mu \pig) F_2^j(k_2,\beta_2)
 \nn \\
&= \hat{\widetilde{S}}_{C_j} \pig( yQQ_{\text{cut}1}^{\frac{-1}{1+\beta_1}}, \beta_1, \mu \pig) \widetilde{F}_1^j (yQQ_{\text{cut}1}^{\frac{-1}{1+\beta_1}},\beta_1 ) 
\hat{\widetilde{D}}_{{C}_j} \pig( yQQ_{\text{cut}2}^{\frac{-1}{1+\beta_2}}, \beta_2, \mu \pig) \widetilde{F}_2^j(yQQ_{\text{cut}2}^{\frac{-1}{1+\beta_2}},\beta_2 ) 
\,,\nn
\end{align}
where
\begin{align} \label{eq:app_laplacep_shape}
\hat{\widetilde{S}}_{C_j} \pig( yQQ_{\text{cut}1}^{\frac{-1}{1+\beta_1}}, \beta_1, \mu \pig) 
&= \int \diff q_1
\exp\Big( -y e^{-\gamma_E} QQ_{\text{cut}1}^{\frac{-1}{1+\beta_1}} q_1 \Big)
\hat{S}_{C_j}( q_1, \beta_1, \mu ) \,, \\
\hat{\widetilde{D}}_{{C}_j} \pig( yQQ_{\text{cut}2}^{\frac{-1}{1+\beta_2}}, \beta_2, \mu \pig) 
&= \int \diff q_2
\exp\Big( -y e^{-\gamma_E} QQ_{\text{cut}2}^{\frac{-1}{1+\beta_2}} q_2 \Big)
\hat{D}_{{C}_j}( q_2, \beta_2, \mu ) \,, \nn \\
\widetilde{F}^j_i \pig( yQQ_{\text{cut}\,i}^{\frac{-1}{1+\beta_i}},\beta_i \pig) 
&= \int \diff k_i \exp\Big( -y e^{-\gamma_E} QQ_{\text{cut}\,i}^{\frac{-1}{1+\beta_i}} k_i^{\frac{2+\beta_i}{1+\beta_i}} \Big) F_i^j(k_i,\beta_i) 
\,. \nn
\end{align}
In these Laplace transforms, the mass dimension of $q_i$ is $\frac{2+\beta_i}{1+\beta_i}$ 
and the three functions on the left-hand side are dimensionless. Using the solutions to the general RG equations of the perturbative CS functions in Laplace space (\ref{eq:sol_rg_cs2}), we find
\begin{align} \label{eq:Pcd_laplace}
\widetilde{P}_j^{\text{CD}}(y,\mu) 
&=  \widetilde{F}^j_1 \pig( yQQ_{\text{cut}1}^{\frac{-1}{1+\beta_1}},\beta_1 \pig)
\hat{\widetilde{S}}_{C_j} \pig( yQQ_{\text{cut}1}^{\frac{-1}{1+\beta_1}}, \beta_1, \Lambda_{\text{cs}1} \pig) 
 \\*
&\times
\widetilde{F}^j_2 \pig( yQQ_{\text{cut}2}^{\frac{-1}{1+\beta_2}},\beta_2 \pig)
\hat{\widetilde{D}}_{{C}_j} \pig( yQQ_{\text{cut}2}^{\frac{-1}{1+\beta_2}}, \beta_2, \Lambda_{\text{cs}2} \pig)
\nn\\
&\times \exp\bigg( -2C_j\frac{2+\beta_1}{1+\beta_1} K(\Lambda_{\text{cs}1}, \mu) + \omega_{S_{C_j}}(\Lambda_{\text{cs}1}, \mu) \bigg) \Bigg( \frac{Q_{\text{cut}1}^{\frac{1}{1+\beta_1}}}{y Q \Lambda^{\frac{2+\beta_1}{1+\beta_1}}_{\text{cs}1}} \Bigg)^{2C_j\omega(\Lambda_{\text{cs}1}, \mu)}\nn\\
&\times \exp\bigg( +2C_j\frac{2+\beta_2}{1+\beta_2} K(\Lambda_{\text{cs}2}, \mu) +  \omega_{D_{{C}_j}}(\Lambda_{\text{cs}2}, \mu) \bigg) \Bigg( \frac{Q_{\text{cut}2}^{\frac{1}{1+\beta_2}}}{y Q \Lambda^{\frac{2+\beta_2}{1+\beta_2}}_{\text{cs}2}} \Bigg)^{-2C_j\omega(\Lambda_{\text{cs}2}, \mu)} 
\,, \nn
\end{align}
where $\Lambda_{{\rm cs}\,i}$ are perturbative scales introduced as the endpoint of the RG evolution of the CS functions. Here the dependence on $\Lambda_{{\rm cs}\,i}$ cancels order-by-order between the boundary condition terms on the first two lines, and the evolution kernel terms on the last two lines. 

In the perturbative region the scales $\Lambda_{{\rm cs}1}=\mu_{{\rm cs}1}$  and $\Lambda_{{\rm cs}2}=\mu_{{\rm cs}2}$ are parametrically different. However when both scales are frozen prior to entering the nonperturbative region we are free to make the simplifying choice of $\Lambda_{{\rm cs}1}=\Lambda_{{\rm cs}2}=\Lambda_{\rm cs}\sim 2\,{\rm GeV}$. For this region the factorization formula provides a way of defining the shape function in the $\overline{\rm MS}$ scheme, through the combinations appearing in the first two lines of \eqn{Pcd_laplace}:
\begin{align} 
\label{eq:Fmsbar1}
\widetilde{F}^{(\overline{\rm MS})j}_1 \pig( yQQ_{\text{cut}1}^{\frac{-1}{1+\beta_1}},\beta_1, \Lambda_{{\rm cs}} \pig) 
&= \widetilde{F}^j_1 \pig( yQQ_{\text{cut}1}^{\frac{-1}{1+\beta_1}},\beta_1 \pig) 
\hat{\widetilde{S}}_{C_j} \pig( yQQ_{\text{cut}1}^{\frac{-1}{1+\beta_1}}, \beta_1, \Lambda_{\text{cs}} \pig) 
 \,, \\
\widetilde{F}^{(\overline{\rm MS})j}_2 \pig( yQQ_{\text{cut}2}^{\frac{-1}{1+\beta_2}},\beta_2, \Lambda_{{\rm cs}} \pig) 
&= \widetilde{F}^j_2 \pig( yQQ_{\text{cut}2}^{\frac{-1}{1+\beta_2}},\beta_2 \pig) 
\hat{\widetilde{D}}_{C_j} \pig( yQQ_{\text{cut}2}^{\frac{-1}{1+\beta_2}}, \beta_2, \Lambda_{\text{cs}} \pig)
\nn \,.
\end{align}


To transform back to the momentum space, we apply the inverse Laplace transform
\begin{align}
P_j^{\text{CD}}(\Dms) = \frac{e^{-\gamma_E}}{2\pi i} \int_{c-i\infty}^{c+i\infty} \diff y \, \exp\big( y e^{-\gamma_E} \Dms \big) \widetilde{P}_j^{\text{CD}}(y)\,,
\end{align}
where $c$ is constant, chosen to be large enough that all the poles of the integrand $\widetilde{P}_j^{\text{CD}}(y)$ are on the left of the integration line. It turns out that it is convenient to separately consider two cases, since the optimal method for carrying out this inverse transform differs in  the perturbative and nonperturbative regions.

\paragraph{Perturbative Region:}

First we consider values of $\Delta m^2$, and thus $y$, in the perturbative regime where nonperturbative effects are power corrections. 
Here the endpoints of the collinear-soft evolution are $\Lambda_{{\rm cs}\,i}=\mu_{{\rm cs}\,i}$ given by the canonical $\Delta m_c$ dependent scales in \eqn{gscs-scales}. 
The dimensionlessness of the Laplace space CS functions implies
\begin{align}
\hat{\widetilde{S}}_{C_j} \pig( yQQ_{\text{cut}1}^{\frac{-1}{1+\beta_1}}, \beta_1, \mu_{{\rm cs}1}\pig) 
&= \hat{\widetilde{S}}_{C_j} \Bigg( \ln\frac{Q_{\text{cut}1}^{\frac{1}{1+\beta_1}}}
 {yQ\mu_{{\rm cs}1}^{\frac{2+\beta_1}{1+\beta_1}}}, \beta_1 ,\alpha_s(\mu_{{\rm cs}1})\Bigg) 
 \,, \\
\hat{\widetilde{D}}_{{C}_j} \pig( yQQ_{\text{cut}2}^{\frac{-1}{1+\beta_2}}, \beta_2, \mu_{{\rm cs}2} \pig) 
&= \hat{\widetilde{D}}_{{C}_j} \Bigg( \ln\frac{Q_{\text{cut}2}^{\frac{1}{1+\beta_2}}}
 {yQ\mu_{{\rm cs}2}^{\frac{2+\beta_2}{1+\beta_2}}}, \beta_2 ,
  \alpha_s(\mu_{{\rm cs}2}) \Bigg) 
 \,. \nn
\end{align}
Using
\be
f\Big(\ln\frac{1}{y}\Big) \Big(\frac{1}{y}\Big)^\eta = f\Big(\frac{\partial}{\partial\eta}\Big) \Big(\frac{1}{y}\Big)^\eta\,,
\ee
for the polynomial functions $f$ that show up in perturbation theory, we find
\begin{align} \label{eq:Pcd_shape_momentum}
& \hat{P}_j^{\text{CD}}(\Dms,\mu) 
 \\
&=
\exp\bigg( -2C_j\frac{2+\beta_1}{1+\beta_1} K(\mu_{\text{cs}1}, \mu) + \omega_{S_{C_j}}(\mu_{{\rm cs}1}, \mu) \bigg) \Bigg( \frac{Q_{\text{cut}1}^{\frac{1}{1+\beta_1}}}{ Q \mu^{\frac{2+\beta_1}{1+\beta_1}}_{\text{cs}1}} \Bigg)^{2C_j\omega(\mu_{{\rm cs}1}, \mu)} \nn\\
&\times \exp\bigg( +2C_j\frac{2+\beta_2}{1+\beta_2} K(\mu_{{\rm cs}2}, \mu) + \omega_{D_{{C}_j}}(\mu_{{\rm cs}2}, \mu) \bigg) \Bigg( \frac{Q_{\text{cut}2}^{\frac{1}{1+\beta_2}}}{ Q \mu^{\frac{2+\beta_2}{1+\beta_2}}_{\text{cs}2}} \Bigg)^{-2C_j\omega(\mu_{{\rm cs}2}, \mu)}
 \nn\\
&\times
\hat{\widetilde{S}}_{C_j} \Big( \frac{\partial}{\partial\eta}+\ln\frac{Q_{\text{cut}1}^{\frac{1}{1+\beta_1}}}{Q\mu_{{\rm cs}1}^{\frac{2+\beta_1}{1+\beta_1}}}, \beta_1 \Big)
\hat{\widetilde{D}}_{{C}_j} \Big( \frac{\partial}{\partial\eta}+\ln\frac{Q_{\text{cut}2}^{\frac{1}{1+\beta_2}}}{Q\mu_{{\rm cs}2}^{\frac{2+\beta_2}{1+\beta_2}}}, \beta_2 \Big) 
\nn\\
&\times \frac{e^{-\gamma_E}}{2\pi i} \int_{c-i\infty}^{c+i\infty} \diff y \,\exp \pig( y e^{-\gamma_E} \Delta m^2 \pig)
\Big(\frac{1}{y} \Big)^{\eta} \bigg|_{\eta = 2C_j \omega(\mu_{{\rm cs}1}, \mu_{{\rm cs}2})} 
\nn\\[4pt]
&=
\exp\bigg( -2C_j\frac{2+\beta_1}{1+\beta_1} K(\mu_{{\rm cs}1}, \mu) + \omega_{S_{C_j}}(\mu_{{\rm cs}1}, \mu) \bigg) \Bigg( \frac{Q_{\text{cut}1}^{\frac{1}{1+\beta_1}}}{ Q \mu^{\frac{2+\beta_1}{1+\beta_1}}_{\text{cs}1}} \Bigg)^{2C_j\omega(\mu_{{\rm cs}1}, \mu)} 
\nn\\
&\times \exp\bigg( +2C_j\frac{2+\beta_2}{1+\beta_2} K(\mu_{{\rm cs}2}, \mu) + \omega_{D_{{C}_j}}(\mu_{{\rm cs}2}, \mu) \bigg) \Bigg( \frac{Q_{\text{cut}2}^{\frac{1}{1+\beta_2}}}{ Q \mu^{\frac{2+\beta_2}{1+\beta_2}}_{\text{cs}2}} \Bigg)^{-2C_j\omega(\mu_{{\rm cs}2}, \mu)}
 \nn\\
&\times
\hat{\widetilde{S}}_{C_j} \Big( \frac{\partial}{\partial\eta}+\ln\frac{Q_{\text{cut}1}^{\frac{1}{1+\beta_1}}}{Q\mu_{{\rm cs}1}^{\frac{2+\beta_1}{1+\beta_1}}}, \beta_1 \Big)
\hat{\widetilde{D}}_{{C}_j} \Big( \frac{\partial}{\partial\eta}+\ln\frac{Q_{\text{cut}2}^{\frac{1}{1+\beta_2}}}{Q\mu_{{\rm cs}2}^{\frac{2+\beta_2}{1+\beta_2}}}, \beta_2 \Big)
\frac{e^{-\gamma_E\eta} (\Delta m^2)^{\eta-1}}{\Gamma(\eta)} \bigg|_{\eta = 2C_j \omega(\mu_{{\rm cs}1}, \mu_{{\rm cs}2})} 
\,. \nn
\end{align}
For the cumulative jet mass in collinear drop, we need to integrate $\Dms$ from 0 to $\Delta m_c^2$, which just replaces $\frac{(\Delta m^2)^{\eta-1}}{\Gamma(\eta)}$ with $\frac{(\Delta m^2_c)^{\eta}}{\Gamma(1+\eta)}$ in Eq.~(\ref{eq:Pcd_shape_momentum}) and gives $\Sigma = \sum_{j=q,g} f_j \hat\Sigma_j$ where 
\begin{align}
\hat \Sigma_j(\Delta m_c^2) &= 
\frac{1}{\sigma_j} 
S^{ee}_{G_j}(Q_{{\rm gs}1},\beta_1,\mu_{{\rm gs}1}) \exp\bigg( \frac{2C_j}{1+\beta_1} K(\mu_{{\rm gs}1},\mu) + \omega_{S_{G_j}}(\mu_{{\rm gs}1},\mu) \bigg)
\Big( \frac{\mu_{{\rm gs}1}}{Q_{{\rm gs}1}} \Big)^{\frac{2C_j}{1+\beta_1}\omega(\mu_{{\rm gs}1},\mu)} \nn\\ 
&\times S^{ee}_{\overline{G}_j}(Q_{{\rm gs}2},\beta_2,\mu_{{\rm gs}2}) \exp\bigg( \frac{-2C_j}{1+\beta_2} K(\mu_{{\rm gs}2},\mu) + \omega_{S_{\overline{G}_j}}(\mu_{{\rm gs}2},\mu) \bigg) \Big( \frac{\mu_{{\rm gs}2}}{Q_{{\rm gs}2}} \Big)^{\frac{-2C_j}{1+\beta_2}\omega(\mu_{{\rm gs}2},\mu)} \nn\\
&\times \exp\bigg( -2C_j\frac{2+\beta_1}{1+\beta_1} K(\mu_{{\rm cs}1}, \mu) + \omega_{S_{C_j}}(\mu_{{\rm cs}1}, \mu) \bigg) \Bigg( \frac{Q_{\text{cut}1}^{\frac{1}{1+\beta_1}}}{ Q \mu^{\frac{2+\beta_1}{1+\beta_1}}_{\text{cs}1}} \Bigg)^{2C_j\omega(\mu_{{\rm cs}1}, \mu)} \nn\\
&\times \exp\bigg( +2C_j\frac{2+\beta_2}{1+\beta_2} K(\mu_{{\rm cs}2}, \mu) + \omega_{D_{{C}_j}}(\mu_{{\rm cs}2}, \mu) \bigg) \Bigg( \frac{Q_{\text{cut}2}^{\frac{1}{1+\beta_2}}}{ Q \mu^{\frac{2+\beta_2}{1+\beta_2}}_{\text{cs}2}} \Bigg)^{-2C_j\omega(\mu_{{\rm cs}2}, \mu)} \nn\\
&\times 
\hat{\widetilde{S}}_{C_j} \bigg( \frac{\partial}{\partial\eta}+\ln\frac{Q_{\text{cut}1}^{\frac{1}{1+\beta_1}}}{Q\mu_{{\rm cs}1}^{\frac{2+\beta_1}{1+\beta_1}}}, \beta_1 \bigg)
\hat{\widetilde{D}}_{{C}_j} \bigg( \frac{\partial}{\partial\eta}+\ln\frac{Q_{\text{cut}2}^{\frac{1}{1+\beta_2}}}{Q\mu_{{\rm cs}2}^{\frac{2+\beta_2}{1+\beta_2}}}, \beta_2 \bigg) 
\nn\\
&\times
\frac{e^{-\gamma_E\eta} (\Delta m^2_c)^{\eta}}{\Gamma(1+\eta)} \bigg|_{\eta = 2C_j \omega(\mu_{{\rm cs}1}, \mu_{{\rm cs}2})}\,.
\end{align}
At LL and NLL accuracy, the boundary function contributions from the GS and CS functions can be set to be unity.  These results were used in the text in Sections~\ref{sec:resum} and~\ref{sec:cumulative}. 

\paragraph{Nonperturbative Region:} 
In the nonperturbative region it is more convenient to carry out the inverse Laplace transform in terms of the $\overline{\rm MS}$ shape functions in \eqn{Fmsbar1}.  It also turns out that it is easier to integrate over $\Dms$ before applying the inverse Laplace transform for the calculation of the cumulative jet mass. The relevant piece of the integration can be written as
\begin{align}
& \int_0^{\Delta m_c^2}\diff \Delta m^2\,
\frac{e^{-\gamma_E}}{2\pi i} \int_{c-i\infty}^{c+i\infty}\!\!\! \diff y \
 e^{ y e^{-\gamma_E} \Delta m^2 }\:
\widetilde{F}^{(\overline{\rm MS})j}_1 \pig( yQQ_{\text{cut}1}^{\frac{-1}{1+\beta_1}},\beta_1 \pig)\,
\widetilde{F}^{(\overline{\rm MS})j}_2 \pig( yQQ_{\text{cut}2}^{\frac{-1}{1+\beta_2}},\beta_2 \pig)\,
\Big(\frac{1}{y} \Big)^{\eta} \nn\\
&= \frac{e^{-\gamma_E}}{2\pi i} \int_{c-i\infty}^{c+i\infty} \frac{\diff y}{e^{-\gamma_E}} \Big( 
 e^{ y e^{-\gamma_E} \Delta m^2_c } - 1
\Big)
\widetilde{F}^{(\overline{\rm MS})j}_1 \pig( yQQ_{\text{cut}1}^{\frac{-1}{1+\beta_1}},\beta_1 \pig)
\widetilde{F}^{(\overline{\rm MS})j}_2 \pig( yQQ_{\text{cut}2}^{\frac{-1}{1+\beta_2}},\beta_2 \pig)
\Big(\frac{1}{y} \Big)^{1+\eta} \,.\nn\\
\end{align}
From the Laplace transform of the shape function \eqn{app_laplacep_shape}, the inverse Laplace transform is given by
\begin{align} \label{eq:LapFmsbar}
\frac{e^{-\gamma_E}}{2\pi i} \int_{c-i\infty}^{c+i\infty}  \diff y\, 
 \exp \pig( y e^{-\gamma_E} \Delta m^2_c \pig) 
  \widetilde{F}^{(\overline{\rm MS})j}_i \pig( yQQ_{\text{cut}\,i}^{\frac{-1}{1+\beta_i}},\beta_i \pig)  
  = F_i^{(\overline{\rm MS})j}(k_i,\beta_i) \bigg|_{k_i^{\frac{2+\beta_i}{1+\beta_i}}= Q^{-1}Q_{{\rm cut}\,i}^{\frac{1}{1+\beta_i}}\Delta m^2_c} \,.
\end{align}
The inverse Laplace transform of $1/y^{1+\eta}$ is given by
\be
\frac{e^{-\gamma_E}}{2\pi i} \int_{c-i\infty}^{c+i\infty}  \diff y\, \exp \pig( y e^{-\gamma_E} \Delta m^2_c \pig) \Big( \frac{1}{y} \Big)^{1+\eta} = e^{-(1+\eta)\gamma_E} \frac{(\Delta m^2_c)^\eta}{\Gamma(1+\eta)} \theta(\Delta m^2_c) \,.
\ee
Using the fact that the inverse Laplace transform of three functions multiplied is their convolution,
\bea
\ml{L}^{-1}\big(f(s)g(s)h(s)\big) = f*g*h(t) = \int_{-\infty}^{\infty} \diff \tau_1 \int_{-\infty}^{\infty} \diff \tau_2 \,
f(t-\tau_1-\tau_2)g(\tau_1)h(\tau_2) \,,
\eea
we have $\Sigma = \sum_{j=q,g} f_j \Sigma_j$ with 
\begin{align}
& \Sigma_{j}(\Delta m_c^2) = \frac{1}{\sigma_j} 
S^{ee}_{G_j}(Q_{{\rm gs}1},\beta_1,\mu_{{\rm gs}1}) \exp\bigg( \frac{2C_j}{1+\beta_1} K(\mu_{{\rm gs}1},\mu) + \omega_{S_{G_j}}(\mu_{{\rm gs}1},\mu) \bigg)
\Big( \frac{\mu_{{\rm gs}1}}{Q_{{\rm gs}1}} \Big)^{\frac{2C_j}{1+\beta_1}\omega(\mu_{{\rm gs}1},\mu)} \nn\\ 
&\times S^{ee}_{\overline{G}_j}(Q_{{\rm gs}2},\beta_2,\mu_{{\rm gs}2}) \exp\bigg( \frac{-2C_j}{1+\beta_2} K(\mu_{{\rm gs}2},\mu) + \omega_{S_{\overline{G}_j}}(\mu_{{\rm gs}2},\mu) \bigg) \Big( \frac{\mu_{{\rm gs}2}}{Q_{{\rm gs}2}} \Big)^{\frac{-2C_j}{1+\beta_2}\omega(\mu_{{\rm gs}2},\mu)} \nn\\
&\times \exp\bigg( -2C_j\frac{2+\beta_1}{1+\beta_1} K(\Lambda_{\rm cs}, \mu) + \omega_{S_{C_j}}(\Lambda_{\rm cs}, \mu) \bigg) \Bigg( \frac{Q_{\text{cut}1}^{\frac{1}{1+\beta_1}}}{ Q \Lambda_{\rm cs}^{\frac{2+\beta_1}{1+\beta_1}}} \Bigg)^{2C_j\omega(\Lambda_{\rm cs}, \mu)} \nn\\
&\times \exp\bigg( +2C_j\frac{2+\beta_2}{1+\beta_2} K(\Lambda_{\rm cs}, \mu) + \omega_{D_{{C}_j}}(\Lambda_{\rm cs}, \mu) \bigg) \Bigg( \frac{Q_{\text{cut}2}^{\frac{1}{1+\beta_2}}}{ Q \Lambda_{\rm cs}^{\frac{2+\beta_2}{1+\beta_2}}} \Bigg)^{-2C_j\omega(\Lambda_{\rm cs}, \mu)} \nn\\
&\times  \int_0^{\infty}\!\!\! \diff k_1 \int_0^{\infty}\!\!\! \diff k_2\, \theta\bigg(\Delta m_c^2 - QQ_{{\rm cut}1}^{-\frac{1}{1+\beta_1}} k_1^{\frac{2+\beta_1}{1+\beta_1}} - QQ_{{\rm cut}2}^{-\frac{1}{1+\beta_2}} k_2^{\frac{2+\beta_2}{1+\beta_2}}\bigg)  
F_1^{(\overline{\rm MS})j}(k_1,\beta_1) 
F_2^{(\overline{\rm MS})j}(k_2,\beta_2) \,,
\end{align}
where we have set $\Lambda_{{\rm cs}1} = \Lambda_{{\rm cs}2}= \Lambda_{\rm cs}$ and thus $\eta = 0$.  This result was used in \sec{NPshapefunctions}.

\section{Treating ISR and MPI in Large $R$ Jets}
\label{app:ISRlargeR}

\begin{figure}[t!]
    \begin{subfigure}[t]{0.49\textwidth}
        \centering
        \includegraphics[height=2.2in]{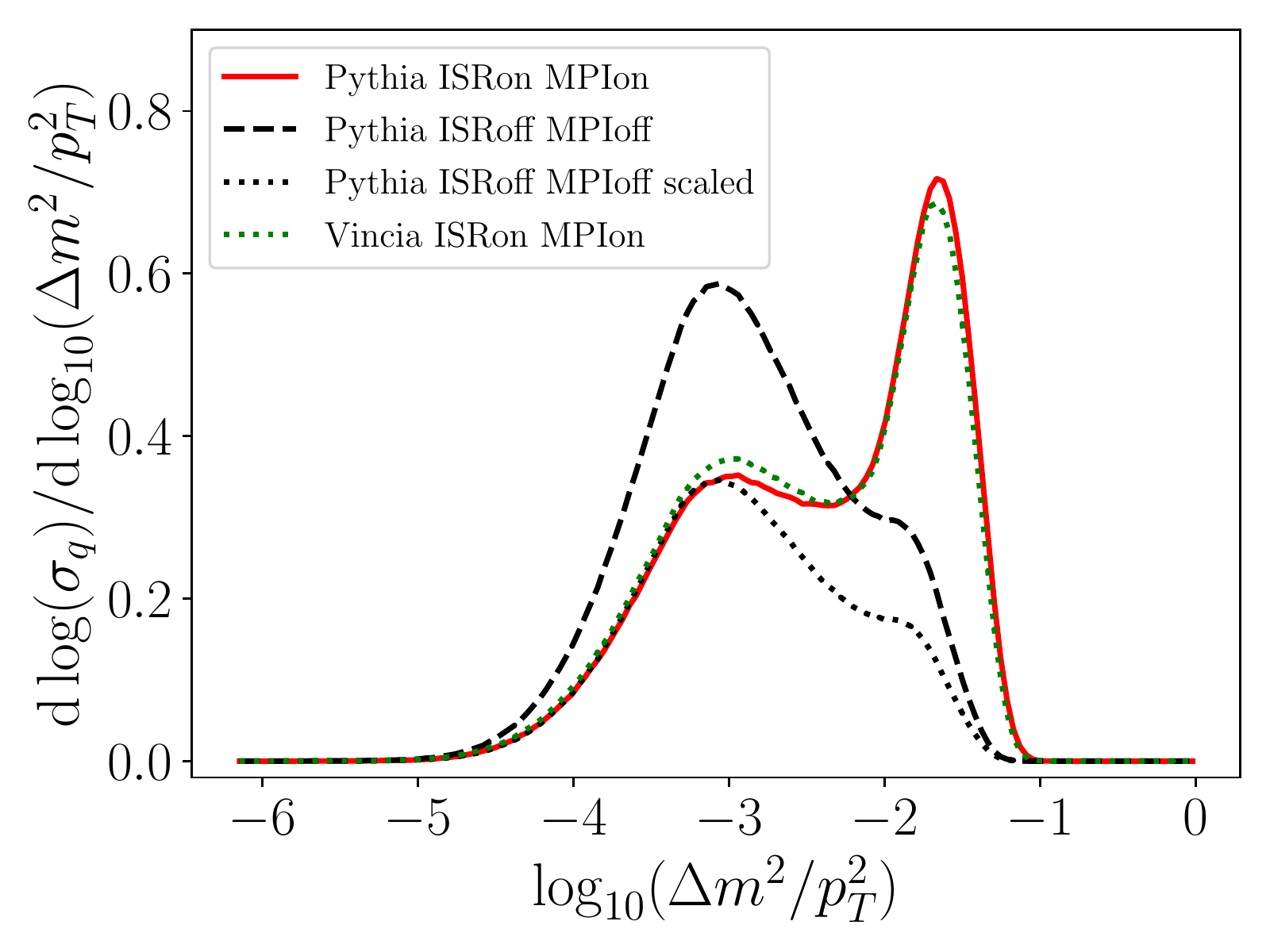}
        \caption{$\diff \sigma_q$ with $z_{{\rm cut}1}=0.02$, $\rho_q^{\rm ISR}=0.59$.}
        \label{fig:quark0.02}
    \end{subfigure}%
    ~
    \begin{subfigure}[t]{0.49\textwidth}
        \centering
        \includegraphics[height=2.2in]{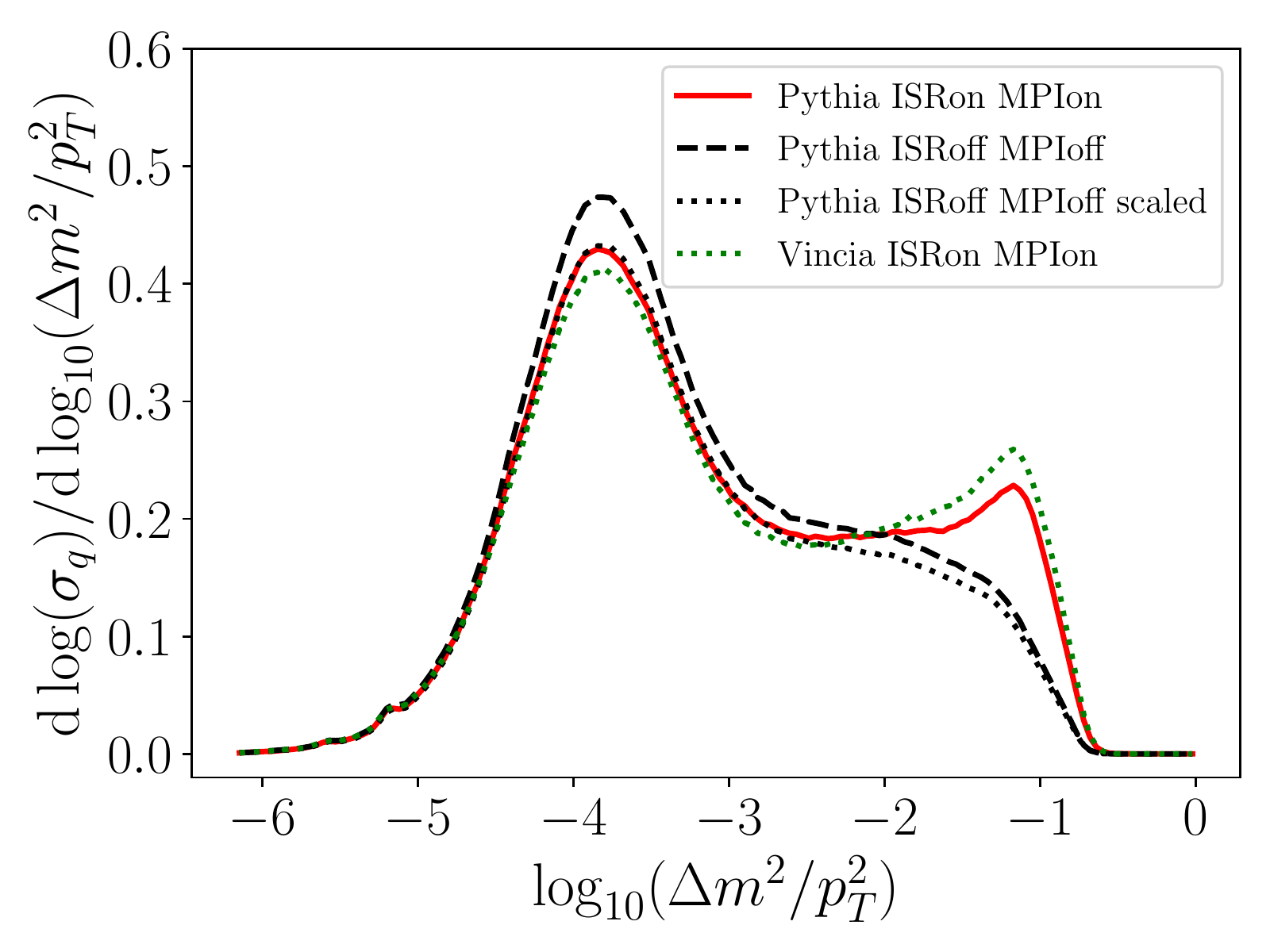}
        \caption{$\diff \sigma_q$ with $z_{{\rm cut}1}=0.1$, $\rho_q^{\rm ISR}=0.91$.}
        \label{fig:quark0.1}
    \end{subfigure}%
    
    \begin{subfigure}[t]{0.49\textwidth}
        \centering
        \includegraphics[height=2.2in]{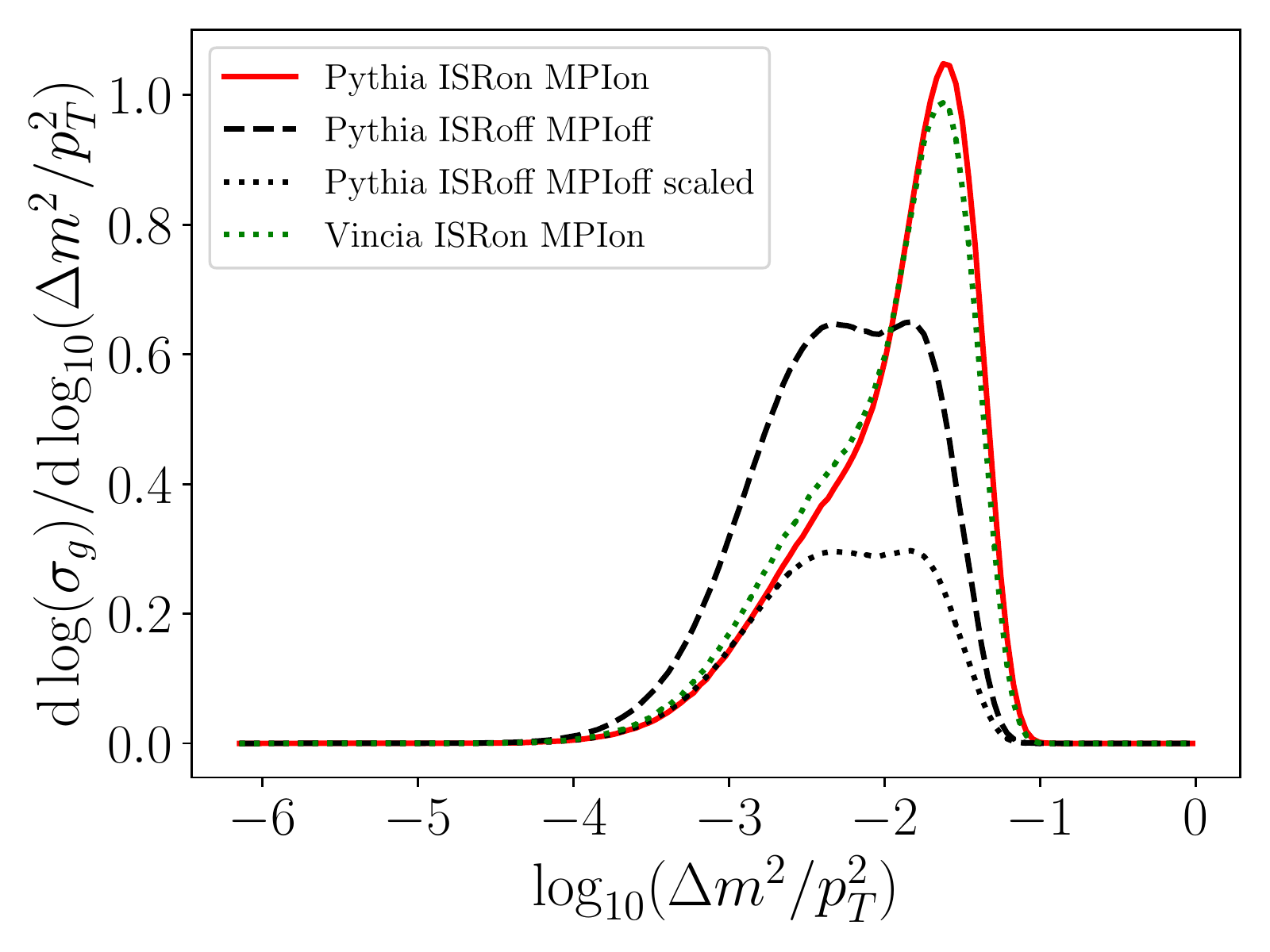}
        \caption{$\diff \sigma_g$ with $z_{{\rm cut}1}=0.02$, $\rho_g^{\rm ISR}=0.46$.}
        \label{fig:gluon0.02}
    \end{subfigure}%
    ~
    \begin{subfigure}[t]{0.49\textwidth}
        \centering
        \includegraphics[height=2.2in]{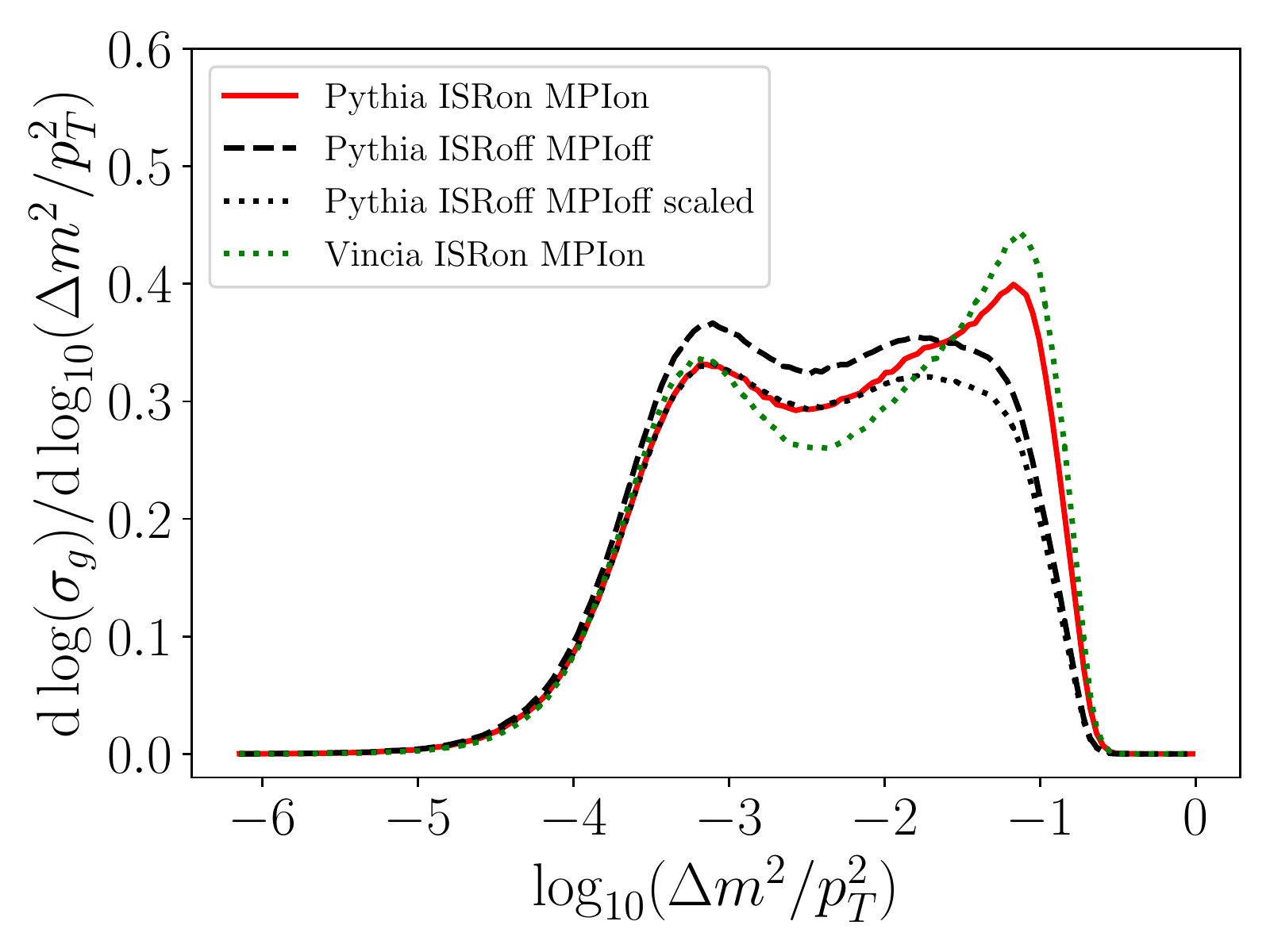}
        \caption{$\diff \sigma_g$ with $z_{{\rm cut}1}=0.1$, $\rho_g^{\rm ISR}=0.91$.}
        \label{fig:gluon0.1}
    \end{subfigure}%
\caption{Monte Carlo results of the differential jet mass cross sections of
quark jet (top row) and gluon jet (bottom row) in collinear drop with $\beta_1=\beta_2=0$, $z_{{\rm cut}1}=0.02,\,z_{{\rm cut}2}=0.08$ (left) and $z_{{\rm cut}1}=0.1,\,z_{{\rm cut}2}=0.4$ (right). The red solid and green dotted lines are results obtained by \Pythia and \Vincia generators respectively, with the ISR and MPI effects included. The black dashed line labels the \Pythia result without the ISR and MPI effects included and the black dotted line indicates the same results but multiplied by a scaling factor. These phenomenological factors are 0.59, 0.91, 0.46 and 0.91 for the four cases (a)--(d).
 }
\label{fig:zcut1_vs_ISR}
\end{figure}

As discussed in the main text, grooming most ISR and MPI effects away requires $z_{{\rm cut}1}^{(a/b)} \gtrsim 0.15$ for jets with large radii $R\sim1$. Such large values significantly reduce the discrimination power of the observables as can be seen from \fig{xi_j}, since well separated $z_{{\rm cut}1}$ and $z_{{\rm cut}2}$ lead to stronger discrimination power.  However, we observe that the ISR and MPI effects in the nonperturbative region can be accounted for by an overall phenomenological scaling factor $\rho_{j}^{\rm ISR}$ that depends on the collinear drop parameters, such that $\Sigma_j^{\rm ISR}(\Delta m_c) \simeq \rho_{j}^{\rm ISR}\,\Sigma_j^{\rm no\, ISR}(\Delta m_c)$. This implies that we can still obtain pure quark and gluon observables in the nonperturbative region in the presence of ISR and MPI effects, just with different values of the linear combination coefficients, e.g. $\xi_q \to \xi_q \,\rho_{(b)q}^{\rm ISR}/\rho_{(a)q}^{\rm ISR}$. (It turns out that the ISR effects dominate over those from MPI, so we only use a superscript ISR for simplicity even though both effects are included.)

To illustrate the idea introduced above, we consider jets with $p_T\in[280,320]$ GeV and $R=0.8$. 
In Fig.~\ref{fig:zcut1_vs_ISR}, we plot Monte Carlo results of the differential jet mass cross sections in collinear drop from \Pythia in three cases: with both ISR and MPI turned on, with both ISR and MPI turned off, and with both turned off but multiplied by an overall $\rho_j^{\rm ISR}$ factor. In addition, we also plot the results from the \Vincia Monte Carlo generator with both ISR and MPI turned on. Two sets of CD parameters are considered here: $\beta_1=\beta_2=0$, $z_{{\rm cut}1}=0.02$, $z_{{\rm cut}2}=0.08$ and $z_{{\rm cut}1}=0.1$, $z_{{\rm cut}2}=0.4$.
In both cases, we see that in the $\log_{10}(\Delta m^2/p_T^2) \leq -3$ region, the ISR and MPI effects are multiplicative. The values of these factors are $\rho_q^{\rm ISR}=0.59, 0.91$ and $\rho_g^{\rm ISR}= 0.46, 0.91$ for the four cases shown in Fig.~\ref{fig:zcut1_vs_ISR}, from left to right.
Experimentally, one could extract these $\rho_j^{\rm ISR}$ factors by comparing jet mass cross sections in the small jet mass region from electron-positron and proton-proton collisions, with the same CD grooming parameters.  

As shown in Fig.~\ref{fig:zcut1_vs_ISR}, even with a small $z_{{\rm cut}1}$ parameter such as $0.02$, the ISR and MPI effects are multiplicative in the kinematic region $\log_{10}(\Delta m^2/p_T^2) \leq -3$. This enables us to lift the third constraint listed in Section~\ref{sec:optimizing}. This multiplicative behavior in the nonperturbative region will not change the $\Delta m_c$ scaling behavior of the cumulative jet mass cross section, which is the most crucial part of our construction of the pure quark and gluon observables. If we focus on constructing pure quark and gluon observables in the kinematic region $\log_{10}(\Delta m_c^2/p_T^2) \leq -3$, the ISR and MPI effects only result in a multiplicative change of the linear combination coefficients:
\begin{align}
\label{eq:xi_isr}
\xi_q^{{\rm ISR}} = \xi_q^{{\rm w/o\ ISR}} \frac{\rho^{{\rm ISR}}_{(b)q}}{\rho^{{\rm ISR}}_{(a)q}} \,,
  \qquad\quad
\xi_g^{{\rm ISR}} = \xi_g^{{\rm w/o\ ISR}} \frac{\rho^{{\rm ISR}}_{(b)g}}{\rho^{{\rm ISR}}_{(a)g}} \,.
\end{align}
Then the quark and gluon contributions to the pure quark and gluon observables with the ISR and MPI effects are expected to have the form 
\begin{align}
\Sigma_{(b)q}^{\rm ISR}(\Delta m_c) - \xi_{g}^{\rm ISR}
\Sigma_{(a)q}^{\rm ISR}(\Delta m_c) \,,\qquad\quad
\Sigma_{(b)g}^{\rm ISR}(\Delta m_c) - \xi_{q}^{\rm ISR}
\Sigma_{(a)b}^{\rm ISR}(\Delta m_c)
\,,
\end{align}
for small $\Delta m_c^2$.

By comparing the \Pythia and \Vincia results of the differential jet mass cross sections, we see that they agree reasonably well in the deep nonperturbative region $\log_{10}(\Delta m^2/p_T^2) < -3$ but differ quite significantly in the intermediate and perturbative regions. This further emphasizes the need for a more dedicated study of the predictions of various Monte Carlo programs for the behavior of soft radiation in the perturbative region of the observables we exploit here.

\bibliography{cdrop}
\end{document}